\documentclass[table]{csmadapter}

\jvol{XX}
\jnum{XX}
\paper{X}
\jmonth{August}
\jname{IEEE CONTROL SYSTEMS}
\pubyear{2024}

\usepackage{todonotes}

\newcount\Comments  
\usepackage{color}
\definecolor{darkgreen}{rgb}{0,0.5,0}
\definecolor{purple}{rgb}{1,0,1}
\definecolor{teal}{rgb}{0,0.4627,0.5804}
\newcommand{\kibitz}[2]{\ifnum\Comments=1\textcolor{#1}{#2}\fi}

\usepackage{xspace}
\newcommand{\acronym}[1]{{#1}\xspace}

\newcommand{\sdbarfigfullwidth}[2]{%
    \end{multicols}
    \let\thefigure\thesfigure
    \renewcommand{\thesfigure}{S\arabic{sfigure}}%
    \sdbarfig{#1}{#2}
    \begin{multicols}{2}
}

\usepackage{amsmath}
\usepackage{bbm}
\usepackage[normalem]{ulem}
\usepackage{dsfont}
\usepackage{soul}
\usepackage[compress]{cite}
\usepackage{float}

\usepackage{tikz}
\usetikzlibrary{positioning}
\usetikzlibrary{arrows.meta,shadows,positioning}
\usetikzlibrary{shapes.multipart}
\tikzset{
  frame/.style={
    rectangle, draw,
    text width=6em, text centered,
    minimum height=4em,drop shadow,fill=white,
    rounded corners,
  },
  line/.style={
    draw, -{Latex},rounded corners=3mm,
  }
}

\begin{document}

\title{Reinforcement Learning Based Oscillation Dampening \stitle{Scaling up Single-Agent RL algorithms \break to a 100 AV highway field operational test}}

\author{
    KATHY JANG\textsuperscript{*,1},
    NATHAN LICHTLÉ\textsuperscript{*,\textdagger,1},
    EUGENE VINITSKY\textsuperscript{**,2},
    ADIT SHAH\textsuperscript{*,2},
    MATTHEW BUNTING\textsuperscript{\textdaggerdbl,3},
    MATTHEW NICE\textsuperscript{\textdaggerdbl,3},
    BENEDETTO PICCOLI\textsuperscript{\textdaggerdbl\textdaggerdbl},
    BENJAMIN SEIBOLD\textsuperscript{\textbardbl},
    DANIEL B.~WORK\textsuperscript{\textdaggerdbl},
    MARIA LAURA DELLE MONACHE\textsuperscript{\textsection},
    JONATHAN SPRINKLE\textsuperscript{\textdaggerdbl},
    JONATHAN W.~LEE\textsuperscript{*,\textbardbl\textbardbl},
    ALEXANDRE M.~BAYEN\textsuperscript{*, \textbardbl\textbardbl, \textsection}
    }
\affil{
    *--University of California, Berkeley, Department of Electrical Engineering and Computer Sciences\\
    **--University of California, Berkeley, Department of Mechanical Engineering\\
    \textsuperscript{\textsection}--University of California, Berkeley, Department of Civil and Environmental Engineering\\
    \textsuperscript{\textbardbl\textbardbl}-- University of California, Berkeley, Institute for Transportation Studies\\
    \textsuperscript{\textdagger}--\'Ecole des Ponts Paristech, Marne la Vallée\\
    \textsuperscript{\textdaggerdbl}--Vanderbilt University, Institute for Software Integrated Systems\\
    \textsuperscript{\textbardbl}--Temple University, Department of Mathematics\\
    \textsuperscript{\textdaggerdbl\textdaggerdbl}--Rutgers University-Camden, Department of Mathematical Sciences\\
    \textsuperscript{1}--These authors contributed equally\\
    \textsuperscript{2}--These authors contributed equally\\
    \textsuperscript{3}--These authors contributed equally\\
}

\maketitle




\dois{}{}

\begin{summary}

\summaryinitial{I}n this article, we explore the technical details of the reinforcement learning (RL) algorithms that were deployed in the largest field test of automated vehicles designed to smooth traffic flow in history as of 2023, uncovering the challenges and breakthroughs that come with developing RL controllers for automated vehicles. We delve into the fundamental concepts behind RL algorithms and their application in the context of self-driving cars, discussing the developmental process from simulation to deployment in detail, from designing simulators to reward function shaping. We present the results in both simulation and deployment, discussing the flow-smoothing benefits of the RL controller. From understanding the basics of Markov decision processes to exploring advanced techniques such as deep RL, our article offers a comprehensive overview and deep dive of the theoretical foundations and practical implementations driving this rapidly evolving field.

We also showcase real-world case studies and alternative research projects that highlight the impact of RL controllers in revolutionizing autonomous driving. From tackling complex urban environments to dealing with unpredictable traffic scenarios, these intelligent controllers are pushing the boundaries of what automated vehicles can achieve. 
Furthermore, we examine the safety considerations and hardware-focused technical details surrounding deployment of RL controllers into automated vehicles. As these algorithms learn and evolve through interactions with the environment, ensuring their behavior aligns with safety standards becomes crucial. We explore the methodologies and frameworks being developed to address these challenges, emphasizing the importance of building reliable control systems for automated vehicles.
\end{summary}

\section{Introduction}

\chapterinitial{A}s the automotive industry continues to evolve, the quest for safer and more efficient self-driving cars has become a top priority. To achieve this, engineers and researchers are turning to state-of-the-art techniques like reinforcement learning to create intelligent control systems that can navigate complex environments with unprecedented precision and adaptability. \acronym{Reinforcement learning} (RL), a subfield of machine learning, offers a promising avenue for training automated vehicles to make optimal decisions and actions in real-time scenarios. Unlike traditional rule-based approaches, RL-based controllers learn through trial and error, progressively refining their behavior based on feedback from the environment. This ability to adapt and improve over time makes RL an invaluable tool for enhancing the performance and reliability of automated vehicles. 

Together with the other articles presented in this special issue, this article constitutes an element of the technical work involved in bringing together objective of the CIRCLES project \cite{Hayat2022holistic}: the \acronym{MegaVanderTest} (MVT), the largest deployment of automated vehicles (AVs) designed to smooth traffic flow in history as of 2023. These AVs are partially automated and limited to longitudinal control. They do not communicate between each other, but communicate with a central server to get information about the downstream state of the highway. This article discusses the following: 

\begin{itemize}
    \item Background material, including a short overview of reinforcement learning, human driver models, and policy gradient algorithms.
    \item Problem formulation of the two classes of RL controllers designed for this experiment: (1) Acceleration-based control and (2) \acronym{Adaptive Cruise Control} (ACC)-based control.
    \item Description of the work moving from the software to hardware platform, deployment onto real roadways, and the MVT field test week.
    \item Result analysis of all controllers in simulation and in deployment
\end{itemize}


\begin{figure*}[b]
    \centering
    \includegraphics[width=\linewidth]{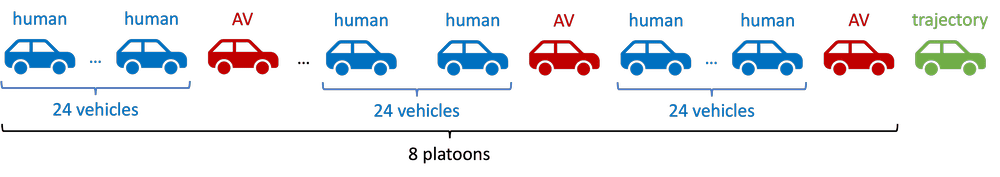}
    \caption{Example setup of the trajectory simulator for evaluation. One vehicle, which we name \emph{trajectory leader}, replays a velocity trajectory from the I-24 Trajectory Dataset \cite{nice2021dataset}. Following it are a combination of IDM-controlled human vehicles and automated vehicles (AVs) all on a single lane. This evaluation setup contains 8 platoons, each consisting of one AV followed by 24 human vehicles. The human vehicles are used to assess the smoothing performances of the AV. During training, only one platoon is simulated.}
    \label{fig:sim_platoon}
\end{figure*}

\section{Background}

\subsection{Reinforcement Learning}
We adhere to the conventional \acronym{reinforcement learning} (RL) framework \cite{sutton2018reinforcement, lillicrap2015continuous}, which seeks to optimize the discounted cumulative rewards within a finite time frame for a \acronym{Partially Observable Markov Decision Process} (POMDP) \cite{cassandra1998survey}. This POMDP can be officially characterized by the tuple $\mathcal{P} = (\mathcal{S},\mathcal{A} ,T ,R, T_0 ,\gamma, \Omega ,\mathcal{O})$, where $\mathcal{S}$ symbolizes the set of states; $\mathcal{A}$ stands for the possible actions; the function $T: \mathcal{S} \times \mathcal{A} \times \mathcal{S} \rightarrow \mathbb{R}$ represents the conditional likelihood of moving to a subsequent state $s'$ given the current state $s$ and the chosen action $a$; $R: \mathcal{S} \times \mathcal{A} \times \mathcal{S} \rightarrow \mathbb{R}$ is the reward function; $T_0 : S \to \mathbb{R}$ is the probability distribution of initial states; and $\gamma \in (0, 1]$ is the discount coefficient applied to the accumulation of rewards. The last two parameters, $\Omega$ and $\mathcal{O}$, are included due to the concealed nature of the state: $\Omega$ refers to the observations of the hidden state, and $\mathcal{O}: \Omega \times \mathcal{S} \rightarrow \mathbb{R}$ stands for the conditional observation probability distribution.

Given a POMDP, one can then sample an initial state $s_0 \sim T_0(\cdot)$ and initial observation of that state $o_0 \sim \mathcal{O}(\cdot \mid s_0)$. The agent, which we denote as a stochastic conditional distribution $\pi_\theta : \mathcal{S} \times \mathcal{A} \to \mathbb{R}$ parametrized by $\theta$, samples an action $a_0 \sim \pi_\theta(\cdot \mid s_0)$ from the observation. That action is then used to generate a next state $s_1 \sim T(\cdot \mid s_0, a_0)$ and observation of the state $o_1 \sim \mathcal{O}(\cdot \mid s_1)$, as well as a reward $r_0 = R(s_0, a_0, s_1)$. Iterating this process until termination (which can occur after a fixed time horizon or a certain termination condition) yields a trajectory $\tau = (s_i, o_i, a_i, r_i)_{i \geq 0}$. Let us define the return of a trajectory as the discounted sum of rewards $G(\tau) = \sum_{i \geq 0} \gamma^i r_i$. The goal for the agent is to maximize the expected return over all trajectories, and to learn an optimal policy $\pi^* = \arg\max_\pi \mathbb{E}_{\tau \sim (\pi, \mathcal{P})} G(\tau)$.
More aspects of RL are explored in sidebars ``\nameref{sidebar:rl}'' and ``\nameref{sidebar:policygradient}''.

\subsection{Human Driver Models}
\label{sec:human_driver_models}
In order to train RL controllers, we simulate mixed-autonomy traffic~\cite{toghi2021cooperative} where RL-controlled AVs interact with human-driven vehicles \cite{wu2017emergent, wu2021flow, vinitsky2018benchmarks}. In order to model human car-following behavior, we use the \acronym{Intelligent Driver Model} (IDM) \cite{kesting2010enhanced, Treiber2000} to govern a vehicle's longitudinal motion according to its leading vehicle. To ensure authenticity, IDM parameters are selected to mimic key non-equilibrium features of real-world traffic, particularly dynamic instabilities (phantom traffic jams) with realistic wave growth and propagating stop-and-go waves caused by human driving behavior \cite{wilson2011car, treiber2017intelligent} or even ACC \cite{gunter2020commercially}. To trigger dynamic instabilities that the model possesses for suitably congested densities, zero-mean Gaussian noise is added to the acceleration in each time step.
Additional safety measures are included to maintain 
acceleration bounds and to avoid collisions---note that pure IDM is mathematically collision-free, but discretized time-stepping and heterogeneous vehicle composition yields potential collision opportunities.

The IDM \cite{Treiber2000} prescribes the acceleration of vehicle $\alpha$ as a function of its space gap (bumper-to-bumper distance to its leader vehicle) $s_\alpha$; its speed $v_\alpha$; and relative speed with the leader, $\Delta v_\alpha$:
\begin{equation} \label{eq:idm}
a_{\text{IDM}} = \frac{dv_\alpha}{dt} = a \bigg[ 1 - \bigg( \frac{v_\alpha}{v_0} \bigg)^\delta - \bigg( \frac{s^*(v_\alpha,\Delta v_\alpha)}{s_\alpha} \bigg)^2 \bigg],
\end{equation}
where $s^*$ is the desired space gap and is given by:
\begin{equation} \label{eq:s_star}
s^*(v_\alpha,\Delta v_\alpha) = s_0 + \max \bigg( 0, v_\alpha T + \frac{v_\alpha \Delta v_\alpha}{2 \sqrt{ab}} \bigg),
\end{equation}
where $s_0$, $v_0$, $T$, $\delta$, $a$, and $b$ are known parameters. Parameter values are in Table~\ref{table:idm}.

\subsection{MegaController / Speed Planner}
\label{section:mega_speed}

The \emph{MegaController} (see \cite{lee2024traffic}) is the outermost framework that addresses the challenge of smoothing traffic flow despite a limited penetration rate of automated vehicles by introducing two key components: the algorithm and the speed planner. The \emph{Speed Planner}, discussed in further detail in its own standalone article \cite{wang2024hierarchical}, is a control framework for dynamic speed advisories in a traffic environment in which both automated and human-driven vehicles coexist. The speed planner is a centralized unit hosted on a server which incorporates various algorithms to handle computationally intensive tasks. The algorithms deployed on individual vehicles serve as operators, following the instructions provided by the centralized planner. The primary function of the Speed Planner is to design target speed profiles with the objective of minimizing vehicle energy consumption and maximizing overall traffic flow efficiency. The implemented Speed Planner is dependent on INRIX, a data source which provides the realtime average speed of road segments globally, including the segments of the \emph{Interstate 24} (I-24) that the RL algorithm is deployed on. INRIX data are typically represented in 500-800m segments, with high variability in segment size; approximately three minutes of latency; and is aggregated over all lanes across 60 seconds. Thus, while representing imperfect data, INRIX data provides a solid foundation upon which to develop prediction modules.

\begin{sidebar}{Short Overview of Reinforcement Learning}
\section[Overview of Reinforcement Learning]{By Kathy Jang and Nathan Lichtlé}
\label{sidebar:rl}
\renewcommand{\thestable}{S\arabic{stable}}
\renewcommand{\thesfigure}{S\arabic{sfigure}}
\emph{Reinforcement learning} has emerged as a powerful paradigm for enabling autonomous systems to learn and optimize their behavior through interaction with their environment. RL offers a unique approach to control system design by incorporating an agent's decision-making process and its impact on the environment. This sidebar provides a concise overview of RL, highlighting its key components and applications in the context of autonomous system control \cite{amari1998natural}.

RL revolves around the concept of an agent learning from its experiences in an environment. The agent takes actions based on its current state and receives feedback in the form of rewards or penalties, which reflect the desirability or undesirability of the agent's actions. By iteratively exploring the environment and adapting its actions based on the received rewards, the agent learns to optimize its behavior over time.

\sdbarfig{
\begin{tikzpicture}[font=\small\sffamily\bfseries,very thick,node distance = 4cm]
\node [frame] (agent) {Agent};
\node [frame, below=1.2cm of agent] (environment) {Environment};
\draw[line] (agent) -- ++ (2.5,0) |- (environment) 
node[right,pos=0.25,align=left] {action\\ $A_t$};
\coordinate[left=12mm of environment] (P);
\draw[thin,dashed] (P|-environment.north) -- (P|-environment.south);
\pgfmathsetmacro{\Ldist}{4mm}
\draw[line] ([yshift=-\Ldist]environment.west) -- 
([yshift=-\Ldist]environment.west -| P) node[midway,above]{$S_{i+1}$};
\draw[line,thick] ([yshift=\Ldist]environment.west) -- ([yshift=\Ldist]environment.west
-|P) node[midway,above]{$R_{i+1}$};
\draw[line] ([yshift=-\Ldist]environment.west -| P) -- ++ (-12mm-\Ldist,0) |- 
([yshift=\Ldist]agent.west) node[left, pos=0.25, align=right] {state\\ $S_t$};
\draw[line,thick] ([yshift=\Ldist]environment.west -| P) -- ++ (-12mm+\Ldist,0) 
|- ([yshift=-\Ldist]agent.west) node[right,pos=0.25,align=left] {reward\\ $R_t$};
\end{tikzpicture}
}{RL MDP depicting how an agent interacts with its environment via actions, observations, and rewards.\label{fig:rlmdp}}

At the core of RL lies the \emph{Markov Decision Process} (MDP) framework, which mathematically formalizes the problem of sequential decision-making under uncertainty. MDPs provide a principled representation of the agent's interaction with the environment, as illustrated in Figure~\ref{fig:rlmdp}, encapsulating the states, actions, transition dynamics, and rewards that govern the learning process. Algorithms such as Q-learning \cite{sd_watkins1992q}, SARSA \cite{sd_rummery1994qu}, and policy gradient methods \cite{sd_suttonpolicy} leverage MDPs to guide the agent's learning and decision-making processes.

The Bellman equation \cite{bellman1954theory}, a central element to understanding the structure of the RL problem, is given by the following: 
\begin{sequation}
    v_\pi(s) = \sum_{a \in \mathcal{A}(s)}\pi(a|s)\sum_{s' \in \mathcal{S}, r\in\mathcal{R}}p(s',r|s,a)(r + \gamma v_\pi(s'))
\end{sequation}
The Bellman equation is used to model the \textit{value function}, which describes the estimated value of a particular state. It operates under a dynamic programming paradigm, meaning that the estimate of the value function is calculated recursively. In practice, RL algorithms leverage other techniques like bootstrapping, function approximation or temporal difference learning such that the value of a state can be estimated without having to calculate all the way to the terminal state. 

One of the distinctive features of RL is its ability to handle complex, high-dimensional state and action spaces. Through the utilization of function approximation techniques, such as neural networks, RL algorithms can effectively learn representations of states and policies, enabling the control of systems with large state spaces or continuous actions \cite{sd_duan2016benchmarking}. Deep RL, which combines deep neural networks with RL, has garnered significant attention in recent years due to its ability to handle complex control tasks, such as autonomous driving and robotics \cite{sd_arulkumaran2017drl}.

RL has found diverse applications in the realm of autonomous systems, including automated vehicles, robotics, and game-playing agents. In the context of automated vehicles, RL controllers can learn to navigate complex traffic scenarios, adapt to changing road conditions, and optimize driving strategies based on safety and efficiency objectives \cite{sd_wu2017lflow}. By leveraging RL, autonomous systems can acquire adaptive and intelligent control policies, making them more capable of handling real-world challenges.

However, RL also presents certain challenges and considerations. The exploration-exploitation trade-off, sample efficiency, and generalization to new environments are areas that continue to be actively researched. Moreover, ensuring the safety, interpretability, and ethical behavior of RL-based controllers remain crucial concerns in their practical deployment.

In conclusion, RL provides a powerful paradigm for developing autonomous system control, enabling agents to learn and optimize their behavior through interaction with the environment. By leveraging the principles of RL, researchers and engineers can advance the capabilities of autonomous systems, paving the way for the realization of intelligent, adaptive, and efficient autonomous systems across various domains.

\end{sidebar}

\begin{sidebar}{Policy Gradient Algorithms}
\section[Policy Gradient Algorithms]{by Kathy Jang and Nathan Lichtlé}\label{sidebar:policygradient} \noindent
\renewcommand{\thestable}{S\arabic{stable}}
\renewcommand{\thesfigure}{S\arabic{sfigure}}

\sdbarinitial{P}\emph{olicy gradient algorithms} are a foundational class of RL methodologies. By directly optimizing the policy through the gradient of the expected, cumulative, discounted reward with respect to the policy parameters, these algorithms seek the optimal policy $\pi^*(a|s)$, which determines the action $a$ that maximizes the expected reward given a state $s$. The algorithm updates the policy parameters $\theta$, often represented as the parameters of a neural network in deep RL applications, which ascend the gradient of the expected reward. This process, illustrated in Figure~\ref{fig:policygradient}, is denoted as $\theta \gets \theta + \alpha \nabla_\theta J(\theta)$ where $\alpha$ is the learning rate, and $J(\theta)$ represents the expected reward under the policy parameterized by $\theta$.

\sdbarfig{\includegraphics[width=19.0pc]{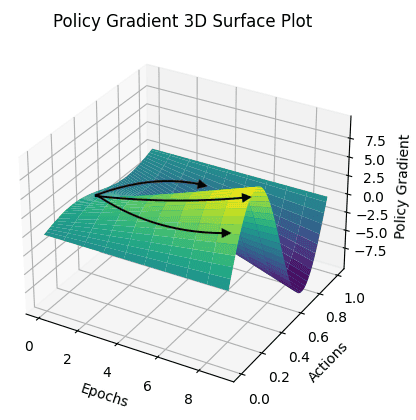}}{Policy Gradient Surface Plot: This 3D surface plot illustrates how policy gradients evolve over epochs and actions, providing insights into the optimization process of a reinforcement learning algorithm. The z-axis represents the magnitude of policy gradients, with colors indicating the strength of gradient values. As training progresses (left to right along the x-axis), the policy gradient landscape changes, guiding the agent's decision-making strategy in reinforcement learning tasks.\label{fig:policygradient}}

The gradient of this expected reward with respect to the policy parameters can be distilled into an expectation of the gradient of the log policy multiplied by the cumulative reward (the return). This can be written as 
\begin{sequation}\nabla_\theta J(\theta) = \mathbb{E}_{\pi_\theta}[\nabla_\theta \log \pi_\theta (a|s) G_t],\end{sequation}
 where $\pi_\theta(a|s)$ represents the policy parameterized by $\theta$, $G_t = \sum_{k=0}^{\infty} \gamma^k r_{t+k+1}$ is the return from the time step $t$, $r_t$ is the reward at time $t$ and $\gamma \in (0,1]$ is the discount factor. This equation forms the basis of the REINFORCE algorithm \cite{sd_williams1992reinforce}, a fundamental policy gradient method. Updates in this algorithm occur episodically, with the gradient ascent step typically implemented after each episode. However, more sophisticated policy gradient algorithms, such as Actor-Critic methods \cite{sd_grondman2012ac}, offer more frequent updates based on estimates of the expected return, which often accelerates convergence.

More advanced policy gradient algorithms introduce the concept of trust region. These algorithms aim to keep the policy update within a certain ``trust region'' to ensure stability and prevent harmful policy changes, as well as increase sample efficiency. They optimize a similar objective to the policy gradient methods, but add a constraint to limit the policy update.
This leads us to \emph{Proximal Policy Optimization} (PPO) \cite{sd_schulman2017proximal}, the RL algorithm employed in this work. PPO is a specific implementation of trust region methods that harmoniously combines state-of-the-art performances on standard benchmarks and complex control tasks with a significant ease in implementation and tuning. PPO aims to compute an update at each step that not only minimizes the cost function but also ensures a modest deviation from the previous policy, optimizing both implementation simplicity and computational efficiency.

\end{sidebar}

\begin{sidebar}{FLOW: the first integration of microsimulation and deep-RL on the cloud}
\section[FLOW]{by Kathy Jang and Nathan Lichtlé}\label{sbar-FLOW}

\renewcommand{\thestable}{S\arabic{stable}}
\renewcommand{\thesfigure}{S\arabic{sfigure}}


\sdbarinitial{F}LOW is a framework that addresses the challenges arising from the rapid growth of automated vehicles in ground traffic \cite{sd_wu2017flow}. FLOW seamlessly integrates deep RL techniques with the traffic microsimulator SUMO, providing a powerful platform for designing, implementing, and evaluating traffic control strategies. FLOW is open-source and free for download at \url{https://github.com/flow-project/flow}.

\subsection{Features of FLOW}

By leveraging advancements in deep RL, FLOW enables the utilization of RL methods, such as policy gradient, for effective traffic control in the presence of complex nonlinear dynamics. It offers a unique opportunity to benchmark the performance of classical controllers, including hand-designed ones, against learned policies or control laws.

\sdbarfig{
\includegraphics[width=9.0pc]{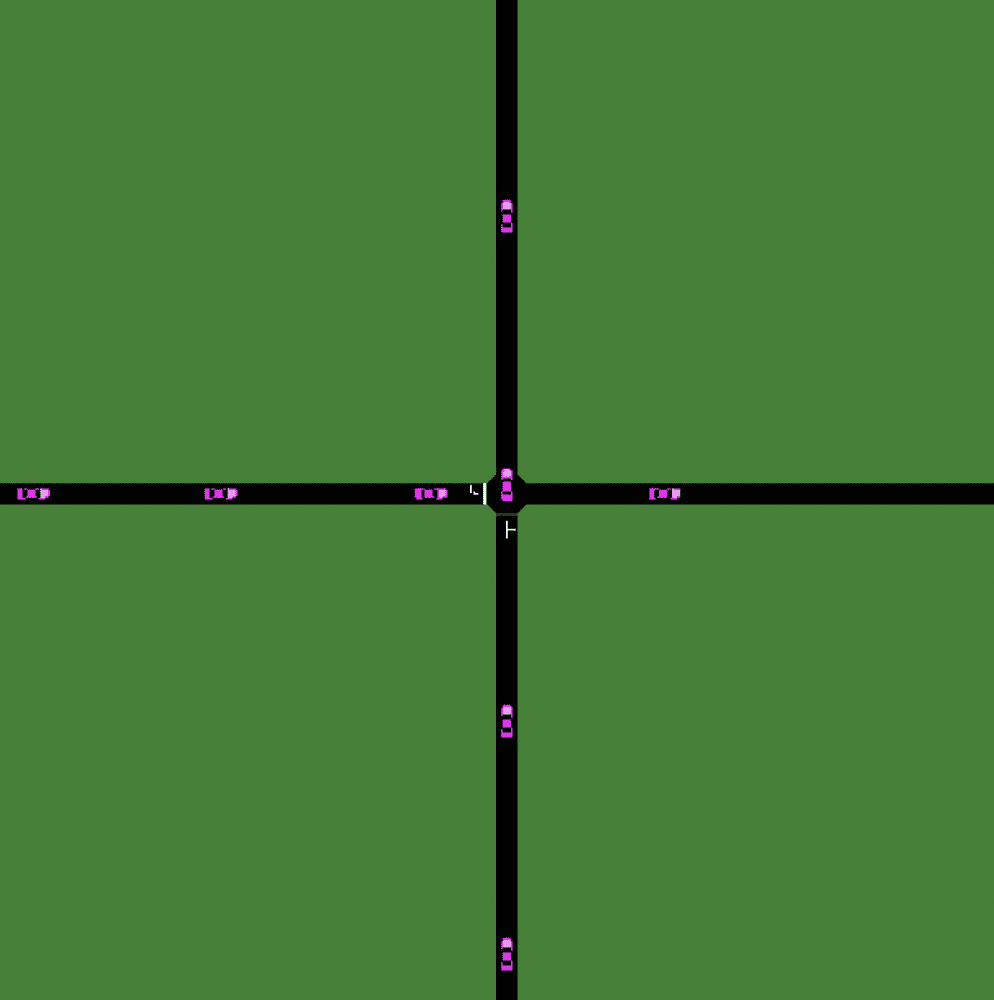}
\includegraphics[width=9.0pc]{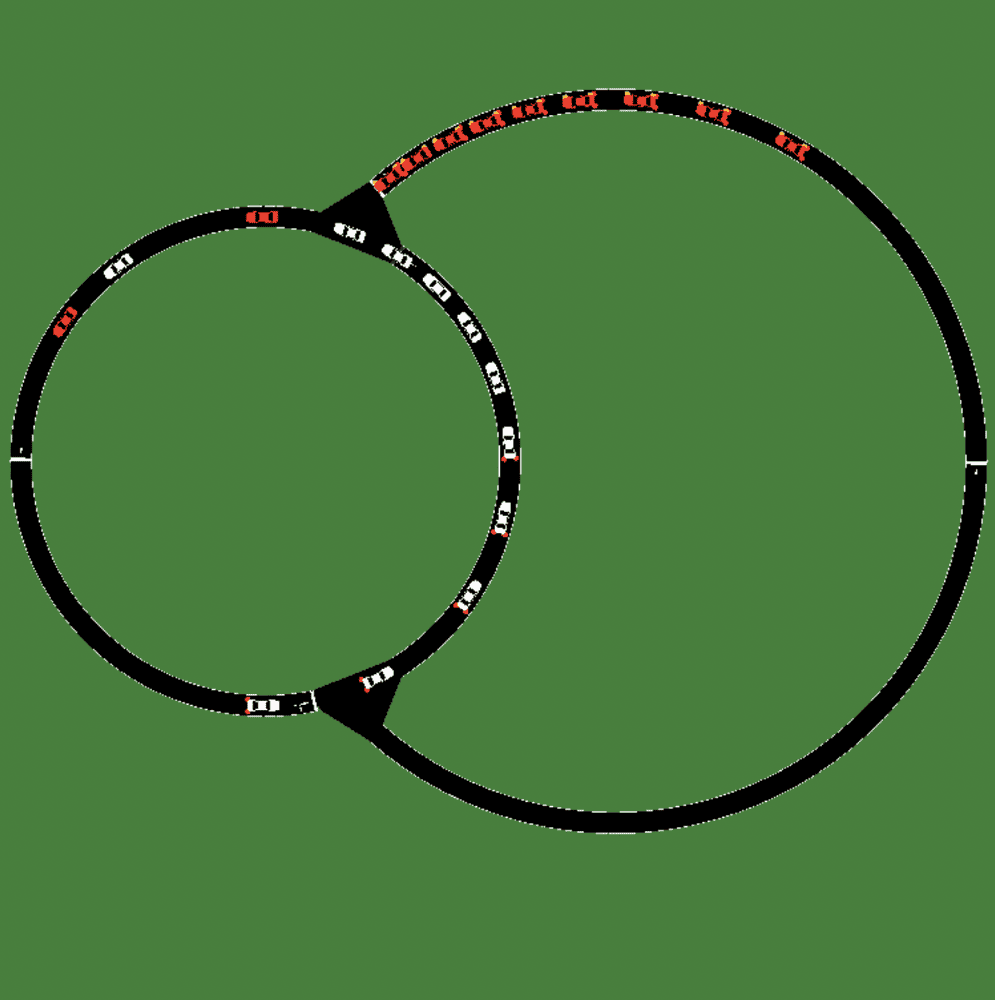}
}{Example of network building blocks that be configured and then trained via the FLOW architecture.\label{fig:flow_simple_networks}}
\sdbarfig{
\includegraphics[width=15.0pc]{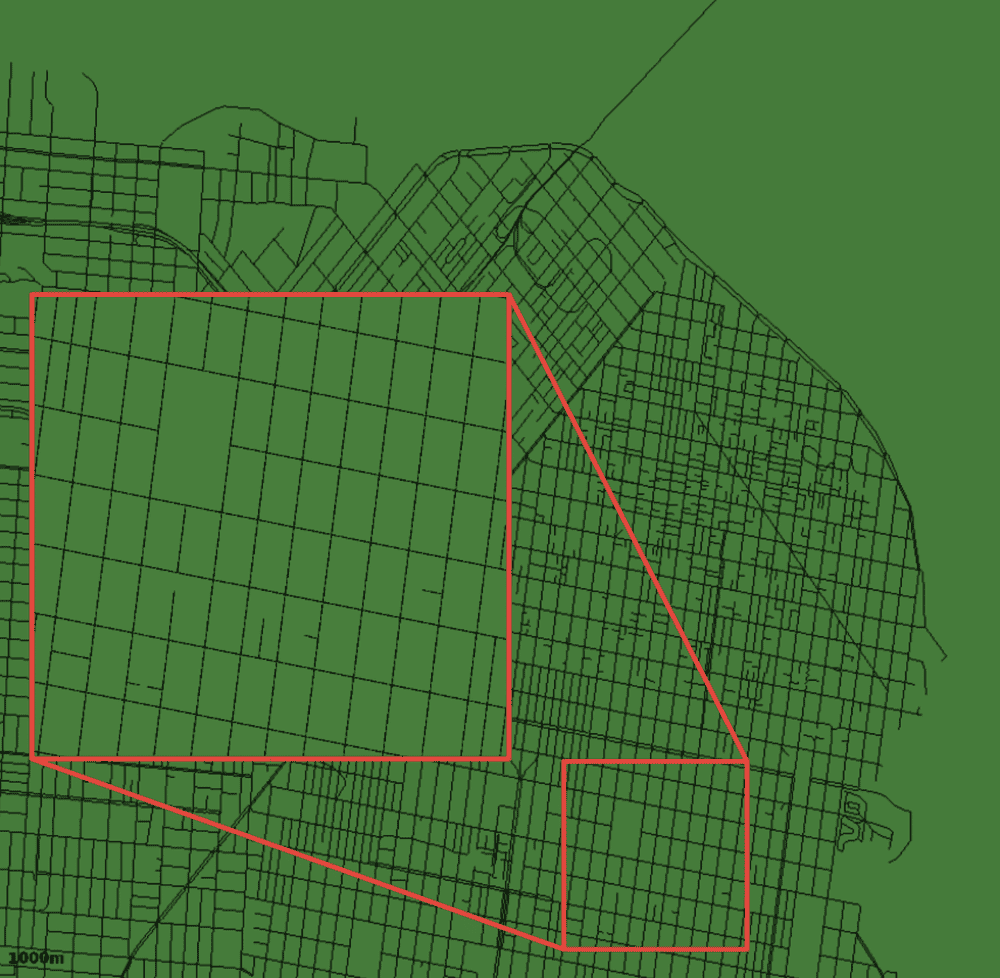}
}{San Francisco map loaded in FLOW from OpenStreetMap data.\label{fig:flow_sf_network}}


FLOW's integration with SUMO and Aimsun as well as a variety of RL libraries such as RLlib and Stable Baselines empowers researchers and practitioners to effortlessly design and simulate a wide range of traffic scenarios. This includes the flexibility to explore different network configurations and vehicle dynamics, allowing for comprehensive evaluation and optimization of traffic control strategies.

FLOW focuses on the challenging problem of controlling mixed-autonomy traffic, involving both automated and human-driven vehicles, in a variety of different traffic settings. FLOW implements some simple scenarios such as an intersection, a ring road, a figure eight, a merge, a bottleneck or a grid available, some of which can be seen in Figure~\ref{fig:flow_simple_networks}, while a complex network created from real map data is shown in Figure~\ref{fig:flow_sf_network}. Several benchmarks are available in \cite{sd_vinitsky2018benchmarks}.

\subsection{Traffic Control using FLOW}

\cite{sd_wu2017flow} trains simple neural network policies within the FLOW framework on a ring road scenario. Through experiments and analysis, they find that while state-of-the-art hand-designed controllers excel within specific conditions, they struggle to generalize to new situations. In contrast, even basic neural network policies trained using FLOW demonstrate remarkable performance in stabilizing traffic across various density settings and exhibit the ability to adapt to out-of-distribution scenarios. 

Similarly, \cite{sd_kreidieh2018dissipating} shows that a low proportion of RL-trained AVs is able to significantly dampen waves in closed and open single-lane networks. \cite{sd_jang2019simulation} trains AVs with a small artificial sensing noise in a roundabout scenario modeled in FLOW and directly transfers the trained policy to a real-world toy network. \cite{sd_vinitsky2020optimizing} creates a bottleneck model of the San Francisco-Oakland Bay Bridge and significantly improves throughput using a low penetration rate of RL-trained AVs. 

FLOW highlights the immense potential of RL in advancing traffic control research and its practical application in real-world autonomous vehicle systems. With its ability to seamlessly combine deep RL techniques with traffic micro-simulation, FLOW represents a significant step forward in developing reliable and adaptive controllers for complex traffic environments.
By providing a comprehensive and accessible computational framework, FLOW enables researchers and engineers to enhance traffic safety, efficiency, and overall control system performance in the era of automated vehicles.

\end{sidebar}

\begin{sidebar}{Wave-smoothing in micro-simulation}
\section[Wave-smoothing in micro-simulation]{by Nathan Lichtlé} \label{sbar-wavesmoothing}

\renewcommand{\thestable}{S\arabic{stable}}
\renewcommand{\thesfigure}{S\arabic{sfigure}}

\sdbarfig{\includegraphics[width=19.0pc]{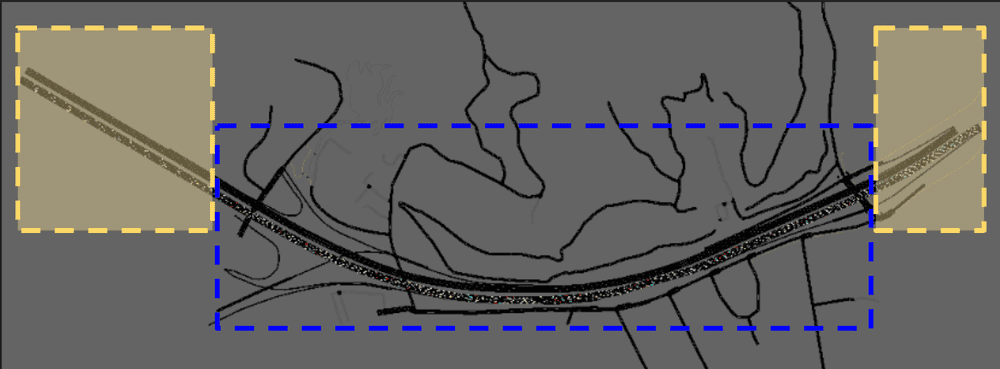}}{The I-210 network simulated within SUMO. The two yellow cells represent the uncontrolled region where AVs behave as humans, while the blue rectangle shows where control is applied and where metrics are computed. Besides, a speed limit is applied over the downstream yellow cell. \label{fig:i210_network}}

 \sdbarinitial{I}n this sidebar we demonstrate optimization of fuel economy in a large, calibrated model of a portion of the Ventura Freeway \cite{lichtlei210}. The location was one of the possible deployment locations of the 10 car test~\cite{lichtle2022deploying}, before the I-24 was chosen. It leverages the connected corridors model built for Caltrans D7 \cite{aimsun_i210_model}, which can be seen in Figure~\ref{fig:i210_network}.

We create synthetic waves in this scenario by introducing a low speed limit at the end of the highway portion. This creates a sharp slowdown which, coupled with human vehicles using a string-unstable car-following model, leads to growing stop-and-go waves that propagate upstream through the traffic. We then introduce into the network a small percentage of AVs, and learn a wave-dampening energy-improving controller using multi-agent RL. Each AV only observes its own speed, the speed of its leader and the space gap between the two, and is rewarded for minimizing energy consumption.

\sdbarfig{\includegraphics[width=18.0pc]{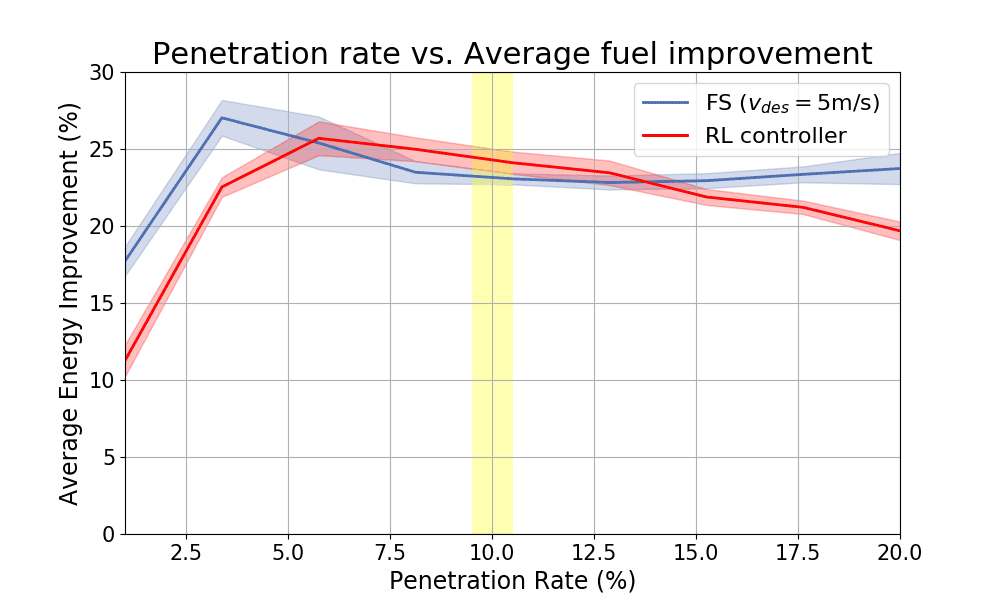}}{Fuel efficiency improvement of the RL controller at its training penetration rate of 10\% (highlighted in yellow) over the uncontrolled human baseline, and generalization to penetration rates outside that range. Fuel improvement is also shown for the FollowerStopper controller with a desired speed of 5m/s. The downstream speed is fixed at 5m/s.\label{fig:i210_sweep_penetration}}

As can be seen in Figure~\ref{fig:i210_sweep_penetration}, when the downstream speed limit is set to 5m/s and that 10\% of the vehicles are AVs, the control yields a 25\% reduction in average fuel consumption reduction for all the vehicles on the highway, and we shown that this does not reduce the throughput of the system compared to the uncontrolled case \cite{lichtlei210}.
Note that the controller was trained solely at a penetration rate of 10\% and a downstream speed limit of 5m/s, which corresponds to the highlighted yellow regions in Figs.~\ref{fig:i210_sweep_penetration} and \ref{fig:i210_envelope}. We compare the RL controller to an explicit \emph{FollowerStopper} (FS) controller (see ``\nameref{sbar-FS}"), which aims to drive smoothly and regulate at a desired speed $v_\text{des} = 5$m/s (equal to the downstream speed limit) while remaining safe. Figure~\ref{fig:i210_sweep_penetration} shows that the RL controller beats FS on its training domain and appears robust to changes in the penetration rate, improving energy efficiency significantly well outside its training domain.

Finally, we demonstrate robustness of the designed controller to the downstream speed limit. Figure~\ref{fig:i210_envelope} shows energy improvements when the RL controller, trained at a speed limit of 5m/s only, is evaluated on a wide range of speed limits outside the training set.
Impressively, the RL controller has learned to generalize: at each speed limit $v_\text{lim}$, it obtains similar performances to FS set with a desired speed $v_\text{des} = v_\text{lim}$. Importantly, the RL controller operates without knowing what the downstream speed limit is, as it only uses local observations, 
and thus does better than FS with less information. In summary, a single RL controller beats many control algorithms that each attempt to drive around the system equilibrium speed given knowledge of the speed limit. 



\sdbarfig{\includegraphics[width=18.0pc]{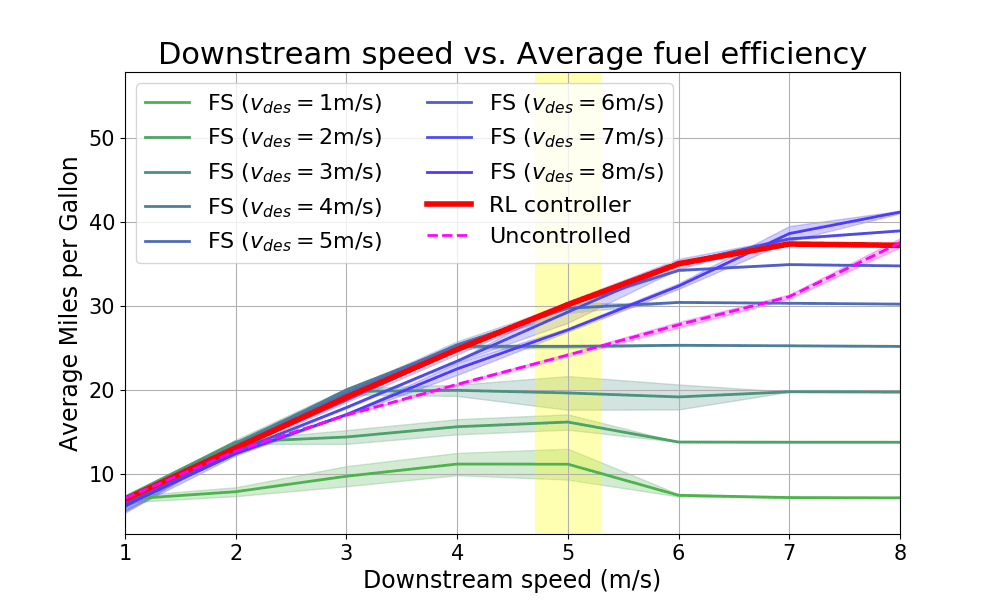}}{Average fuel efficiency of the RL controller on its training downstream speed of 5m/s (highlighted in yellow), and generalization to speeds outside that range. Miles per gallon fuel consumption is also shown for the FS controller with desired speed ranging from 1m/s to 8m/s and for the uncontrolled human baseline, as a function of the downstream speed. All plots are computed using a fixed penetration rate of 10\%. \label{fig:i210_envelope}}

\end{sidebar}

\begin{sidebar}{FollowerStopper}
\section[FollowerStopper]{by Nathan Lichtlé}\label{sbar-FS}

\sdbarinitial{F}\emph{ollowerStopper} (FS) is an explicit car-following control introduced in \cite{followerstopper}, which has demonstrated efficient wave-smoothing in experimental field tests. It outputs a velocity command $v_\text{cmd}$ as a function of ego speed $v$, leader speed $v_\ell$ and space gap $h$, and aims to drive at a desired speed $v_\text{des}$ which is a control parameter. However it may command a lower speed $v_\text{cmd} < v_\text{des}$ whenever safety requires it. The velocity command is given as a function of as follows:

\begin{sequation}
    v_\text{cmd} = \begin{cases}
        0 & \text{ if } \Delta x \leq \Delta x_1 \\
        v \frac{\Delta x - \Delta x_1}{\Delta x_2 - \Delta x_1} & \text{ if } \Delta x_1 < \Delta x \leq \Delta x_2 \\
        v + (U-v) \frac{\Delta x - \Delta x_2}{\Delta x_3 - \Delta x_2} & \text{ if } \Delta x_2 < \Delta x \leq \Delta x_3 \\
        v_\text{des} & \text{ if } \Delta x_3 < \Delta x
    \end{cases}
\end{sequation}

where the four cases respectively correspond to a stopping region where a speed of 0 is commanded, two adaptation regions where some weighted average of ego speed and leader speed is commanded, and a safe region where the desired speed is commanded, where the $\Delta x_i$'s dictate the transition thresholds between the 4 regions and depend on the difference between ego speed and leader speed:

\begin{sequation}
    \Delta x_k = \Delta x_k^0 + \frac{1}{2d_k} (\min (v_\ell - v, 0)^2, \; k=1,2,3.
\end{sequation}

In this work, we use the default parameters provided in \cite{followerstopper}: $\Delta x_1^0 = 4.5$m, $\Delta x_2^0=5.25$m, $\Delta x_3^0=6.0$m, $d_1 = 1.5 \frac{\text{m}}{\text{s}^2}$, $d_2=1.0\frac{\text{m}}{\text{s}^2}$ and $d_3=0.5\frac{\text{m}}{\text{s}^2}$.

\cite{followerstopper} demonstrates the smoothing effect of FS in a controlled experiment. 21 human drivers are instructed to drive on a circular road, invariably leading to the formation of stop-and-go waves. After some time, one of the vehicles on the ring road becomes controlled using FS, with different set speeds. The authors show that FS is able to successfully dampen the waves, reducing the speed variance by 50\% to 80\%, and consequently decreasing the fuel consumption of the vehicles by 20\% to 40\% using a single controlled vehicle.

\end{sidebar}

\section{Simulation / Problem Formulation}
\label{sec:problem-formulation}

In this article, we explore the avenues in which AVs can improve upon fuel economy for the entirety of the vehicles on the road, and not just for themselves. One effective focus to achieve this goal is targeting stop-and-go waves. Stop-and-go waves are a phenomenon in which high density traffic can cause vehicles to stop and restart without any apparent reason. This phenomena is commonly experienced in all kinds of driving scenarios, including highways and local roads. They are easy to form, as shown in a variety of literature and past experiments \cite{sugiyama2008traffic,cui2017stabilizing}. We use AVs as a tool in attempting to solve this problem by leveraging their understanding of human-driving and their access to nonlocal traffic information, thus improving driving efficiency for themselves and for surrounding vehicles. The success of this approach in contained scenarios has been documented in past work \cite{stern2018dissipation, Hayat2023theory}, in which a single AV can be shown to impact traffic conditions significantly, seeing improvements of up to 49\% in a ring road setting \cite{wu2017flowold}. 

Prior to the MVT, we conducted an experiment with 4 RL-controlled AVs deployed on the I-24 \cite{lichtle2022deploying}. Energy savings of about 11\%  were obtained in simulation, with AVs having the most impact at lower speeds, where waves are more likely to occur. Besides, AV trajectories and their corresponding leader trajectories were replayed in simulation, demonstrating a low sim-to-real gap and increased energy efficiency.

With the intention of deploying our AVs onto a four-mile stretch of I-24, instead of simulating the traffic scenario on a simulation remake of the I-24 (which would be both highly complex and time-inefficient), algorithms were trained using a data-driven, one-lane simulator introduced in \cite{lichtle2022deploying}; see also in ``\nameref{sbar-SIM}''.

\begin{sidebar}{Leveraging Field Operational Test Data to Build a Fast Traffic Simulator}
\section[Leveraging Field Operational Test Data to Build a Fast Traffic Simulator]{by Nathan Lichtlé and Kathy Jang}
\label{sbar-SIM}

\sdbarfigfullwidth{\includegraphics[width=38.0pc]{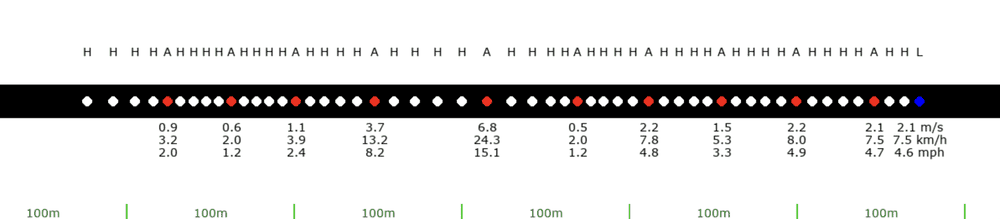}}{A render from the simulator. Each circle represents one vehicle. The leading blue vehicle (L) replays a trajectory from the dataset \cite{icradata} 
, the white vehicles (H) 
use IDM behavior 
and the red vehicles (A) are AVs. In this screenshot we can observe two large waves propagating through the traffic. \label{fig:sim_render}}

\noindent \sdbarinitial{C}\emph{reating} a robust and representative simulation environment is crucial for the successful implementation of wave-smoothing controllers in automated vehicles, ensuring their efficient operation across diverse traffic conditions. While we have previously employed large traffic micro-simulations in the past (see Sidebar~``\nameref{sbar-FLOW}'', based on SUMO), they require careful and extensive calibration of factors such as network parameters, road conditions, and both lane-changing and longitudinal driving models. This is essential to capture the intricate dynamics of the targeted highway, especially the nuances of wave dynamics. One other major limitation is that these simulations can be time-consuming, given the hundreds or thousands of vehicles interacting.

Prioritizing speed and efficiency for fast turnover in training new RL algorithms, we built a new simulator for CIRCLES. This simulator is built on real human driver data collected from the I-24 highway~\cite{nice2021dataset} \cite{icradata}, enabling it to mirror real-world, albeit simplified, wave dynamics. This approach, assuming the existence of trajectory data for the highway of interest, not only does not require any calibration but also allows for very fast simulations. Consequently, it has since been extensively leveraged for the design, training and evaluation of the different types of controls in CIRCLES.

\subsection{Data Acquisition}
Trajectory data is recorded~\cite{bunting2021libpanda} on a 14.5-kilometer segment of I-24, near Nashville, Tennessee, using an instrumented vehicle. The data includes vehicle \acronym{Controller Area Network} (CAN) data and GPS information, capturing speed, relative speed of the lead vehicle, acceleration, and space gap (distance from front bumper to rear bumper of leading vehicle).


\subsection{Dataset Analysis}
The trajectory dataset covers a wide range of traffic conditions, including stationary congested traffic and high-speed free-flow traffic, with diverse acceleration patterns. Training data retain both low and high-speed instances to ensure competent controller behavior during transitions. An example dataset trajectory 
is displayed in Figure~\ref{fig:example_dataset_traj_7050}.

\sdbarfig{\includegraphics[width=19.0pc]{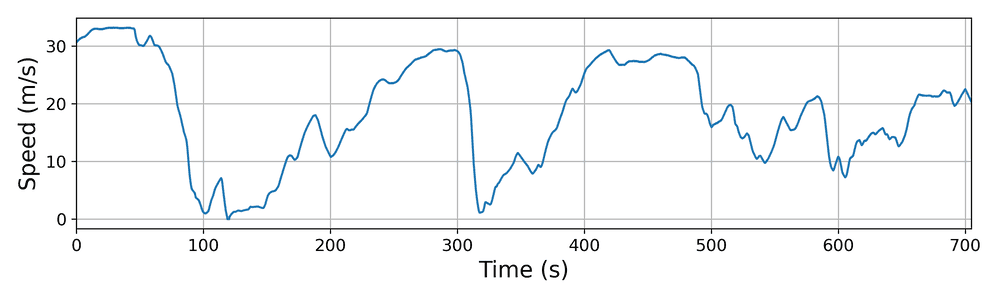}}{Speed vs. time for one of the dataset trajectories (corresponding to the ego trajectory of one of our drivers), exhibiting large acceleration and breaking patterns that can typically lead to stop-and-go waves.\label{fig:example_dataset_traj_7050}}


\subsection{Training and Evaluation Framework}
The simulation employs a single-lane environment where a mixed platoon of \emph{human} vehicles and \acronym{automated vehicles} (AVs) follow a real-world trajectory from the dataset. A render from the simulator can be seen in Figure~\ref{fig:sim_render}. Human vehicles use the \acronym{Intelligent Driver Model} (IDM) (see Sec.~``\nameref{sec:human_driver_models}'') with string-unstable parameters to ensure realistic driving dynamics and wave propagation in congestion. The simulator achieves high efficiency, allowing faster training compared to micro-simulation, and sidesteps the need for extensive calibration of large micro-simulations. A data-based lane-changing model was also developed for the simulator, allowing for assessing the robustness of the different controls to lane-changes.

\end{sidebar}

A simulation environment was established based on human driver data~\cite{nice2021dataset} gathered on a segment of the \mbox{I-24} highway southeast of Nashville, Tennessee. This data, collected from an instrumented vehicle, included details such as vehicle speed, leader speed, instantaneous acceleration, and space gap. The collected data varied in factors such as the time of day, day of the week, direction of travel, and levels of traffic congestion. \acronym{Controller Area Network} (CAN) data and GPS data from the drives are stored in csv files. The data were then refined by extracting relevant data from the CAN file~\cite{bhadani2022strym}, adjusting it to match the GPS time, and downsampling and interpolating it to align with the GPS time. The resulting dataset~\cite{nice2021dataset} encapsulates a wide range of traffic conditions from stationary congested traffic to maximum speed free-flow traffic, including many stop-and-go driving patterns. An example speed profile of such a trajectory is shown in Sidebar~``\nameref{sbar-SIM}''. This allows us to simulate realistic driving dynamics while avoiding the computational cost of a full micro-simulation of the highway. The training and evaluation framework was developed using a single-lane training environment where an AV and human-driven vehicles follow the trajectory data recorded from human drivers, and as shown in \cite{lichtle2022deploying}, the simulation to reality gap is minimal. Human-driven vehicles in the simulation are modeled using IDM \cite{albeaik2022limitations}, using string-unstable hyperparameters so that the vehicles in the platoon are able to propagate the wave patterns that the AV attempts to smooth out. Figure~\ref{fig:sim_platoon} shows an example setup of the simulation. Furthermore, we equip the simulator with a data-driven lane-changing model in order to assess the robustness of the controls to cut-ins and cut-outs. Some trajectories were used for training, different ones were kept for evaluation, and real-world deployment served as the test set. 

\begin{table*}[]
\setlength{\tabcolsep}{5.25pt} 
\begin{tabular}{|l|cccccc|cccccc|}
\hline
\textbf{}              & \multicolumn{6}{c|}{\textbf{Speed Planner A}}                                                                                                                                                                                            & \multicolumn{6}{c|}{\textbf{Speed Planner B}}                                                                                                                                                                                            \\ \hline
\textbf{BOTTLENECK}    & \multicolumn{2}{c|}{Fuel Economy}                                                 & \multicolumn{2}{c|}{Throughput}                                                      & \multicolumn{2}{c|}{Speed}                                    & \multicolumn{2}{c|}{Fuel Economy}                                                 & \multicolumn{2}{c|}{Throughput}                                                      & \multicolumn{2}{c|}{Speed}                                    \\ \hline
\textbf{}              & \multicolumn{1}{c|}{Val}   & \multicolumn{1}{c|}{\cellcolor[HTML]{FFFFFF}Pct}     & \multicolumn{1}{c|}{Val}      & \multicolumn{1}{c|}{\cellcolor[HTML]{FFFFFF}Pct}     & \multicolumn{1}{c|}{Val}   & \cellcolor[HTML]{FFFFFF}Pct      & \multicolumn{1}{c|}{Val}   & \multicolumn{1}{c|}{\cellcolor[HTML]{FFFFFF}Pct}     & \multicolumn{1}{c|}{Val}      & \multicolumn{1}{c|}{\cellcolor[HTML]{FFFFFF}Pct}     & \multicolumn{1}{c|}{Val}   & \cellcolor[HTML]{FFFFFF}Pct      \\ \hline
\textbf{Baseline}      & \multicolumn{1}{c|}{33.61} & \multicolumn{1}{c|}{\cellcolor[HTML]{FFFFFF}0.00\%}  & \multicolumn{1}{c|}{524.75}   & \multicolumn{1}{c|}{\cellcolor[HTML]{FFFFFF}0.00\%}  & \multicolumn{1}{c|}{11.64} & \cellcolor[HTML]{FFFFFF}0.00\%   & \multicolumn{1}{c|}{33.61} & \multicolumn{1}{c|}{\cellcolor[HTML]{FFFFFF}0.00\%}  & \multicolumn{1}{c|}{524.75}   & \multicolumn{1}{c|}{\cellcolor[HTML]{FFFFFF}0.00\%}  & \multicolumn{1}{c|}{11.64} & \cellcolor[HTML]{FFFFFF}0.00\%   \\ \hline
\textbf{Microaccel A}  & \multicolumn{1}{c|}{35}    & \multicolumn{1}{c|}{\cellcolor[HTML]{BDE4D1}3.97\%}  & \multicolumn{1}{c|}{593.81}   & \multicolumn{1}{c|}{\cellcolor[HTML]{57BB8A}13.16\%} & \multicolumn{1}{c|}{12.09} & \cellcolor[HTML]{BFE5D2}3.87\%   & \multicolumn{1}{c|}{35.34} & \multicolumn{1}{c|}{\cellcolor[HTML]{ADDEC6}4.90\%}  & \multicolumn{1}{c|}{591.78}   & \multicolumn{1}{c|}{\cellcolor[HTML]{57BB8A}12.77\%} & \multicolumn{1}{c|}{12.13} & \cellcolor[HTML]{B9E3CE}4.21\%   \\ \hline
\textbf{Microaccel B}  & \multicolumn{1}{c|}{33.41} & \multicolumn{1}{c|}{\cellcolor[HTML]{FDF7F6}-0.60\%} & \multicolumn{1}{c|}{593.81}   & \multicolumn{1}{c|}{\cellcolor[HTML]{57BB8A}13.16\%} & \multicolumn{1}{c|}{12.57} & \cellcolor[HTML]{79C9A2}7.99\%   & \multicolumn{1}{c|}{34.23} & \multicolumn{1}{c|}{\cellcolor[HTML]{E1F3EA}1.81\%}  & \multicolumn{1}{c|}{595.86}   & \multicolumn{1}{c|}{\cellcolor[HTML]{57BB8A}13.55\%} & \multicolumn{1}{c|}{12.59} & \cellcolor[HTML]{76C8A0}8.16\%   \\ \hline
\textbf{Microaccel C}  & \multicolumn{1}{c|}{35}    & \multicolumn{1}{c|}{\cellcolor[HTML]{BDE4D1}3.97\%}  & \multicolumn{1}{c|}{593.81}   & \multicolumn{1}{c|}{\cellcolor[HTML]{57BB8A}13.16\%} & \multicolumn{1}{c|}{12.09} & \cellcolor[HTML]{BFE5D2}3.87\%   & \multicolumn{1}{c|}{35.34} & \multicolumn{1}{c|}{\cellcolor[HTML]{ADDEC6}4.90\%}  & \multicolumn{1}{c|}{591.78}   & \multicolumn{1}{c|}{\cellcolor[HTML]{57BB8A}12.77\%} & \multicolumn{1}{c|}{12.13} & \cellcolor[HTML]{B9E3CE}4.21\%   \\ \hline
\textbf{Microaccel D}  & \multicolumn{1}{c|}{35}    & \multicolumn{1}{c|}{\cellcolor[HTML]{BDE4D1}3.97\%}  & \multicolumn{1}{c|}{593.81}   & \multicolumn{1}{c|}{\cellcolor[HTML]{57BB8A}13.16\%} & \multicolumn{1}{c|}{12.09} & \cellcolor[HTML]{BFE5D2}3.87\%   & \multicolumn{1}{c|}{35.34} & \multicolumn{1}{c|}{\cellcolor[HTML]{ADDEC6}4.90\%}  & \multicolumn{1}{c|}{591.78}   & \multicolumn{1}{c|}{\cellcolor[HTML]{57BB8A}12.77\%} & \multicolumn{1}{c|}{12.13} & \cellcolor[HTML]{B9E3CE}4.21\%   \\ \hline
\textbf{Microaccel E}  & \multicolumn{1}{c|}{35}    & \multicolumn{1}{c|}{\cellcolor[HTML]{BDE4D1}3.97\%}  & \multicolumn{1}{c|}{593.81}   & \multicolumn{1}{c|}{\cellcolor[HTML]{57BB8A}13.16\%} & \multicolumn{1}{c|}{12.09} & \cellcolor[HTML]{BFE5D2}3.87\%   & \multicolumn{1}{c|}{35.34} & \multicolumn{1}{c|}{\cellcolor[HTML]{ADDEC6}4.90\%}  & \multicolumn{1}{c|}{591.78}   & \multicolumn{1}{c|}{\cellcolor[HTML]{57BB8A}12.77\%} & \multicolumn{1}{c|}{12.13} & \cellcolor[HTML]{B9E3CE}4.21\%   \\ \hline
\textbf{Microaccel F}  & \multicolumn{1}{c|}{33.98} & \multicolumn{1}{c|}{\cellcolor[HTML]{EDF8F3}1.09\%}  & \multicolumn{1}{c|}{589.76}   & \multicolumn{1}{c|}{\cellcolor[HTML]{57BB8A}12.39\%} & \multicolumn{1}{c|}{12.01} & \cellcolor[HTML]{CAEADA}3.18\%   & \multicolumn{1}{c|}{34.04} & \multicolumn{1}{c|}{\cellcolor[HTML]{EAF7F1}1.26\%}  & \multicolumn{1}{c|}{593.81}   & \multicolumn{1}{c|}{\cellcolor[HTML]{57BB8A}13.16\%} & \multicolumn{1}{c|}{12.03} & \cellcolor[HTML]{C7E9D8}3.35\%   \\ \hline
\textbf{RL A}          & \multicolumn{1}{c|}{34.81} & \multicolumn{1}{c|}{\cellcolor[HTML]{C6E8D7}3.45\%}  & \multicolumn{1}{c|}{573.56}   & \multicolumn{1}{c|}{\cellcolor[HTML]{63C093}9.30\%}  & \multicolumn{1}{c|}{12.18} & \cellcolor[HTML]{B2E0C9}4.64\%   & \multicolumn{1}{c|}{35.1}  & \multicolumn{1}{c|}{\cellcolor[HTML]{B8E3CE}4.25\%}  & \multicolumn{1}{c|}{573.56}   & \multicolumn{1}{c|}{\cellcolor[HTML]{63C093}9.30\%}  & \multicolumn{1}{c|}{12.16} & \cellcolor[HTML]{B4E1CB}4.47\%   \\ \hline
\textbf{RL B}          & \multicolumn{1}{c|}{34.78} & \multicolumn{1}{c|}{\cellcolor[HTML]{C7E9D8}3.36\%}  & \multicolumn{1}{c|}{573.56}   & \multicolumn{1}{c|}{\cellcolor[HTML]{63C093}9.30\%}  & \multicolumn{1}{c|}{12.18} & \cellcolor[HTML]{B2E0C9}4.64\%   & \multicolumn{1}{c|}{34.95} & \multicolumn{1}{c|}{\cellcolor[HTML]{BFE5D3}3.83\%}  & \multicolumn{1}{c|}{573.56}   & \multicolumn{1}{c|}{\cellcolor[HTML]{63C093}9.30\%}  & \multicolumn{1}{c|}{12.17} & \cellcolor[HTML]{B3E1CA}4.55\%   \\ \hline
\textbf{RL C}          & \multicolumn{1}{c|}{33.97} & \multicolumn{1}{c|}{\cellcolor[HTML]{EEF8F3}1.06\%}  & \multicolumn{1}{c|}{571.03}   & \multicolumn{1}{c|}{\cellcolor[HTML]{6BC498}8.82\%}  & \multicolumn{1}{c|}{12.2}  & \cellcolor[HTML]{AFDFC7}4.81\%   & \multicolumn{1}{c|}{33.92} & \multicolumn{1}{c|}{\cellcolor[HTML]{F0F9F5}0.91\%}  & \multicolumn{1}{c|}{591.78}   & \multicolumn{1}{c|}{\cellcolor[HTML]{57BB8A}12.77\%} & \multicolumn{1}{c|}{12.08} & \cellcolor[HTML]{C0E6D3}3.78\%   \\ \hline
\textbf{High-speed RL} & \multicolumn{1}{c|}{33.91} & \multicolumn{1}{c|}{\cellcolor[HTML]{F1F9F5}0.88\%}  & \multicolumn{1}{c|}{593.81}   & \multicolumn{1}{c|}{\cellcolor[HTML]{57BB8A}13.16\%} & \multicolumn{1}{c|}{12.01} & \cellcolor[HTML]{CAEADA}3.18\%   & \multicolumn{1}{c|}{33.91} & \multicolumn{1}{c|}{\cellcolor[HTML]{F1F9F5}0.88\%}  & \multicolumn{1}{c|}{593.81}   & \multicolumn{1}{c|}{\cellcolor[HTML]{57BB8A}13.16\%} & \multicolumn{1}{c|}{12.01} & \cellcolor[HTML]{CAEADA}3.18\%   \\ \hline
\textbf{Low-speed RL}  & \multicolumn{1}{c|}{36.03} & \multicolumn{1}{c|}{\cellcolor[HTML]{8FD2B1}6.72\%}  & \multicolumn{1}{c|}{602.05}   & \multicolumn{1}{c|}{\cellcolor[HTML]{57BB8A}14.73\%} & \multicolumn{1}{c|}{12.65} & \cellcolor[HTML]{6EC49A}8.68\%   & \multicolumn{1}{c|}{36.62} & \multicolumn{1}{c|}{\cellcolor[HTML]{75C89F}8.22\%}  & \multicolumn{1}{c|}{600}      & \multicolumn{1}{c|}{\cellcolor[HTML]{57BB8A}14.34\%} & \multicolumn{1}{c|}{12.58} & \cellcolor[HTML]{78C9A1}8.08\%   \\ \hline
\textbf{Deployed}     & \multicolumn{2}{c|}{34.62}                                                        & \multicolumn{2}{c|}{602.09}                                                          & \multicolumn{2}{c|}{12.71}                                    & \multicolumn{2}{c|}{34.63}                                                        & \multicolumn{2}{c|}{602.05}                                                          & \multicolumn{2}{c|}{12.71}                                    \\ \hline \hline
\textbf{SHOCKWAVE}     & \multicolumn{2}{c|}{Fuel Economy}                                                 & \multicolumn{2}{c|}{Throughput}                                                      & \multicolumn{2}{c|}{Speed}                                    & \multicolumn{2}{c|}{Fuel Economy}                                                 & \multicolumn{2}{c|}{Throughput}                                                      & \multicolumn{2}{c|}{Speed}                                    \\ \hline
\textbf{}              & \multicolumn{1}{c|}{Val}   & \multicolumn{1}{c|}{\cellcolor[HTML]{FFFFFF}Pct}     & \multicolumn{1}{c|}{Val}      & \multicolumn{1}{c|}{\cellcolor[HTML]{FFFFFF}Pct}     & \multicolumn{1}{c|}{Val}   & \cellcolor[HTML]{FFFFFF}Pct      & \multicolumn{1}{c|}{Val}   & \multicolumn{1}{c|}{\cellcolor[HTML]{FFFFFF}Pct}     & \multicolumn{1}{c|}{Val}      & \multicolumn{1}{c|}{\cellcolor[HTML]{FFFFFF}Pct}     & \multicolumn{1}{c|}{Val}   & \cellcolor[HTML]{FFFFFF}Pct      \\ \hline
\textbf{Baseline}      & \multicolumn{1}{c|}{37.92} & \multicolumn{1}{c|}{\cellcolor[HTML]{FFFFFF}0.00\%}  & \multicolumn{1}{c|}{2,038.55} & \multicolumn{1}{c|}{\cellcolor[HTML]{FFFFFF}0.00\%}  & \multicolumn{1}{c|}{40.13} & \cellcolor[HTML]{FFFFFF}0.00\%   & \multicolumn{1}{c|}{37.92} & \multicolumn{1}{c|}{\cellcolor[HTML]{FFFFFF}0.00\%}  & \multicolumn{1}{c|}{2,038.55} & \multicolumn{1}{c|}{\cellcolor[HTML]{FFFFFF}0.00\%}  & \multicolumn{1}{c|}{40.13} & \cellcolor[HTML]{FFFFFF}0.00\%   \\ \hline
\textbf{Microaccel A}  & \multicolumn{1}{c|}{42.47} & \multicolumn{1}{c|}{\cellcolor[HTML]{57BB8A}10.71\%} & \multicolumn{1}{c|}{2,062.57} & \multicolumn{1}{c|}{\cellcolor[HTML]{ECF7F2}1.18\%}  & \multicolumn{1}{c|}{39.83} & \cellcolor[HTML]{FDF5F4}-0.75\%  & \multicolumn{1}{c|}{41.82} & \multicolumn{1}{c|}{\cellcolor[HTML]{63C092}9.33\%}  & \multicolumn{1}{c|}{2,036.83} & \multicolumn{1}{c|}{\cellcolor[HTML]{FEFDFD}-0.08\%} & \multicolumn{1}{c|}{39.98} & \cellcolor[HTML]{FEFAF9}-0.37\%  \\ \hline
\textbf{Microaccel B}  & \multicolumn{1}{c|}{41.61} & \multicolumn{1}{c|}{\cellcolor[HTML]{6BC398}8.87\%}  & \multicolumn{1}{c|}{2,084.85} & \multicolumn{1}{c|}{\cellcolor[HTML]{D9F0E5}2.27\%}  & \multicolumn{1}{c|}{41.28} & \cellcolor[HTML]{CFECDE}2.87\%   & \multicolumn{1}{c|}{39.73} & \multicolumn{1}{c|}{\cellcolor[HTML]{B3E1CA}4.56\%}  & \multicolumn{1}{c|}{2,043.91} & \multicolumn{1}{c|}{\cellcolor[HTML]{FBFEFC}0.26\%}  & \multicolumn{1}{c|}{43.21} & \cellcolor[HTML]{7FCBA6}7.68\%   \\ \hline
\textbf{Microaccel C}  & \multicolumn{1}{c|}{42.47} & \multicolumn{1}{c|}{\cellcolor[HTML]{57BB8A}10.71\%} & \multicolumn{1}{c|}{2,062.57} & \multicolumn{1}{c|}{\cellcolor[HTML]{ECF7F2}1.18\%}  & \multicolumn{1}{c|}{39.83} & \cellcolor[HTML]{FDF5F4}-0.75\%  & \multicolumn{1}{c|}{41.82} & \multicolumn{1}{c|}{\cellcolor[HTML]{63C092}9.33\%}  & \multicolumn{1}{c|}{2,036.83} & \multicolumn{1}{c|}{\cellcolor[HTML]{FEFDFD}-0.08\%} & \multicolumn{1}{c|}{39.98} & \cellcolor[HTML]{FEFAF9}-0.37\%  \\ \hline
\textbf{Microaccel D}  & \multicolumn{1}{c|}{42.47} & \multicolumn{1}{c|}{\cellcolor[HTML]{57BB8A}10.71\%} & \multicolumn{1}{c|}{2,062.57} & \multicolumn{1}{c|}{\cellcolor[HTML]{ECF7F2}1.18\%}  & \multicolumn{1}{c|}{39.83} & \cellcolor[HTML]{FDF5F4}-0.75\%  & \multicolumn{1}{c|}{41.82} & \multicolumn{1}{c|}{\cellcolor[HTML]{63C092}9.33\%}  & \multicolumn{1}{c|}{2,036.83} & \multicolumn{1}{c|}{\cellcolor[HTML]{FEFDFD}-0.08\%} & \multicolumn{1}{c|}{39.98} & \cellcolor[HTML]{FEFAF9}-0.37\%  \\ \hline
\textbf{Microaccel E}  & \multicolumn{1}{c|}{42.47} & \multicolumn{1}{c|}{\cellcolor[HTML]{57BB8A}10.71\%} & \multicolumn{1}{c|}{2,062.57} & \multicolumn{1}{c|}{\cellcolor[HTML]{ECF7F2}1.18\%}  & \multicolumn{1}{c|}{39.83} & \cellcolor[HTML]{FDF5F4}-0.75\%  & \multicolumn{1}{c|}{41.82} & \multicolumn{1}{c|}{\cellcolor[HTML]{63C092}9.33\%}  & \multicolumn{1}{c|}{2,036.83} & \multicolumn{1}{c|}{\cellcolor[HTML]{FEFDFD}-0.08\%} & \multicolumn{1}{c|}{39.98} & \cellcolor[HTML]{FEFAF9}-0.37\%  \\ \hline
\textbf{Microaccel F}  & \multicolumn{1}{c|}{44.58} & \multicolumn{1}{c|}{\cellcolor[HTML]{57BB8A}14.94\%} & \multicolumn{1}{c|}{2,045.59} & \multicolumn{1}{c|}{\cellcolor[HTML]{FAFDFB}0.35\%}  & \multicolumn{1}{c|}{40.31} & \cellcolor[HTML]{F8FCFA}0.45\%   & \multicolumn{1}{c|}{43.19} & \multicolumn{1}{c|}{\cellcolor[HTML]{57BB8A}12.20\%} & \multicolumn{1}{c|}{2,059.33} & \multicolumn{1}{c|}{\cellcolor[HTML]{EEF9F4}1.02\%}  & \multicolumn{1}{c|}{40.28} & \cellcolor[HTML]{F9FDFB}0.37\%   \\ \hline
\textbf{RL A}          & \multicolumn{1}{c|}{43.71} & \multicolumn{1}{c|}{\cellcolor[HTML]{57BB8A}13.25\%} & \multicolumn{1}{c|}{2,011.67} & \multicolumn{1}{c|}{\cellcolor[HTML]{FBEDEC}-1.32\%} & \multicolumn{1}{c|}{39.45} & \cellcolor[HTML]{FAE8E7}-1.69\%  & \multicolumn{1}{c|}{44.25} & \multicolumn{1}{c|}{\cellcolor[HTML]{57BB8A}14.31\%} & \multicolumn{1}{c|}{2,015.90} & \multicolumn{1}{c|}{\cellcolor[HTML]{FCF0EF}-1.11\%} & \multicolumn{1}{c|}{39.44} & \cellcolor[HTML]{FAE8E6}-1.72\%  \\ \hline
\textbf{RL B}          & \multicolumn{1}{c|}{42.91} & \multicolumn{1}{c|}{\cellcolor[HTML]{57BB8A}11.63\%} & \multicolumn{1}{c|}{2,008.93} & \multicolumn{1}{c|}{\cellcolor[HTML]{FBEBEA}-1.45\%} & \multicolumn{1}{c|}{39.38} & \cellcolor[HTML]{FAE6E4}-1.87\%  & \multicolumn{1}{c|}{43.28} & \multicolumn{1}{c|}{\cellcolor[HTML]{57BB8A}12.38\%} & \multicolumn{1}{c|}{2,016.03} & \multicolumn{1}{c|}{\cellcolor[HTML]{FCF0EF}-1.10\%} & \multicolumn{1}{c|}{39.38} & \cellcolor[HTML]{FAE6E4}-1.87\%  \\ \hline
\textbf{RL C}          & \multicolumn{1}{c|}{37.72} & \multicolumn{1}{c|}{\cellcolor[HTML]{FDF8F7}-0.53\%} & \multicolumn{1}{c|}{2,040.71} & \multicolumn{1}{c|}{\cellcolor[HTML]{FEFFFE}0.11\%}  & \multicolumn{1}{c|}{40.02} & \cellcolor[HTML]{FEFBFB}-0.27\%  & \multicolumn{1}{c|}{39.19} & \multicolumn{1}{c|}{\cellcolor[HTML]{C9E9DA}3.24\%}  & \multicolumn{1}{c|}{2,043.93} & \multicolumn{1}{c|}{\cellcolor[HTML]{FBFEFC}0.26\%}  & \multicolumn{1}{c|}{40.4}  & \cellcolor[HTML]{F4FBF8}0.67\%   \\ \hline
\textbf{High-speed RL} & \multicolumn{1}{c|}{37.39} & \multicolumn{1}{c|}{\cellcolor[HTML]{FBECEB}-1.42\%} & \multicolumn{1}{c|}{2,045.98} & \multicolumn{1}{c|}{\cellcolor[HTML]{F9FDFB}0.36\%}  & \multicolumn{1}{c|}{40.19} & \cellcolor[HTML]{FDFEFE}0.15\%   & \multicolumn{1}{c|}{36.9}  & \multicolumn{1}{c|}{\cellcolor[HTML]{F8DAD8}-2.76\%} & \multicolumn{1}{c|}{2,055.17} & \multicolumn{1}{c|}{\cellcolor[HTML]{F2FAF6}0.82\%}  & \multicolumn{1}{c|}{40.28} & \cellcolor[HTML]{F9FDFB}0.37\%   \\ \hline
\textbf{Low-speed RL}  & \multicolumn{1}{c|}{46.51} & \multicolumn{1}{c|}{\cellcolor[HTML]{57BB8A}18.47\%} & \multicolumn{1}{c|}{2,080.09} & \multicolumn{1}{c|}{\cellcolor[HTML]{DDF2E8}2.04\%}  & \multicolumn{1}{c|}{39.45} & \cellcolor[HTML]{FAE8E7}-1.69\%  & \multicolumn{1}{c|}{49.36} & \multicolumn{1}{c|}{\cellcolor[HTML]{57BB8A}23.18\%} & \multicolumn{1}{c|}{2,096.52} & \multicolumn{1}{c|}{\cellcolor[HTML]{D0ECDE}2.84\%}  & \multicolumn{1}{c|}{39.25} & \cellcolor[HTML]{F9E2E0}-2.19\%  \\ \hline
\textbf{Deployed}     & \multicolumn{2}{c|}{45.96}                                                        & \multicolumn{2}{c|}{2,072.91}                                                        & \multicolumn{2}{c|}{39.27}                                    & \multicolumn{2}{c|}{49.22}                                                        & \multicolumn{2}{c|}{2,094.57}                                                        & \multicolumn{2}{c|}{39.21}                                    \\ \hline \hline
\textbf{FREEFLOW}      & \multicolumn{2}{c|}{Fuel Economy}                                                 & \multicolumn{2}{c|}{Throughput}                                                      & \multicolumn{2}{c|}{Speed}                                    & \multicolumn{2}{c|}{Fuel Economy}                                                 & \multicolumn{2}{c|}{Throughput}                                                      & \multicolumn{2}{c|}{Speed}                                    \\ \hline
\textbf{}              & \multicolumn{1}{c|}{Val}   & \multicolumn{1}{c|}{\cellcolor[HTML]{FFFFFF}Pct}     & \multicolumn{1}{c|}{Val}      & \multicolumn{1}{c|}{\cellcolor[HTML]{FFFFFF}Pct}     & \multicolumn{1}{c|}{Val}   & \cellcolor[HTML]{FFFFFF}Pct      & \multicolumn{1}{c|}{Val}   & \multicolumn{1}{c|}{\cellcolor[HTML]{FFFFFF}Pct}     & \multicolumn{1}{c|}{Val}      & \multicolumn{1}{c|}{\cellcolor[HTML]{FFFFFF}Pct}     & \multicolumn{1}{c|}{Val}   & \cellcolor[HTML]{FFFFFF}Pct      \\ \hline
\textbf{Baseline}      & \multicolumn{1}{c|}{36.85} & \multicolumn{1}{c|}{\cellcolor[HTML]{FFFFFF}0.00\%}  & \multicolumn{1}{c|}{2,136.49} & \multicolumn{1}{c|}{\cellcolor[HTML]{FFFFFF}0.00\%}  & \multicolumn{1}{c|}{71.9}  & \cellcolor[HTML]{FFFFFF}0.00\%   & \multicolumn{1}{c|}{36.85} & \multicolumn{1}{c|}{\cellcolor[HTML]{FFFFFF}0.00\%}  & \multicolumn{1}{c|}{2,136.49} & \multicolumn{1}{c|}{\cellcolor[HTML]{FFFFFF}0.00\%}  & \multicolumn{1}{c|}{71.9}  & \cellcolor[HTML]{FFFFFF}0.00\%   \\ \hline
\textbf{Microaccel A}  & \multicolumn{1}{c|}{43.11} & \multicolumn{1}{c|}{\cellcolor[HTML]{57BB8A}14.52\%} & \multicolumn{1}{c|}{2,247.74} & \multicolumn{1}{c|}{\cellcolor[HTML]{A8DCC3}5.21\%}  & \multicolumn{1}{c|}{63.61} & \cellcolor[HTML]{E67C73}-11.53\% & \multicolumn{1}{c|}{43.1}  & \multicolumn{1}{c|}{\cellcolor[HTML]{57BB8A}14.50\%} & \multicolumn{1}{c|}{2,246.04} & \multicolumn{1}{c|}{\cellcolor[HTML]{A9DDC4}5.13\%}  & \multicolumn{1}{c|}{63.54} & \cellcolor[HTML]{E67C73}-11.63\% \\ \hline
\textbf{Microaccel B}  & \multicolumn{1}{c|}{42.75} & \multicolumn{1}{c|}{\cellcolor[HTML]{57BB8A}13.80\%} & \multicolumn{1}{c|}{2,262.47} & \multicolumn{1}{c|}{\cellcolor[HTML]{9CD7BB}5.90\%}  & \multicolumn{1}{c|}{63.54} & \cellcolor[HTML]{E67C73}-11.63\% & \multicolumn{1}{c|}{42.74} & \multicolumn{1}{c|}{\cellcolor[HTML]{57BB8A}13.78\%} & \multicolumn{1}{c|}{2,232.41} & \multicolumn{1}{c|}{\cellcolor[HTML]{B4E1CB}4.49\%}  & \multicolumn{1}{c|}{62.75} & \cellcolor[HTML]{E67C73}-12.73\% \\ \hline
\textbf{Microaccel C}  & \multicolumn{1}{c|}{43.11} & \multicolumn{1}{c|}{\cellcolor[HTML]{57BB8A}14.52\%} & \multicolumn{1}{c|}{2,247.74} & \multicolumn{1}{c|}{\cellcolor[HTML]{A8DCC3}5.21\%}  & \multicolumn{1}{c|}{63.61} & \cellcolor[HTML]{E67C73}-11.53\% & \multicolumn{1}{c|}{43.1}  & \multicolumn{1}{c|}{\cellcolor[HTML]{57BB8A}14.50\%} & \multicolumn{1}{c|}{2,246.04} & \multicolumn{1}{c|}{\cellcolor[HTML]{A9DDC4}5.13\%}  & \multicolumn{1}{c|}{63.54} & \cellcolor[HTML]{E67C73}-11.63\% \\ \hline
\textbf{Microaccel D}  & \multicolumn{1}{c|}{43.11} & \multicolumn{1}{c|}{\cellcolor[HTML]{57BB8A}14.52\%} & \multicolumn{1}{c|}{2,247.74} & \multicolumn{1}{c|}{\cellcolor[HTML]{A8DCC3}5.21\%}  & \multicolumn{1}{c|}{63.61} & \cellcolor[HTML]{E67C73}-11.53\% & \multicolumn{1}{c|}{43.1}  & \multicolumn{1}{c|}{\cellcolor[HTML]{57BB8A}14.50\%} & \multicolumn{1}{c|}{2,246.04} & \multicolumn{1}{c|}{\cellcolor[HTML]{A9DDC4}5.13\%}  & \multicolumn{1}{c|}{63.54} & \cellcolor[HTML]{E67C73}-11.63\% \\ \hline
\textbf{Microaccel E}  & \multicolumn{1}{c|}{43.11} & \multicolumn{1}{c|}{\cellcolor[HTML]{57BB8A}14.52\%} & \multicolumn{1}{c|}{2,247.74} & \multicolumn{1}{c|}{\cellcolor[HTML]{A8DCC3}5.21\%}  & \multicolumn{1}{c|}{63.61} & \cellcolor[HTML]{E67C73}-11.53\% & \multicolumn{1}{c|}{43.1}  & \multicolumn{1}{c|}{\cellcolor[HTML]{57BB8A}14.50\%} & \multicolumn{1}{c|}{2,246.04} & \multicolumn{1}{c|}{\cellcolor[HTML]{A9DDC4}5.13\%}  & \multicolumn{1}{c|}{63.54} & \cellcolor[HTML]{E67C73}-11.63\% \\ \hline
\textbf{Microaccel F}  & \multicolumn{1}{c|}{43.77} & \multicolumn{1}{c|}{\cellcolor[HTML]{57BB8A}15.81\%} & \multicolumn{1}{c|}{2,149.98} & \multicolumn{1}{c|}{\cellcolor[HTML]{F5FBF8}0.63\%}  & \multicolumn{1}{c|}{60.97} & \cellcolor[HTML]{E67C73}-15.20\% & \multicolumn{1}{c|}{43.7}  & \multicolumn{1}{c|}{\cellcolor[HTML]{57BB8A}15.68\%} & \multicolumn{1}{c|}{2,143.64} & \multicolumn{1}{c|}{\cellcolor[HTML]{FAFDFC}0.33\%}  & \multicolumn{1}{c|}{60.84} & \cellcolor[HTML]{E67C73}-15.38\% \\ \hline
\textbf{RL A}          & \multicolumn{1}{c|}{49.39} & \multicolumn{1}{c|}{\cellcolor[HTML]{57BB8A}25.39\%} & \multicolumn{1}{c|}{2,193.21} & \multicolumn{1}{c|}{\cellcolor[HTML]{D3EDE0}2.65\%}  & \multicolumn{1}{c|}{55.81} & \cellcolor[HTML]{E67C73}-22.38\% & \multicolumn{1}{c|}{50.67} & \multicolumn{1}{c|}{\cellcolor[HTML]{57BB8A}27.27\%} & \multicolumn{1}{c|}{2,189.98} & \multicolumn{1}{c|}{\cellcolor[HTML]{D5EEE2}2.50\%}  & \multicolumn{1}{c|}{53.79} & \cellcolor[HTML]{E67C73}-25.19\% \\ \hline
\textbf{RL B}          & \multicolumn{1}{c|}{49.31} & \multicolumn{1}{c|}{\cellcolor[HTML]{57BB8A}25.27\%} & \multicolumn{1}{c|}{2,210.67} & \multicolumn{1}{c|}{\cellcolor[HTML]{C5E8D7}3.47\%}  & \multicolumn{1}{c|}{56.41} & \cellcolor[HTML]{E67C73}-21.54\% & \multicolumn{1}{c|}{54.8}  & \multicolumn{1}{c|}{\cellcolor[HTML]{57BB8A}32.76\%} & \multicolumn{1}{c|}{2,186.25} & \multicolumn{1}{c|}{\cellcolor[HTML]{D8F0E4}2.33\%}  & \multicolumn{1}{c|}{54.8}  & \cellcolor[HTML]{E67C73}-23.78\% \\ \hline
\textbf{RL C}          & \multicolumn{1}{c|}{36.87} & \multicolumn{1}{c|}{\cellcolor[HTML]{FFFFFF}0.05\%}  & \multicolumn{1}{c|}{2,139.03} & \multicolumn{1}{c|}{\cellcolor[HTML]{FEFFFE}0.12\%}  & \multicolumn{1}{c|}{71.28} & \cellcolor[HTML]{FCF3F2}-0.86\%  & \multicolumn{1}{c|}{36.87} & \multicolumn{1}{c|}{\cellcolor[HTML]{FFFFFF}0.05\%}  & \multicolumn{1}{c|}{2,139.03} & \multicolumn{1}{c|}{\cellcolor[HTML]{FEFFFE}0.12\%}  & \multicolumn{1}{c|}{71.28} & \cellcolor[HTML]{FCF3F2}-0.86\%  \\ \hline
\textbf{High-speed RL} & \multicolumn{1}{c|}{36.84} & \multicolumn{1}{c|}{\cellcolor[HTML]{FEFEFE}-0.03\%} & \multicolumn{1}{c|}{2,170.76} & \multicolumn{1}{c|}{\cellcolor[HTML]{E5F5ED}1.60\%}  & \multicolumn{1}{c|}{72.09} & \cellcolor[HTML]{FBFEFC}0.26\%   & \multicolumn{1}{c|}{36.78} & \multicolumn{1}{c|}{\cellcolor[HTML]{FEFCFC}-0.19\%} & \multicolumn{1}{c|}{2,168.24} & \multicolumn{1}{c|}{\cellcolor[HTML]{E7F5EE}1.49\%}  & \multicolumn{1}{c|}{72.21} & \cellcolor[HTML]{F8FDFA}0.43\%   \\ \hline
\textbf{Low-speed RL}  & \multicolumn{1}{c|}{49.22} & \multicolumn{1}{c|}{\cellcolor[HTML]{57BB8A}25.13\%} & \multicolumn{1}{c|}{2,186.74} & \multicolumn{1}{c|}{\cellcolor[HTML]{D8F0E4}2.35\%}  & \multicolumn{1}{c|}{52.11} & \cellcolor[HTML]{E67C73}-27.52\% & \multicolumn{1}{c|}{52.25} & \multicolumn{1}{c|}{\cellcolor[HTML]{57BB8A}29.47\%} & \multicolumn{1}{c|}{2,166.56} & \multicolumn{1}{c|}{\cellcolor[HTML]{E8F6EF}1.41\%}  & \multicolumn{1}{c|}{47.25} & \cellcolor[HTML]{E67C73}-34.28\% \\ \hline
\textbf{Deployed}     & \multicolumn{1}{c|}{36.84} & \multicolumn{1}{c|}{\cellcolor[HTML]{FEFEFE}-0.03\%} & \multicolumn{1}{c|}{2,170.76} & \multicolumn{1}{c|}{\cellcolor[HTML]{E5F5ED}1.60\%}  & \multicolumn{1}{c|}{72.09} & \cellcolor[HTML]{FBFEFC}0.26\%   & \multicolumn{1}{c|}{36.78} & \multicolumn{1}{c|}{\cellcolor[HTML]{FEFCFC}-0.19\%} & \multicolumn{1}{c|}{2,168.24} & \multicolumn{1}{c|}{\cellcolor[HTML]{E7F5EE}1.49\%}  & \multicolumn{1}{c|}{72.21} & \cellcolor[HTML]{F8FDFA}0.43\%   \\ \hline
\end{tabular}
\vspace{2mm}
\caption{Performance results across a variety of metrics. The final speed planner was a combination of Speed Planner 1 and Speed Planner 2, which are depicted here. The controllers are tested across two different scenarios, shown in Column 1: a bottleneck and a shockwave scenario. The three controllers that are being tested here include (1) the high-speed controller, (2) the low-speed controller, and (3) the final controller, which is described in Section~\ref{sec:final-obs-space} to be a combination of (1) and (2). The evaluations were computed by running each controller on simulation with two leader trajectories (See \nameref{sbar-SIM}), one in a congested setting and the second in a bottleneck setting. Evaluations are run with a 4\% penetration rate of AVs. These evaluations were ultimately used to determine which controllers to use during the MVT. For more details on the selection process, see 
\cite{lee2024traffic}.
}
\label{tab:freeze_evaluations}
\end{table*}


\renewcommand{\thestable}{S\arabic{stable}}
\renewcommand{\thesfigure}{S\arabic{sfigure}}

\begin{sidebar}{Alternative Algorithms: Imitation Learning}
\section[Alternative Algorithms: Imitation Learning]{by Zhe Fu}\label{sbar:imitation}

\sdbarfigfullwidth{
    \includegraphics[width=38pc]{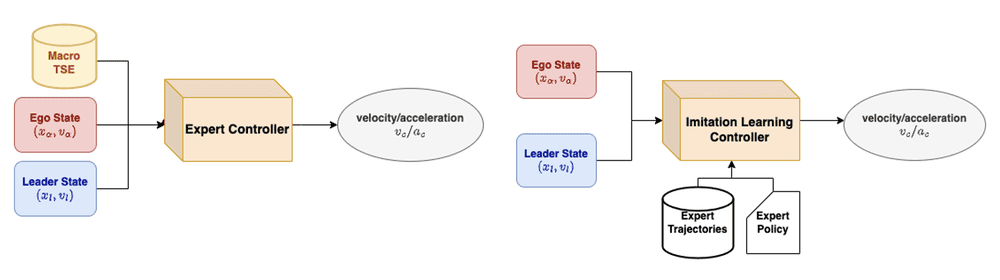}}
    {The structure comparison of expert and imitation control strategies. \textbf{Left:} The expert controller, which is using global sensing as the input. \textbf{Right:} IL controller, which eliminates the real-time macro TSE as the input.
    \label{fig:imitation_illustration}
    }

 
\sdbarinitial{I}\emph{n} this section, we share results on an alternative \textit{imitation learning} (IL) based algorithm that was a candidate for the MVT.
This algorithm aims to bridge the gap between global/semi sensing (where global real-time traffic state estimation (TSE), e.g. INRIX, are available) and local sensing (where only ego vehicle and preceding vehicle information is available), thus enabling AVs to reduce their dependence on real-time downstream information. Through proper usage of IL, we robustly map behaviors generated by global sensing to locally observable traffic features.

IL shares the same fundamental structure as RL, discussed technically in ``\nameref{sidebar:rl}``. In this setting, IL optimizes: $\theta := \text{argmin}_\theta \left[ \mathbb{E}_{\mathcal{D}} \left[ \mathcal{L}(\pi_\theta(s_i), a_i) \right] \right]$, where $\mathcal{L}(\cdot, \cdot)$ is a distance metric between the predicted and expert action.

Most IL techniques consider human behaviors as demonstrations, and this approach can be problematic in dense traffic settings, where human drivers often contribute to traffic congestion rather than alleviate it. For this reason, we deviate from conventional methods. Rather than imitating human driving behaviors, we select our feedback control strategy under the global sensing paradigm, introduced in \cite{fu2023cooperative}, as the expert to emulate. In this context, we leverage IL not to reproduce driving behaviors, but to introduce flexibility and decentralization while preserving behavioral alignment with the experts' potential actions.

IL is prone to covariate shift, wherein long sequences of actions exacerbate the discrepancies between demonstrations and observed states. In traffic settings, this can result in vehicles drifting off-road after prolonged periods of time~\cite{bojarski2016end} or failing to maintain safe car-following distances~\cite{bansal2018chauffeurnet}. To address the covariate shift problem, we utilize the DAgger algorithm~\cite{ross2010efficient, ross2011reduction}, an online IL procedure, correct for covariate shift, for generating desired behavior from demonstrations. These demonstrations comprise the ``Expert Trajectories'' derived from our expert control strategies. In order to generate those ``Expert Trajectories,'' we deploy the expert controller in a simulation using historical traffic data, thereby collecting expert's responses to different states. Throughout the training process, we would consult the ``Expert Policy'' -- our feedback control strategy -- when encountering unvisited states, in order to prevent distribution shift. The structures and the differences between our IL controller and the expert controller are shown in Fig~\ref{fig:imitation_illustration}. The ultimate goal of this process is to devise a policy that produces satisfactory mappings from local states to actions that mirror those of the expert.

To evaluate the effectiveness of our proposed IL controller, we applied the control strategies to the same simulation environment described in the main text across 10 scenarios, each reflecting varying degrees of traffic congestion. The outcomes show that the IL strategy can significantly enhance energy efficiency using only local sensing. In all 10 scenarios, the expert with global sensing leads to an average energy savings of $16.57\%$, whereas the IL strategy yields an average of $14.50\%$. We highlight two specific outcomes, which demonstrate our controllers' ability to mitigate wave effects in both moderate (see Fig~\ref{fig:moderate_traffic}) and heavy traffic conditions (see Fig~\ref{fig:heavy_traffic}). Further details and discussions are available in \cite{kernel}. 

\end{sidebar}

\begin{sidebar}{Alternative Algorithms: Imitation Learning (continued)}

\sdbarfigfullwidth{
    \includegraphics[width=38pc]{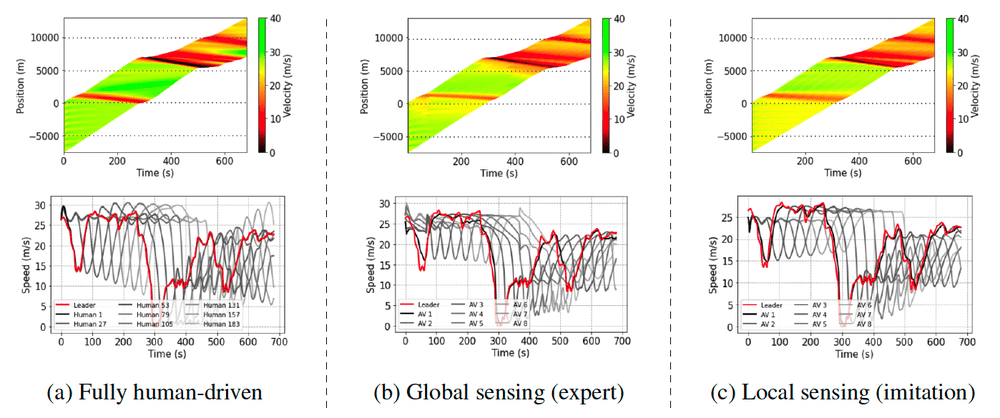}}
    {Performance of expert and IL controllers in the presence of \textbf{moderate/infrequent} oscillations in speeds. Compared with the fully human-driven baseline in \textbf{(a)}, AVs in both expert \textbf{(b)} and IL strategies \textbf{(c)} succeed in reducing the magnitude of oscillations in driving speeds resulting from aggressive human driving behaviors. Interestingly, the IL strategies incrementally alleviate the shockwave effects by accepting larger headways and slowing down in a smoother fashion.  
    \label{fig:moderate_traffic}
    }

\sdbarfigfullwidth{
    \includegraphics[width=38pc]{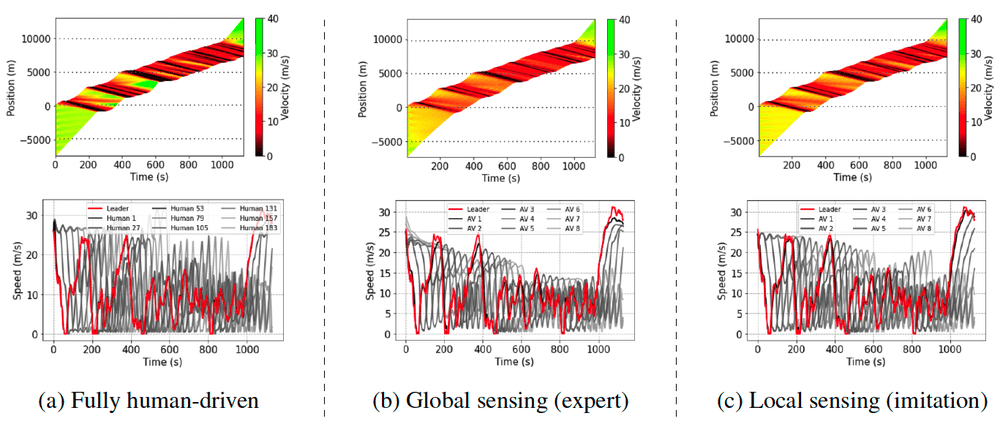}}
    {Performance of expert and IL controllers in the presence of \textbf{sharp and frequent} oscillations in speeds. Compared with the fully human-driven baseline in \textbf{(a)}, both global sensing expert \textbf{(b)} and the IL \textbf{(c)} do well in reducing the intensity of oscillations, with the expert policy slightly outperforming that which imitates it. 
    \label{fig:heavy_traffic}
    }

\end{sidebar}



\begin{sidebar}{Alternative Algorithms: Hierarchical RL}
\section[Alternative Algorithms: Hierarchical RL]{by Abdul Rahman Kreidieh}\label{sbar:hierarchical}

\renewcommand{\thestable}{S\arabic{stable}}
\renewcommand{\thesfigure}{S\arabic{sfigure}}

\sdbarfig{\includegraphics[width=18pc]{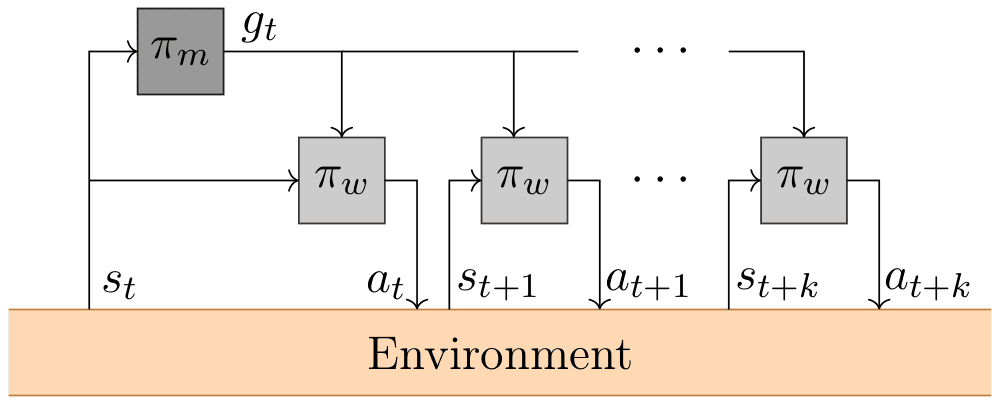}}{Explored hierarchical model.\label{fig:hrl-model}}

\sdbarinitial{A} common drawback of standard RL strategies is their inability to solve long-term planning and control tasks, particularly in the presence of limited or delayed feedback. This is a common problem in mixed-autonomy traffic control tasks, as meaningful events develop across long periods of time as, for instance, oscillations in traffic states propagate slowly through a network. Accordingly, actions by AVs often have a delayed effect on metrics of improvement or success and, at times, require hundreds of steps to meaningfully affect the reward or feedback signals assigned to learning policies. The delayed nature of this feedback exacerbates the credit assignment problem within such tasks, making reasoning on whether a specific action contributed to or mitigated the propagation of congestion difficult.

\acronym{Hierarchical reinforcement learning} (HRL) techniques provide natural methods for inducing structured exploration in these difficult, long-horizon tasks. By decomposing challenging tasks into simpler sub-problems using a hierarchy of policies, such methods refine exploration from the perspective of higher-level policies, enabling agents to search for optimal solutions across a range of temporally extended sequences~\cite{sutton1999between}. This simplifies learning for such policies over large timescales, thereby allowing hierarchies, particularly with well-defined lower-level behaviors, to outperform their non-hierarchical counterparts in a variety of interesting and difficult problems~\cite{sutton1999between, kulkarni2016hierarchical, vezhnevets2017feudal, nachum2018data}.

Through prior work, we have explored the use of feudal, or goal-conditioned, hierarchical models for solving similar mixed-autonomy tasks via RL~\cite{dayan1993feudal, vezhnevets2017feudal, nachum2018data}.
Within such a hierarchy (see Figure~\ref{fig:hrl-model}), a higher-level manager policy $\pi_m$ assigns goals $g_t \sim \pi_m (s_t)$ to a \emph{universal}~\cite{schaul2015universal} worker policy $\pi_w$ at the lower level. The worker policy then performs primitive actions $a_t \sim \pi_w(s_t, g_t)$ to achieve these goals for a meta-interval $k$, at which point a new goal is assigned.

The goals assigned by a manager denote a desired state, defined via the worker's goal-conditioned reward function  $r_w(s_t, g_t, s_{t+1})$. For this study, we use a reward function similar to the ones presented in~\cite{2017-TOG-deepLoco,vezhnevets2017feudal,nachum2018data}, in which goals characterize desired deviations in the state from the perspective of an agent. For automated vehicle (AV) control problems, we define the goal as the desired speeds for each AV. The reward function used to instill such a behavior is $r_w(s_t, g_t, s_{t+1}) := -\lVert s'_t + g_t - s'_{t+1}\rVert_2$, where $s'_t$ is the subset of the state for which goals are assigned. The objective of the worker is to find the optimal parameters $\theta_w^* := \text{argmax}_\theta [J_w(\theta)]$ that solve for the goal-conditioned objective~\cite{nasiriany2019planning, liu2022goal}:
\begin{equation}
    J_w = \mathbb{E}_{s\sim p_\pi} \left[ \sum_{t=0}^k \gamma^t r_w(s_t, g_t,s_{t+1}) \right]
\end{equation}

\sdbarfigfullwidth{\includegraphics[width=.9\linewidth]{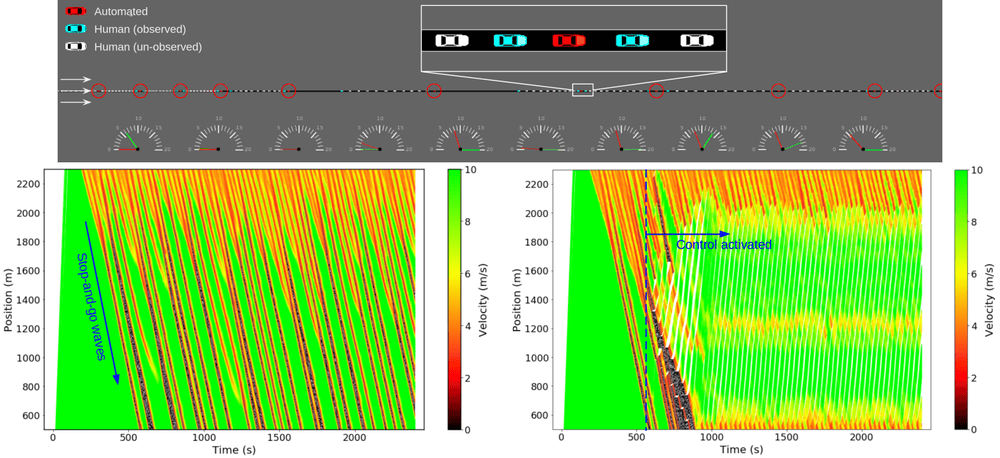}}{This environment is an open-network extension of the ring road task in which downstream traffic instabilities resulting simulated via a slower-moving downstream segment produce congestion in the form of stop-and-go waves (left figure). The policy learned using this approach (right figure) allows AVs to form gaps sufficiently large to prevent stop-and-go waves from propagating. \label{fig:hrl-highway-env}}
\end{sidebar}

\begin{sidebar}{Alternative Algorithms: Hierarchical RL (continued)}
\renewcommand{\thestable}{S\arabic{stable}}
\renewcommand{\thesfigure}{S\arabic{sfigure}}

Concurrently, the manager jointly searches the space of viable goals and receives a reward $r_m(s_t)$ extrinsically defined by the underlying task, solving the original RL objective $J_m$:
\begin{equation}
    J_m = \mathbb{E}_{s\sim p_\pi} \left[ \sum_{t=0}^T \left[ \gamma^t r_m(s_t) \right] \right]
\end{equation}

We validate the efficacy of this approach on two mixed-autonomy traffic control tasks similar to those presented in this article, whereby automated vehicles attempt to mitigate the propagation of stop-and-go traffic. A brief description of each result is presented in Figure~\ref{fig:hrl-highway-env}. To learn more about this study, we refer the reader to the reference provided here.

\sdbarfig{\includegraphics[width=18pc]{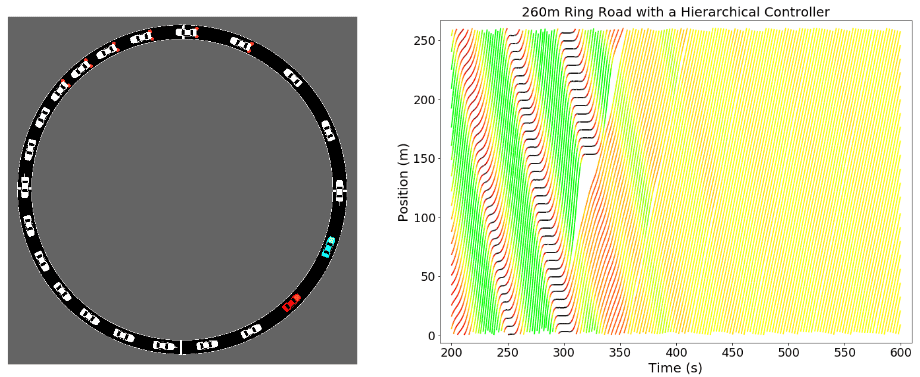}}{This environment is a replication of the task presented in~\cite{wu2017flowold}, in which a single vehicle attempts to stabilize a variable length ring road. The policy learned via this method succeeds in assigning minimal spacings to dissipate stop-and-go waves while not temporarily hindering throughput.\label{fig:ring-env}}

\end{sidebar}

Initial controller ideas sprouted from previous work~\cite{lichtle2022deploying, bunting2021libpanda} on a vehicle which supported acceleration-based control. In ``\nameref{sbar-wavesmoothing}'', we delve into previous work that focused on the development of said acceleration-based controllers on the I-210 highway, prior to knowing the MVT would occur on the I-24. The opportunity to get access to an additional vehicle-type which supported ACC-based control, prompted a pivot from our initial idea of an acceleration-based controller to also develop and support an ACC-based controller. Discussion and results will be provided on the performance and efficacy of each controller. Ultimately, of the two RL controllers presented here, the ACC-based controller was deployed for the MVT and is the controller whose results are presented here. In addition, we present the results of two different alternative algorithms that were developed during CIRCLES but were not ultimately deployed. For details, view ''\nameref{sbar:imitation}'' and ''\nameref{sbar:hierarchical}''.

In this article and the following sections, we discuss the development and results of two controllers:
\begin{enumerate}
    \item The acceleration-based controller
    \item The ACC-based controller
\end{enumerate}


\subsection{Common Problem Formulation}
Due to the considerable overlap in algorithm design between the acceleration-based controller and the ACC-based controller, the parameters that are common to both controllers will be discussed here.

We use single-agent \emph{Proximal Policy Optimization}~\cite{schulman2017proximal} (PPO) with an augmented value function as our training algorithm. PPO is a well-known policy gradient algorithm known for sample efficiency and performance. For more background on the technical details of policy gradient algorithms, view ``\nameref{sidebar:policygradient}.''

The energy models that build up the MPG-based reward function described in ''~\nameref{sec:final-reward-function}'' and Table ~\ref{table:controller-variations} that are used to train RL controllers are discussed in ''\nameref{sbar-energy}'' and 
\cite{khoudari2023reducing}.

Both the acceleration-based and ACC-based controllers use the same IDM parameters, as shown in Table~\ref{table:idm} and adhere to the dynamics prescribed by \ref{eq:idm}. 

\begin{table}[]
\centering
\begin{tabular}{|l|c|c|c|c|c|c|c|}
\hline
& \multicolumn{7}{c|}{Intelligent Driver Model} \\ \hline
\textbf{Parameter} & $v_0$ & $T$ & $a$ & $b$ & $\delta$ & $s_0$ & $\epsilon$ \\ \hline
\textbf{Value} & $35$ & $1.24$ & $1.3$ & $2.0$ & $4$ & $2$ & $\mathcal{N}(0,0)$ \\ \hline
\end{tabular}
\vspace{2mm}
\caption{Selected IDM parameters (see Eq.~\ref{eq:idm}-\ref{eq:s_star}).}
\label{table:idm}
\end{table}

\subsection{Acceleration-based RL Problem Formulation}
\label{sec:accel-formulation}

We had the opportunity to gain access to a vehicle-type which could be deployed for the MVT and supported ACC-based control, so we pivoted from our initial idea of an acceleration-based controller to also develop and support an RL ACC-based controller. However, we still discuss problem formulation of the acceleration-based controller and compare simulation performance. 

\begin{figure*}
\centering
\begin{tikzpicture}
    \node[draw, rectangle, align=left] (A) at (0,0) {
        \textbf{Observations} \\
        $\bullet$ AV speed \\
        $\bullet$ Leader speed \\
        $\bullet$ Space gap \\
        $\bullet$ Leader speed history \\
        $\bullet$ Speed planner \\
        $\bullet$ Gap-closing threshold \\
        $\bullet$ Failsafe threshold
    };
    \node[draw, rectangle, align=left] (B) at (5,0) {
        \textbf{Costs} \\
        $\bullet$ AV + Platoon energy \\
        $\bullet$ AV acceleration \\
        $\bullet$ Gap-closing usage \\
        $\bullet$ Failsafe usage \\
        $\bullet$ Space gap
    };
    \node[draw, rectangle, align=center] (C) at (5,3) {\textbf{Neural Network} \\ $a_t^\text{raw} = \mathcal{F}(o_t)$};
    \node[draw, rectangle, align=left] (D) at (10,2) {\textbf{Wrappers} \\ $\bullet$ Gap-closing \\ $\bullet$ Failsafe};
    \node[draw, rectangle, align=left] (E) at (10,0) {
        \textbf{Environment} \\
        $\bullet$ Trajectory-based simulator \\
        $\bullet$ Lane-changing model
    };

    \draw[->] (C) -| node[pos=0.05,above right]{$a_t^\text{raw}$} (D);
    \draw[->] (D) -- node[anchor=west]{$a_t$} (E);
    \draw[->] (E.west) -- (B);
    \draw[->] (E.south) -- node[anchor=west]{$s_t \; (s_{t+1})$} ++(0, -1.4) -| (A);
    \draw[->] (B) -- node[anchor=west] {$r_t$ \emph{(optimize using RL)}} (C);
    \draw[->] (A) |- node[pos=0.13,above left]{$o_t$} (C);
\end{tikzpicture}
\caption{Diagram summarizing the design of the acceleration-based controller. The environment is in state $s_t$, from which we obtain a vector of observations $o_t$ which is fed into the neural network to get a raw acceleration $a_t^\text{raw}$. This acceleration is wrapped by a gap-closing and a failsafe term if necessary,  the resulting acceleration $a_t$ is applied to the AV and the whole simulation is updated. This leads to a new state $s_{t+1}$, and the process is repeated. Additionally, the environment computes a reward $r_t$ from the action, which is used to optimize the neural network.}
\label{fig:accel_controller_diagram}
\end{figure*}
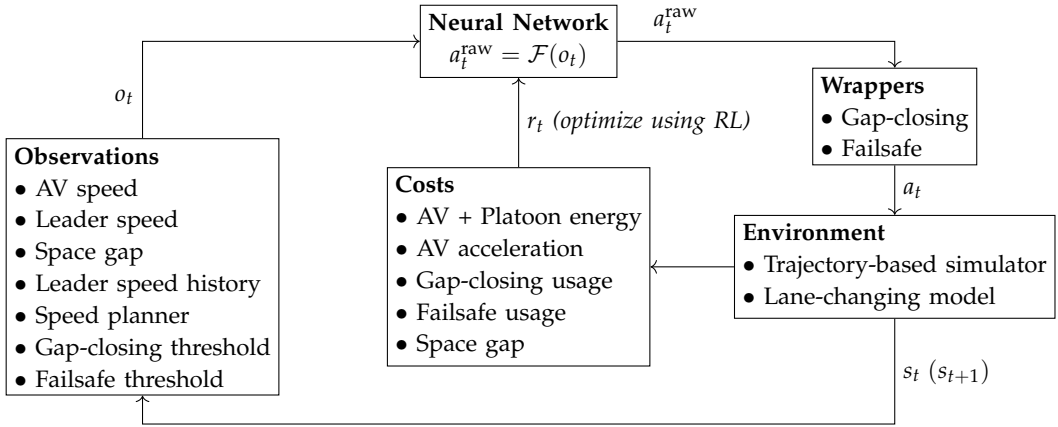


\subsubsection{Observation Space}
\label{sec:accel_obs_space}
The observation space at time $t$ consists of:
\begin{itemize}
    \item $v_t$, AV speed
    \item $v_t^\text{lead}$, AV leader vehicle speed
    \item $h_t$, space gap (bumper-to-bumper distance) between AV and leader
\end{itemize}

The "leader" vehicle is defined to be the vehicle in front of the AV, and can be seen in Figure~\ref{fig:sim_platoon}. Note that all distances are  in $m$, speeds in $m/s$, and accelerations in $m/s^2$. Furthermore, two gap thresholds are also included: $h_t^\text{min}$, which is the failsafe threshold below which the vehicle will always brake, and $h_t^\text{max}$, the gap-closing threshold above which the vehicle will always accelerate. These thresholds are described in the following action space section. We also include the history of the ego vehicle's speed over the last 0.5 seconds. Finally, the observation space includes traffic information from the Speed Planner. This consists of the current target speed $v_t^\text{sp}$, as well as the target speeds 200m, 500m, and 1km downstream of the AV's current position. Note that the AV controller only observes its leading vehicle and that there is no explicit communication between AVs. All observations provided to the RL agent are rescaled to the range $[-1,1]$.

\subsubsection{Action Space}

This section contains several numerical values for parameters, whose origin are not explained in the text. They come from vehicle specifications, estimation, modeling and are empirical or design-based. They are shown here for completeness.

The output of the policy is an instantaneous acceleration in the range $a_t \in [-3, 1.5] m/s^2$, chosen to maintain a comfortable driving experience. We assume that the acceleration is instantly achieved when the command is passed to the vehicle. This acceleration output is then wrapped by a gap-closing component and a failsafe component, to ensure and add a soft constraint on safety and reasonable behavior. The gap above which the gap-closing wrapper will activate is $h_t^\text{max} = \displaystyle \max(120, 6 v_t^\text{av})$, resulting in the control being overridden and the AV forced to accelerate if its space gap becomes larger than $120$m, or its time gap larger than $6$s. $120$m is chosen as an arbitrary large space gap value. Note that in the absence of a leader, the space gap is set to some arbitrarily large value and the AV will accelerate. In freeflow, the AV behavior will be to drive at the road speed limit.

The resulting acceleration is further wrapped by a low-gap failsafe, to encourage safe following distances and prevent collisions. Note that this is enacted in simulation to ensure better training behavior; during deployment on the road, many other failsafes are present on the car. We define the \acronym{time-to-collision} (TTC) $\Delta^\text{TTC}_t$ of the ego and leader vehicles at time $t$ as
\begin{equation}
    \Delta^\text{TTC}_t = \begin{cases} 
\displaystyle \frac{h_t}{v_t^\text{diff}} & \text{if } v_t^\text{diff} > 0, \\
+\infty & \text{otherwise,}
\end{cases}
\end{equation}
where
\begin{equation}
v_t^\text{diff} = \left[ v_t^\text{av}\left(1 + \frac{4}{30}\right) + 1\right] - v_t^\text{lead}.
\end{equation}
In case this modified TTC goes below 6 seconds, the failsafe will trigger, overriding the outputted acceleration and forcing the AV to brake. The failsafe triggering threshold can be computed $h_t^\text{min} = 6 v_t^\text{diff}$. The reasoning for using $v_t^\text{av}\left(1 + \frac{4}{30}\right) + 1$ instead of simply $v_t^\text{av}$ is that if $v_t^\text{diff}$ was set equal to $v_t^\text{av} - v_t^\text{lead}$, and say both the ego and lead vehicles are driving at 30$m/s$, then the failsafe triggering threshold would be $h_t^\text{min} = 0$ which wouldn't be triggered even if both vehicles maintained a space gap as small as 1m. We cannot allow such a dangerous situation to occur, consequently we artificially increase the AV speed for additional safety precautions. With the new formulation, in our example above the minimal gap would now be $h_t^\text{min} = 30$m. This, when driving at 30 $m/s$, corresponds to a time gap of $1$s which is far more reasonable (and is also how we empirically chose the coefficients). Note that the $+1$ term helps to also slightly exaggerate the AV speed when $v_t^\text{av}$ is very low, ensuring robustness at both low and high speeds.

The final RL acceleration is given by:
\begin{equation}
a_t^\text{out} =  \begin{cases} 
      -3 & \text{if } \Delta^\text{TTC}_t \leq 6 \quad (\text{iff. } \; h_t \leq h_t^\text{min} ), \\
      1.5 & \text{if } \Delta^\text{TTC}_t > 6 \text{ and } h_t \geq h_t^\text{max}, \\
      a_t & \text{otherwise}, 
   \end{cases}
   \label{eq:wrapped_accel}
 \end{equation}
 which is further clipped to ensure that the speed $v^\text{av}_t$ remains within the speed limit boundaries $[0, 35]\frac{\text{m}}{\text{s}}$.

\subsubsection{Reward Function}

The primary objective of the MVT was to improve the overall energy consumption of the stop-and-go traffic. This could be modeled as a reward of $0$ for all time steps, except at the last time step where it would be the negative overall fuel economy of all the vehicles (that is, the total miles driven divided by total gallons of gasoline consumed). However, that would lead to a highly sparse reward which is hard to optimize. Instead, we considered proxies that can be optimized at each time step; ultimately we minimize the instantaneous fuel consumption of the AV and of a platoon of human-driven vehicles following it at each time step. Over the course of the simulation, this reward integrates exactly to the total fuel consumed; since the reward is computed over a fixed length of road, it is effectively rewarding the overall fuel economy. For driver comfort, and as an additional proxy for energy consumption, we also minimize the squared instantaneous accelerations. Furthermore, we penalize going outside of the failsafe and gap-closing thresholds---that is, we penalize the controller's output being overridden by the wrappers, to entice the policy not to rely on them. Having these wrappers, or penalizing large gaps in another way (we also penalize time gap), is important to maintain reasonable, socially-acceptable gaps and to avoid the degenerate energy-optimal solution: to simply come to a stop. 

This is formalized as the reward function $r_t$, which is given by:
\begin{align*}
    \displaystyle r_t = & - c_1 \frac{1}{n} \sum_{i=1}^{n} E_t^i - c_2 (a_t^\text{out})^2 - c_3 \mathds{1} \left[{h_t \notin [h_t^\text{min}, h_t^\text{max}]} \right] \\ & - c_4\frac{h_t}{v_t^\text{av}}\mathds{1} \left[h_t > 10 \land v_t^\text{av} > 1\right]
\end{align*}

where $E_t^i$ is the instantaneous fuel consumption of vehicle $i$ at time $t$, index $i=1$ corresponds to the AV, and indexes $i=2$ to $i=n$ correspond to the following $n-1$ human-driven vehicles. $\mathbb{1}$ is defined such that $\mathbb{1} \left[\mathcal{P}\right] = 1 \text{ if } \mathcal P \text{ is true, } 0 \text{ otherwise}$.

\begin{sidebar}{Energy Models Compatible with Optimization}
\section[Energy Models Compatible with Optimization]{by Nour Khoudari and Benjamin Seibold}\label{sbar-energy}

\renewcommand{\thestable}{S\arabic{stable}}
\renewcommand{\thesfigure}{S\arabic{sfigure}}

\sdbarinitial{V}ehicle energy models that accurately reproduce the dynamics of a specific vehicle tend to be complex and possess discontinuities (for instance at gear shifts). This renders their use in energy optimization challenging, because optimizers may get trapped in local minima that are exploiting the gear shifting characteristics of one specific vehicle. However, for real traffic composed of similar but not identical vehicles, such behavior is undesirable. Thus, optimization-friendly energy model frameworks are needed in the design and training of controllers.

\sdbarfig{\includegraphics[width=19.0pc]{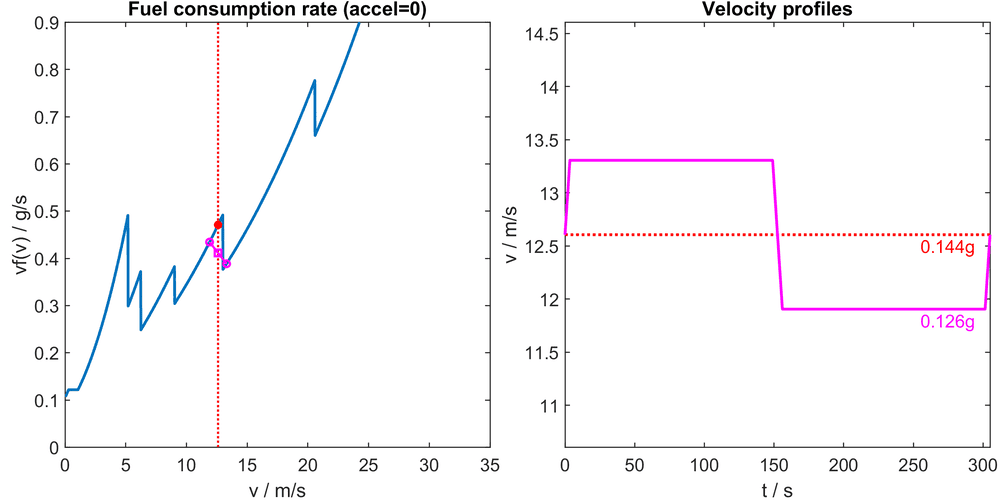}}{Fuel consumption rate as function of speed (for zero acceleration) of an energy model with discontinuities due to gear shifts (left), showing that constant speed driving is not fuel-optimal driving (right).\label{fig:fuel_consumption_rate_with_velocity_profiles}}

Figure~\ref{fig:fuel_consumption_rate_with_velocity_profiles} shows how an energy model with realistic gear shifting will have the tendency to yield unsteady profiles (pink profile) that are specific to one vehicle, rather than producing smooth steady driving (red profile) that is more desirable for traffic composed of different vehicles.

To address these needs, energy models are developed and used \cite{Energy} that are of the form $f(v,a,\theta)$ where the fuel consumption rate function $f$ possesses desirable ``convexity” properties ($v$ is speed, $a$ acceleration, and $\theta$ the road grade). The models average out the non-convex behavior of the vehicle-specific accurate models via carefully constructed fitted polynomial functions. Besides yielding a significant computational speed-up, these simplified models also are more representative of an average energy profile over many makes and models from a given vehicle class.

\sdbarfig{\includegraphics[width=19.0pc]{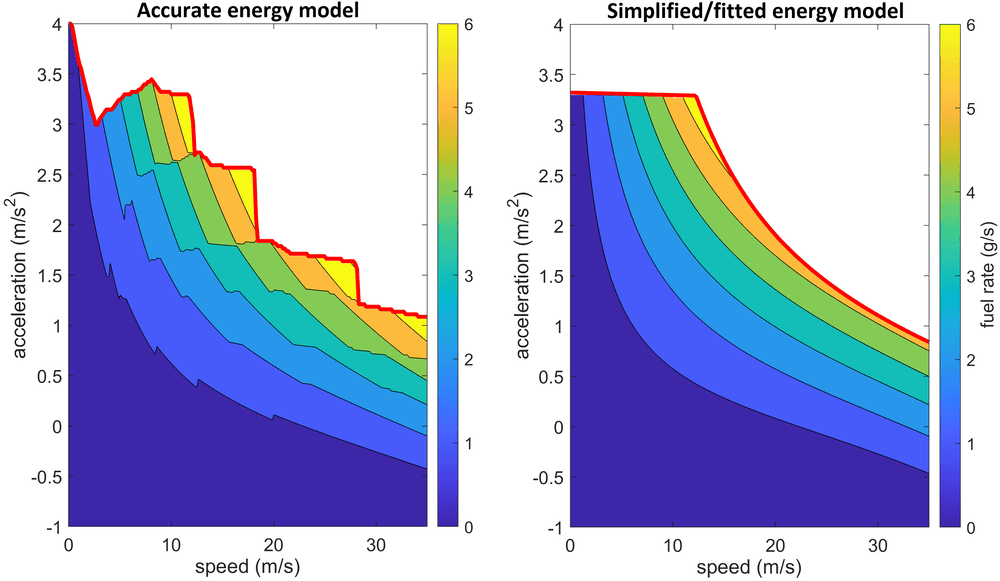}}{Contours of (a) a vehicle-specific energy model with discontinuities at gear shifts, and (b) the fitted/averaged simplified energy model that has desirable convexity properties.\label{fig:contours_energy_model}}

Figure~\ref{fig:contours_energy_model} compares the fuel rate contours of an energy model with discontinuities at gear shifts (left) against the contours of the simplified model used here (right), showing the absence of the gear shift jumps in the latter. The fitted energy function is not convex in the sense of having a positive definite Hessian matrix (vehicle energy models are very far from possessing such a structure), but rather: for a fixed road grade, $f$ is convex in the $a$-direction for a constant $v$, and $f/v$ is convex with respect to $v$ for constant $a$ (with a minor exception of the fuel-cut boundary). Mathematically, this notion of ``convexity” provides precisely the properties that imply that constant speed solutions are equivalent to fuel-optimal trajectories (under reasonable constraints).

\end{sidebar}

\subsection{Acceleration-based RL Simulation Results}
\label{sec:accel_sim_res}

\begin{pullquote}
For the first time, microsimulation has enabled the training of deep reinforcement learning policies capable of dampening large scale traffic oscillations in a field operational test.
\end{pullquote}

After training, we have a controller that provides acceleration based on AV speed, leader speed, space gap, and other parameters outlined in the ``\nameref{sec:accel_obs_space}" section. The control combines the trained neural network with the failsafe and gap-closing components detailed in Eq.~(\ref{eq:wrapped_accel}). 

During the analysis phase, we extensively explore the dynamics of a leading vehicle replaying a preset trajectory from a dataset, trailed by a series of platoons. Each platoon comprises a single AV (either IDM-controlled or RL-controlled) followed by 19 IDM-controlled human vehicles, which translates to a 5\% penetration rate of AVs. 

\begin{figure*}
    \centering
    \includegraphics[width=0.99\linewidth]{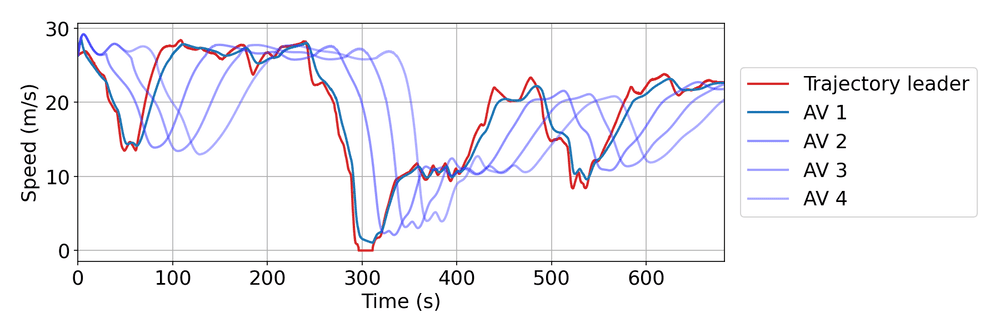}
    \caption{Trajectory leader vehicle replaying a trajectory from the dataset, followed by 10 platoons each composed of 1 AV and 19 IDM human vehicles, which corresponds to 200 vehicles (not including the trajectory leader) and a 5\% AV penetration rate. The plot displays the speed of the trajectory leader as well as the first 4 AVs in the platoon. The remaining ones are omitted for visibility but follow a similar trend.}
    \label{fig:accel_result1}
\end{figure*}

Figure~\ref{fig:accel_result1} portrays the speed trends of the trajectory leader and the AVs in the first four platoons. While the remaining vehicles are not explicitly shown for clarity, they follow a consistent trend with the leading AVs. A significant observation around the $t=300$s mark is the trajectory leader's momentary stop at the trough of the wave. In contrast, successive AVs not only avoid halting but also never decelerate to speeds as low as the preceding vehicle. This pattern repeats in the subsequent wave's crest, where the AVs do not reach speeds as elevated as the leader. Such behavior of the AVs demonstrates a pronounced smoothing effect, which diminishes the speed variance.

\begin{figure}
    \centering
    \includegraphics[width=0.95\linewidth]{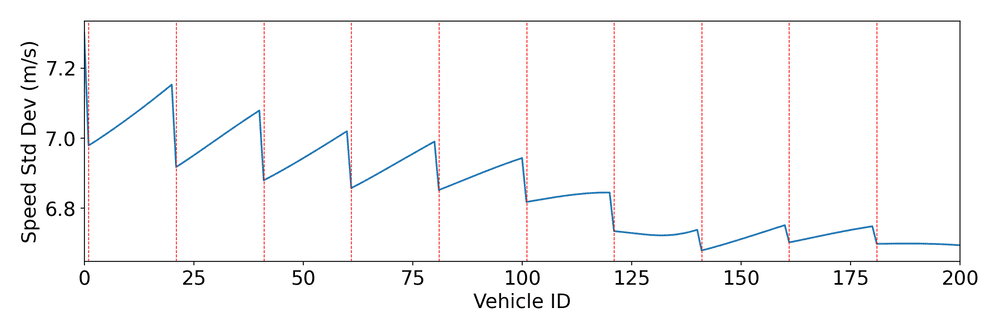}
    \includegraphics[width=0.95\linewidth]{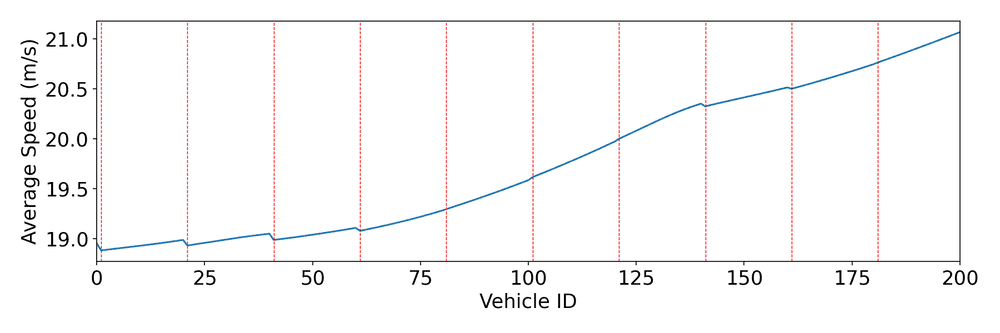}
    \includegraphics[width=0.95\linewidth]{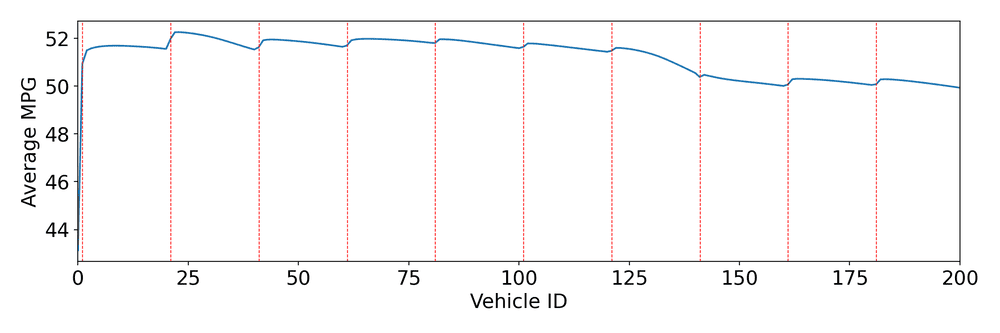}
    \includegraphics[width=0.95\linewidth]{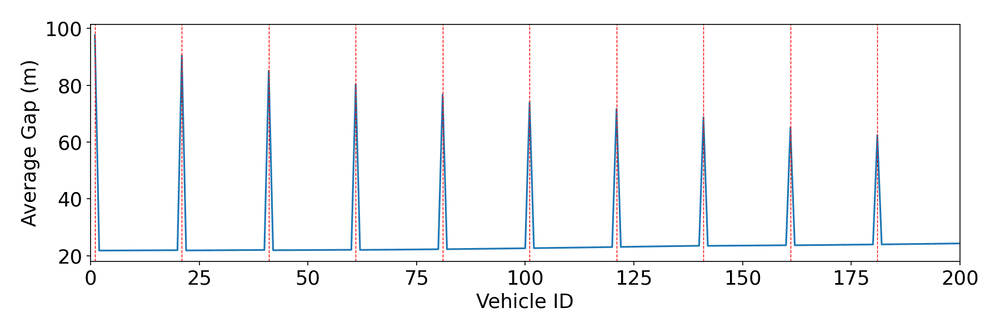}
    \caption{Trajectory leader vehicle replaying a trajectory from the dataset (displayed in Figure~\ref{fig:accel_result1}), followed by 10 platoons each composed of 1 AV and 19 IDM human vehicles, which corresponds to 200 vehicles (not including the trajectory leader) and a 5\% AV penetration rate. We plot metrics aggregated over the whole simulation as a function of vehicle ID, 0 being the trajectory leader, 200 the last vehicle in the platoon, and $1 + 20k$ for $k \in \{ 0, 1, \dots, 9\}$ are AVs (indicated by vertical red lines). From top to bottom, the aggregated metrics are speed variance, speed average, miles-per-gallon average and space gap average. Large spikes correspond to the start of a platoon.}
    \label{fig:accel_result2}
\end{figure}

Further analysis, as seen in Figure~\ref{fig:accel_result2}, provides an aggregated metrics overview throughout the simulation relative to the vehicle ID. Every subsequent AV manifests a decrease in speed standard deviation, corroborating the dampening of waves as intuited from Figure\ref{fig:accel_result1}. Furthermore, these AVs elevate the system's average speed while considerably augmenting the cumulative energy efficiency across all vehicles. They achieve this by creating larger gaps than the IDM vehicles, as evidenced in the bottom subfigure. It's also noteworthy that the average gaps established by AVs reduce progressively along the platoons, culminating in waves becoming increasingly attenuated.

\begin{figure}
    \centering
    \includegraphics[width=0.75\linewidth]{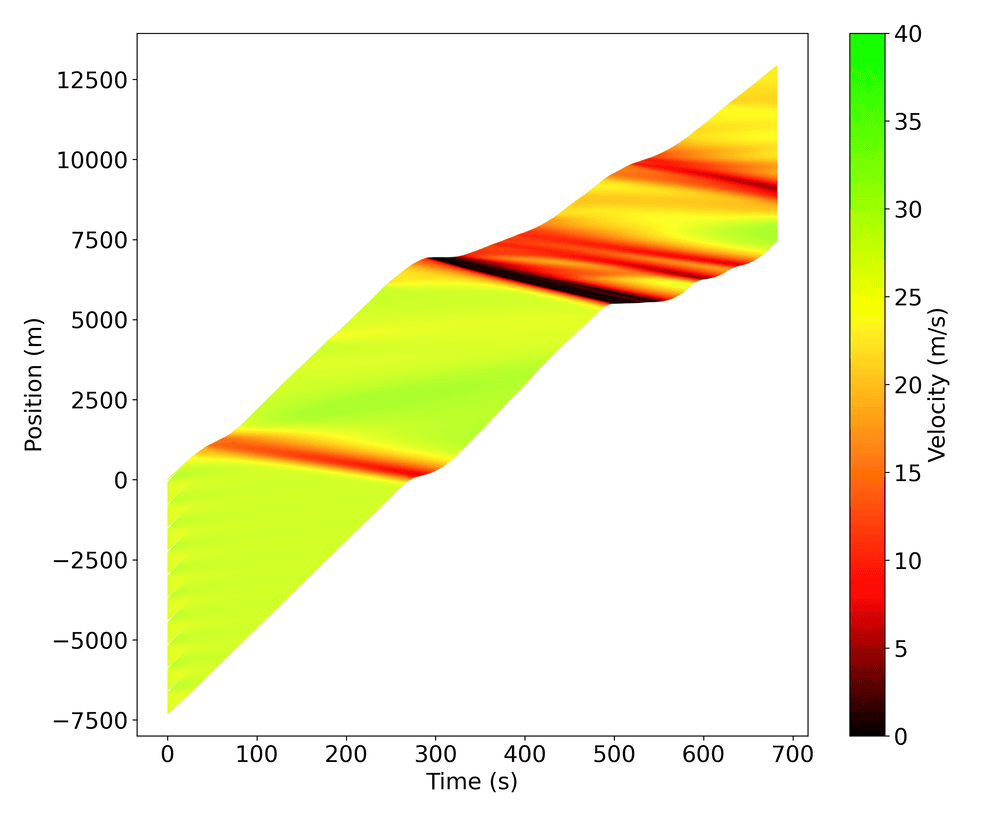}
    \includegraphics[width=0.75\linewidth]{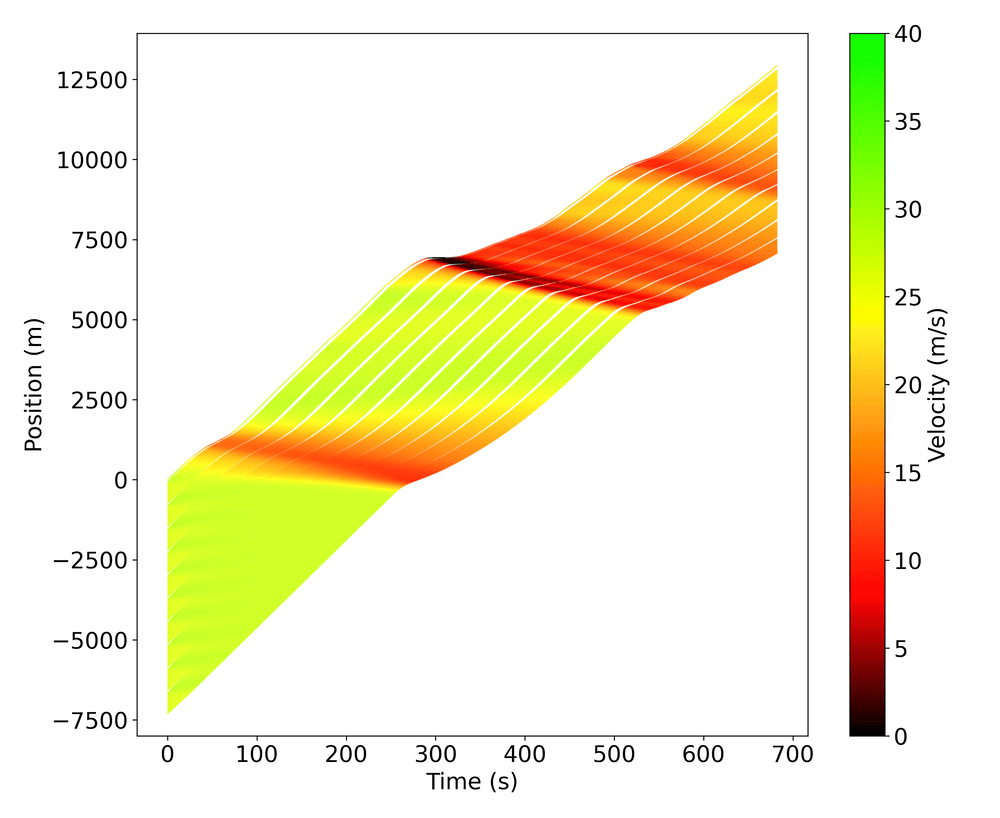}
    \caption{Trajectory leader vehicle replaying a trajectory from the dataset (displayed in Figure~\ref{fig:accel_result1}), followed by 10 platoons each composed of 1 AV and 19 IDM human vehicles, which corresponds to 200 vehicles (not including the trajectory leader) and a 5\% AV penetration rate. We plot a time-space diagram when AVs are IDM-controlled (top) and when they are RL-controlled (bottom). Colors correspond to vehicle speeds. The black color region corresponds to the part where the trajectory leader comes to a stop around the $t=300$s mark (see Figure~\ref{fig:accel_result1}), which the AVs manage to dampen (bottom). On this particular trajectory, the AVs improve the fuel efficiency of all of the 200 vehicles by 12.67\%, as evidenced by the smoothed out colors.}
    \label{fig:accel_result3}
\end{figure}

In Figure~\ref{fig:accel_result3}, a time-space diagram provides a juxtaposition between IDM-controlled AVs and those controlled by RL. The portion depicted in black signifies the trajectory leader's halt at the $t=300$s mark, as previously seen in Figure~\ref{fig:accel_result1}. It becomes evident that RL-controlled AVs significantly dampen this effect, yielding more uniformity in trajectory patterns when compared to their IDM counterparts, as evidenced by the smoother color gradient in the RL-controlled time-space diagram.

\begin{figure}
    \centering
    \includegraphics[width=0.99\linewidth]{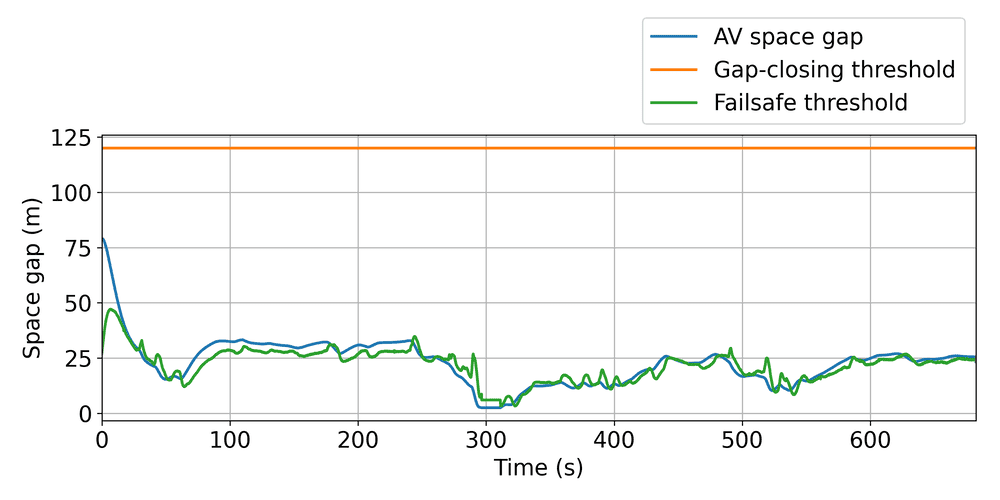}
    \includegraphics[width=0.99\linewidth]{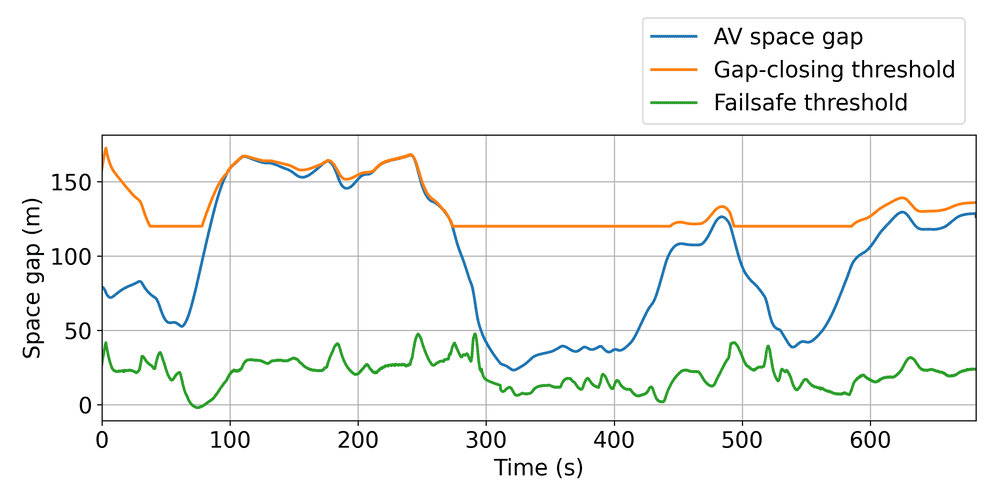}
    \caption{Trajectory leader vehicle replaying a trajectory from the dataset (displayed in Figure~\ref{fig:accel_result1}), followed by 10 platoons each composed of 1 AV and 19 IDM human vehicles, which corresponds to 200 vehicles (not including the trajectory leader) and a 5\% AV penetration rate. We plot the space gap of the first AV in the platoon, in the case where it is IDM-controlled (top) and RL-controlled (bottom). The orange line represents the gap above which our gap-closing wrapper triggers, while the green line represents the gap below which our failsafe wrapper triggers. Note that the wrappers are not applied on IDM but are still shown for comparison.}
    \label{fig:accel_result4}
\end{figure}

Lastly, the space gaps for the foremost AV in the platoon are illustrated in Figure~\ref{fig:accel_result4} for both IDM and RL-controlled AVs. IDM predominantly maintains a gap equivalent to the minimal failsafe gap permissible for RL control. On the contrary, RL-controlled AVs shows a tendency for opening larger gaps, providing them with the flexibility to accommodate slow-downs and thus absorb waves, underlining the sophisticated adaptability of RL in these scenarios.

\subsection{ACC-based RL Problem Formulation}
\label{sec:accformulation}

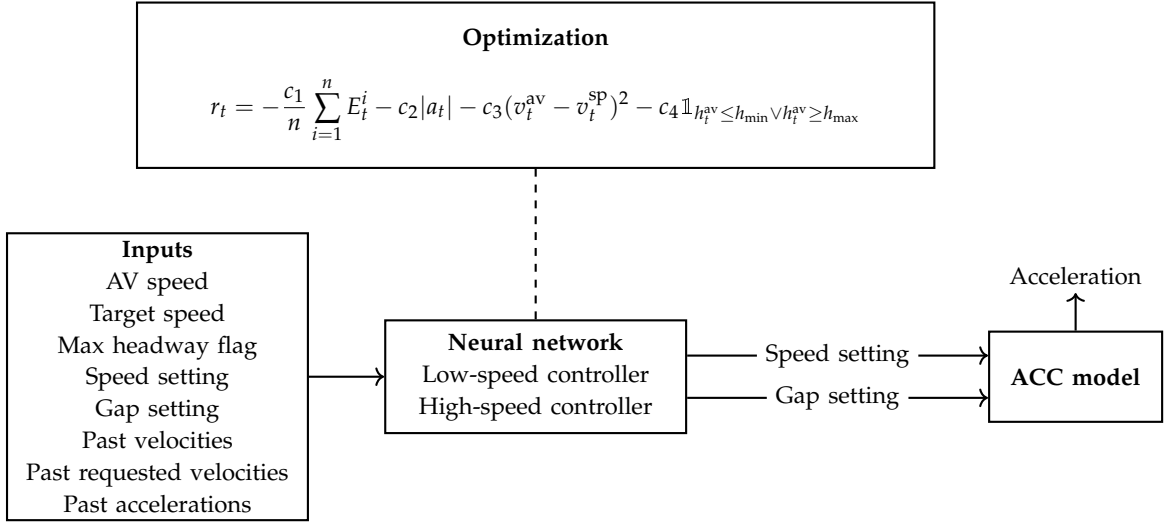
\begin{figure*}
\centering
\begin{tikzpicture}[thick, every text node part/.style={align=center}]

\node [draw, rectangle,
    minimum width=4cm,
    minimum height=3cm,
]  (inputs) {
\textbf{Inputs} \\
AV speed \\
Target speed \\
Max headway flag \\
Speed setting \\
Gap setting \\
Past velocities \\
Past requested velocities \\
Past accelerations
};
 
\node [draw,
    minimum width=4cm, 
    minimum height=1.5cm,
    right=1cm of inputs
] (nn) {\textbf{Neural network} \\ Low-speed controller \\ High-speed controller};
 
\node [draw,
    minimum width=2.3cm, 
    minimum height=1.2cm,
    right=4cm of nn
] (acc) {\textbf{ACC model}};

\node [text centered, above=.5cm of acc
] (accel) {Acceleration};
 
\node [draw,
    minimum width=10.6cm, 
    minimum height=2.2cm,
    above=2cm of nn
] (optim) {\textbf{Optimization} \\ \\ $\displaystyle r_t = - \frac{c_1}n \sum_{i=1}^{n} E_t^i - c_2 |a_t| - c_3 (v_t^\text{av} - v_t^\text{sp})^2 - c_4 \mathbbm{1}_{{h_t^\text{av} \leq h_\text{min} \lor h_t^\text{av} \geq h_\text{max}} }$};
  
\draw[->] (inputs.east) -- (nn.west);

\path ([yshift=8] nn.east) -- node[] (speedsetting) {Speed setting} ([yshift=8] acc.west);
\draw[->] ([yshift=8] nn.east) -- (speedsetting) -- ([yshift=8] acc.west);

\path ([yshift=-8] nn.east) -- node[] (gapsetting) {Gap setting} ([yshift=-8] acc.west);
\draw[->] ([yshift=-8] nn.east) -- (gapsetting) -- ([yshift=-8] acc.west);

\draw[->] (acc.north) -- (accel.south);
\draw[-.,dashed] (nn.north) -- (optim.south);
 
\end{tikzpicture}
\caption{Diagram summarizing the design of the ACC-based controller. The observations described are the inputs into the controller. The controller outputs a speed setting and a gap setting, which are then input into the Rogue's native ACC model, which finally outputs an acceleration. The neural network is optimized via the reward function described in this section.}
\label{fig:acc_controller_diagram}
\end{figure*}

The second of the two major classes of algorithms that were designed is the ACC-based RL controller. While both controllers ultimately control the driving behavior of the AV with the goal of minimizing fuel consumption, they do this in different ways. While the output of the acceleration-based RL controller is simply a scalar indicating what acceleration to actuate, the output of the ACC-based RL controller controls the inputs to the vehicle's native ACC system, and contains two values: (1) the speed setting and (2) the gap setting, both of which are described below in Section ''\nameref{sec:acc-action-space}''.

The final rendition of the controller that was ultimately deployed during the MVT is a combination of two controllers, with a transition being triggered at 60 mph. They are referred to as the low-speed controller and high-speed controller, with the low-speed controller operating when the vehicle is driving below 60 mph and the high-speed controller operating when the vehicle is driving above 60 mph.

In \cite{lee2024traffic},
we describe additional details regarding control interfaces with each vehicle. When interacting through ACC-based control, an approximate model was used to convert desired changes in speed into set point changes for the ACC, which was replicated for RL training purposes. Interaction through ACC-based control reduces the theoretical controllability of the system. However, it has the benefit that it broadens the implementation pathway to include more vehicle types, and depends on stock safety behaviors in order to protect vehicle occupants.


In Table~\ref{table:controller-variations}, we enumerate the progression of the controller until its final stage, including all the intermediate observation spaces, reward functions, and results that prompted design changes, which led to the final controller. For the final problem formulation, refer to the sections named \nameref{sec:final-obs-space} and \nameref{sec:final-reward-function}. The action space remains the same throughout all these variations. The switch between acceleration-based control and ACC-based control can be noted between versions v4 and v5. This also marks the progression to "undetectable" controllers, when lead vehicle distance was removed from the controller's observation space (although lead vehicle distance was still used by the car for its own execution of the ACC), in order to support as many different vehicle types as possible for the final field test. 
ACC-based control with no lead vehicle information provided the opportunity to field control on nearly 100 vehicles, instead of acceleration-based control on only a few vehicles.

\begin{figure}[h]
    \centerline{\includegraphics[width=18.0pc]{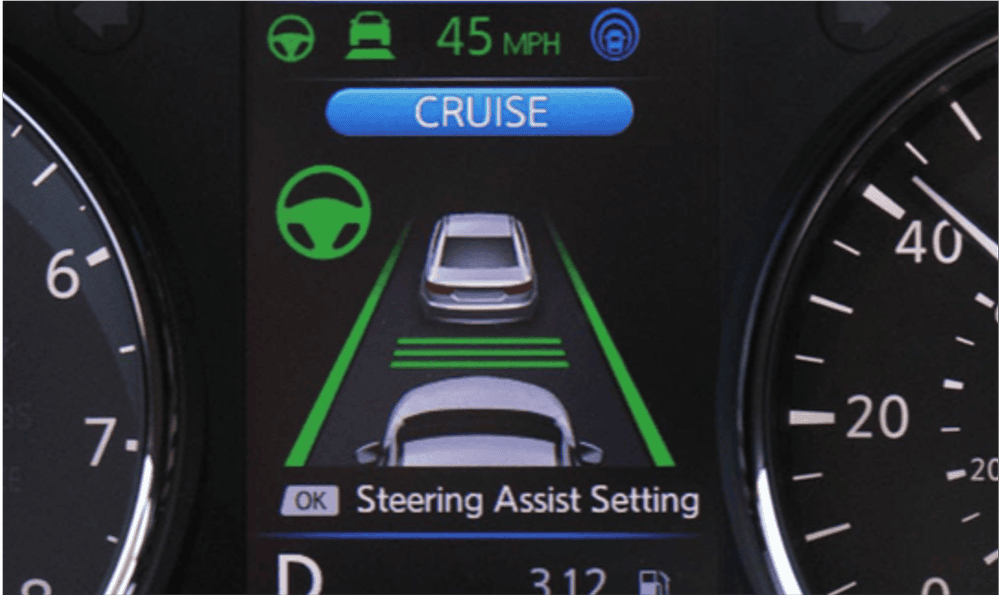}}
    \caption{Dashboard view while the RL controller is being run. The speed setting is set at 45 MPH and the gap setting is set at three bars. 
    }
    \label{fig:nissan_dashboard}
\end{figure}

\begin{table*}[]
\centering
\begin{tabular}{|m{1.2cm}|m{2.2cm}|m{3cm}|m{4.3cm}|m{3cm}|}
\hline
\multicolumn{5}{|c|}{Controllers v1-v5} \\ \hline

\textbf{Version} & \textbf{Action Space} & 
\textbf{Observation Space} & \textbf{Reward Function} & \textbf{Results} \\ \hline

v1
& \vspace{2mm}
    Acceleration
& \vspace{2mm}
\begin{itemize}
    \item $v^{av}$, AV speed
    \item $v^{l}$, leader speed
    \item $h$, space gap
\end{itemize} 
& $$r_t = -c_0 E_t - c_1 a_t^{av2}$$
& Simulated cars crash in order to avoid penalties, opening large space gaps to optimize for the energy model  \\ \hline

v2
& \vspace{2mm}
    Acceleration
& \vspace{2mm}
\begin{itemize}
    \item $v^{av}$, AV speed
    \item $v^{l}$, leader speed
    \item $h$, space gap
    \item Failsafe threshold
    \item Gap-closing threshold
\end{itemize} 
& $$r_t = 1-c_0 E_t - c_1 a_t^{av2} - c_2P_t$$
, where $P_t = \mathbb{1}_{h_t^{av} \leq h_{min} \lor h_t^{av} \geq h_{max}}$
& \vspace{2mm} Continued simulation crashes and large space gaps; most controllers worsen MPG; for controllers with AV energy benefits, system energy consumption does not improve \vspace{2mm} \\ \hline

v3
& 
    Acceleration
& 
\begin{itemize}
    \item $v^{av}$, AV speed
    \item $v^{l}$, leader speed
    \item $h$, space gap
    \item Previous leader speeds
    \item Failsafe threshold
    \item Gap-closing threshold
\end{itemize}
& $$r_t = 1-c_0 \sum_{i=1}^n E_t^i  - c_1 a_t^{av2} - c_2I_t$$
In controller v3, instantaneous energy is taken as a function of an entire platoon of vehicles (the AV and the 24 human vehicles behind it) instead of just the AV. The horizon is also increased for controller v3.
& \vspace{2mm} Eliminated simulated crashes and stopping behavior; MPG gains significantly above baseline (System MPG: \textbf{+20\%}, AV MPG: \textbf{+27\%}); agent maintains maximum distance allowed by gap-closing; reward function aligned with MPG \& comfort objectives, RL successfully optimizing reward function \vspace{2mm} \\ \hline

v4
& \vspace{2mm}
    Acceleration
& \vspace{2mm}
\begin{itemize}
    \item $v^{av}$, AV speed
    \item $v^{l}$, leader speed
    \item $h$, space gap
    \item Previous leader speeds
    \item Failsafe threshold
    \item Gap-closing threshold
    \item Target speed (speed planner)
\end{itemize}
& $$r_t = 1-c_0 \sum_{i=1}^n E_t^i - c_1 a_t^{av2} - c_2I_t - c_3 h_t$$
A small gap penalty is introduced to discourage the controller from staying too close to the gap-closing threshold.
& Reasonable space gaps,  MPG gains substantially above baseline (System MPG: \textbf{+23\%}, AV MPG: \textbf{+27\%})
 \\ \hline

 v5 (Deployed)
& \vspace{2mm}
\begin{itemize}
    \item ACC speed setting
    \item ACC gap setting
\end{itemize}
& \vspace{2mm}
\begin{itemize}
    \item $v^{av}$, AV speed
    \item Previous accelerations$^*$
    \item Previous velocities$^*$
    \item Previous requested velocities$^*$
    \item Target speed
    \item Minicar flag
    \item Max headway flag
    \item Speed setting
    \item Gap setting
\end{itemize}
& $r_t = 1 - c_1 a_t^2 - c_2(v_t^{av} - v_t^{sp})^2 - \frac{c_3}{n} \sum_{i=1}^n E_t^i - c_4 \mathbb{1}_{h_t^{av} \leq h_{min} \lor h_t^{av} \geq h_{max}},$
& Reasonable space gaps,  MPG gains substantially above baseline (System MPG: \textbf{+23\%}). \break $^*$Since the deployed controller is a combination of two RL controllers, these states are slightly different. View \ref{sec:final-obs-space} for these distinctions. For more details on performance, see Table~\ref{tab:freeze_evaluations}.
 \\ \hline

\end{tabular}
\vspace{2mm}
\caption{A description of the progression of controllers, from the first rendition to the final controller that was deployed at the MVT. High-level information on the action space, observation space, reward function, and results are given for each version of the controller.}
\label{table:controller-variations}
\end{table*}

\subsubsection{Action Space}
\label{sec:acc-action-space}

The final action space of the ACC-based controller we deployed consists of:
\begin{enumerate}
    \item The speed setting, which dictates the maximum speed that the ACC can drive at;
    \item The gap setting, which takes one, two, or three bars, with each bar indicating a higher allowable gap. Each bar corresponds roughly to constant time gaps of 1.2, 1.5, and 2.0 seconds.
\end{enumerate}
Depicted in Figure~\ref{fig:nissan_dashboard}, the speed setting appears as the green speed number at the top, and the gap setting appears as green horizontal bars between the two vehicles. In implementation, the controller actions are realized as requested settings, sent to software which sends commands through the in-vehicle network~\cite{bunting2021libpanda,elmadani2021can,nice2023middleware}, and ultimately are realized by physical systems like the engine and wheel torques.


\subsubsection{Observation Space}
\label{sec:final-obs-space}
The observation space of the ACC-based RL controller is as follows:
\begin{itemize}
    \item $v$, velocity of the AV;
    \item $v_s$, target speed given by the Speed Planner;
    \item $h_{max}$, the maximum headeway recommended by the Speed Planner;
    \item $l$, a flag that indicates whether there is a leader vehicle within approximately 80 meters;
    \item $s$, the current ACC speed setting;
    \item $g$, the current ACC gap setting.
\end{itemize}

 The observation spaces of the low-speed and high-speed controller are slightly different. In addition to the base state listed above, the high-speed controller contains the following states:
\begin{itemize}
    \item $\overrightarrow{a_{prev}}$, the accelerations during the previous 6 time steps.
\end{itemize}

The low-speed controller contains the following states in addition to the base state:
\begin{itemize}
    \item $v_{s200}$, target speed given by the Speed Planner 200m downstream;
    \item $v_{s500}$, target speed given by the Speed Planner 500m downstream;
    \item $v_{s1000}$, target speed given by the Speed Planner 1000m downstream;
    \item $\overrightarrow{a_{prev}}$, the accelerations during the previous 5 time steps;
    \item $\overrightarrow{v_{prev}}$, the velocities during the previous 10 time steps;
    \item $\overrightarrow{r_{prev}}$, the requested velocities during the previous 10 time steps.
\end{itemize}

\subsubsection{Reward Function}
\label{sec:final-reward-function}
The final reward function for the ACC-based controller was:
$$r_t = 1 - c_1 a_t^2 - c_2(v_t^{av} - v_t^{sp})^2 - \frac{c_3}{n} \sum_{i=1}^n E_t^i - c_4 \mathbb{1}_{h_t^{av} \leq h_{min} \lor h_t^{av} \geq h_{max}},$$
where $c_1-c_4$ are coefficients, $a_t^2$ is an acceleration penalty; $(v_t^{av} - v_t^{sp})^2$ is a squared penalty on the difference between the Speed Planner's suggested speed and the actual speed; $\frac{1}{n}\sum_{i=1}^n E_t^i$ is instantaneous fuel consumption; and the last indicator term is an intervention penalty that is invoked if the space gap is less than the minimum space gap (and thus invokes the failsafe) or greater than the max space gap (and thus invokes gap closing mode).

\subsection{ACC-based RL Simulation Results}
The deployed ACC-based RL controller interfaces with the Rogue's ACC system by toggling its speed and gap settings. Note that the version of the controller that is deployed during the MVT interfaces with several hardware layers, detailed in ''\nameref{sec:validation}'', so deployment results are discussed in a separate section. We now proceed to analysis of the trained controller.

A series of candidate controllers, including some non-RL controllers, were tested and evaluated, with their results on display in Table~\ref{tab:freeze_evaluations}. Controllers are evaluated in three different traffic scenarios: a) Bottleneck, b) Shockwave, and c) Freeflow. They are also evaluated across two different speed planners, here labeled "Speed Planner A" and "Speed Planner B," the combination of which was ultimately deployed. Fuel economy (described in ''\nameref{sbar-energy}''), throughput, and speed are measured in this evaluation. The other controller, titled "Microaccel," is also featured in this issue 
\cite{hayat2023traffic}.
The performance metrics on display inform the choice to create the deployed RL controller, a mixture of the "low-speed controller" and "high-speed controller", also exhibited in this table. The low-speed controller achieves higher performance than the high-speed controller across all metrics in the bottleneck scenario. Compared to the IDM baseline and maximizing across the two speed planners, it achieves $8.22\%$ gains in fuel economy, $14.73\%$ gains in throughput, and $8.68\%$ gains in speed for  and in fuel economy and throughput in the shockwave scenario. In the shockwave scenario, the low-speed RL controller achieves, over the two speed planners, $22.96\%$ gains in fuel economy and $2.75\%$ gains in throughput. It decreases speed by $2.14\%$. In the freeflow scenario, the low-speed controller saw difficulty with maintaining traffic speeds, thus resulting in a $-27.52\%$ decrease in speed. The high-speed RL controller slightly improves the speed and throughput by $1.6\%$ and $0.26\%$, respectively. Thus, the deployed controller (henceforth simply referred to as the RL controller) is composed of the low-speed controller below 60 mph and the high-speed controller above 60 mph.

\begin{figure}
    \centering
    \includegraphics[width=.6\linewidth]{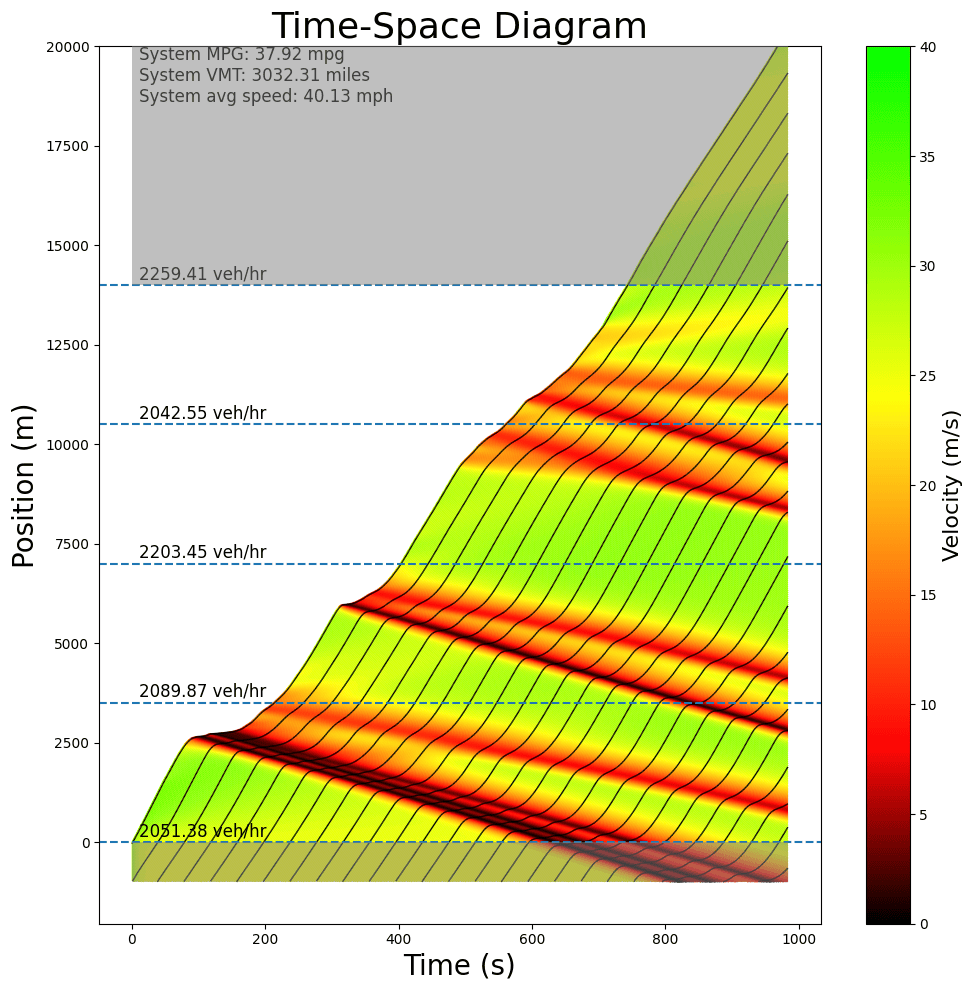}
    \includegraphics[width=.6\linewidth]{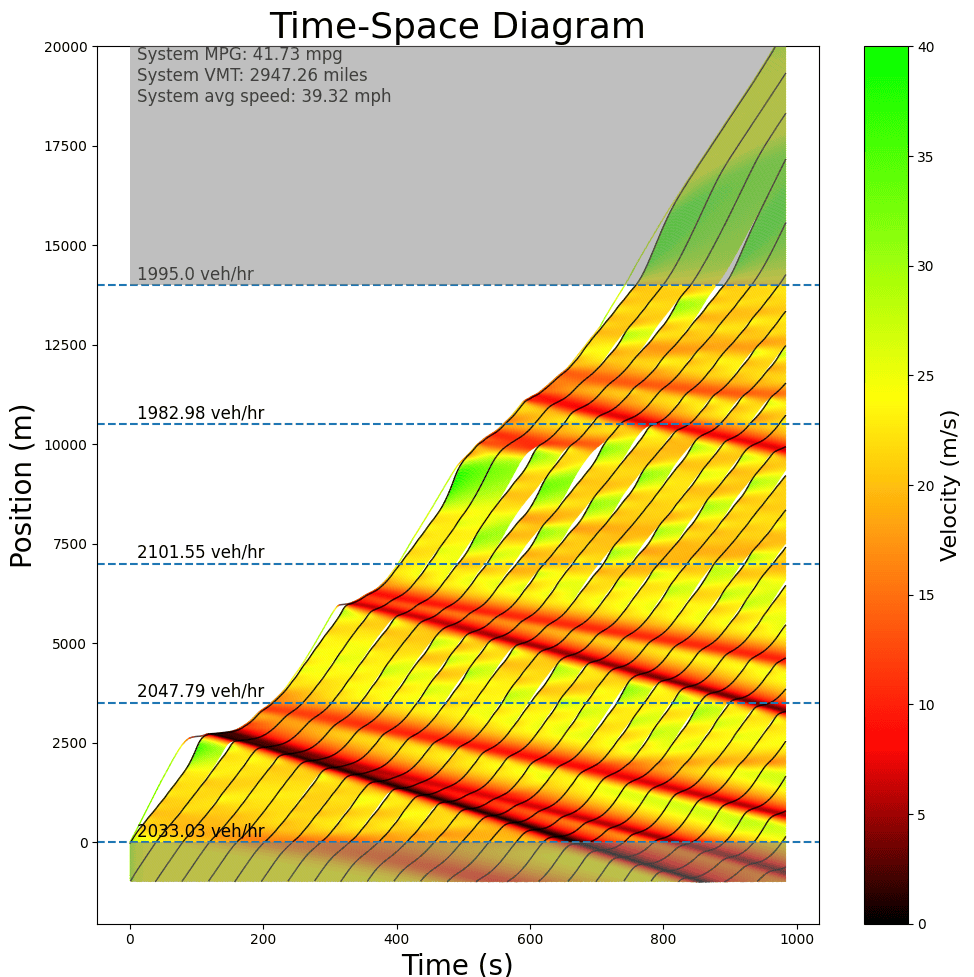}
    \caption{Time-space diagrams of a congested traffic scenario with real data from the I-24. The top depicts the IDM baseline with 0\% penetration rate of AVs. The bottom depicts a 4\% penetration rate of the RL controller. Between the baseline and the RL controller, a 9.1\% increase in MPG is measured. The black lines depict AVs which can be seen opening gaps in the bottom diagram.}
    \label{fig:tsd_acc_sim}
\end{figure}

In general, flow smoothing behavior is evident in the deployed controller. As shown in Figure~\ref{fig:tsd_acc_sim}, which compares the time-space diagrams of the IDM baseline with the RL controller of a congested scenario with real data taken from the I-24, a 9.1\% increase in MPG can be observed. Notably, in evaluation with RL-controlled AVs, the deep red "stop" portions of the stop-and-go-waves are shorter, and there is a smaller deviation in speed. Large fluctuations in speed are contributors to poor fuel economy, so the stabilization of speed explains the performance improvement.

\section{Migrating Control from Simulation to Vehicles}


\begin{sidebar}{Robot Operating System (ROS) for Control Systems Development in Automated Vehicles}
\section[Robot Operating System (ROS) for Control Systems Development in Automated Vehicles]{by Matthew Nice and Matthew Bunting} \label{sbar-ROS}


\renewcommand{\thestable}{S\arabic{stable}}
\renewcommand{\thesfigure}{S\arabic{sfigure}}




\sdbarfig{\includegraphics[width=18.0pc]{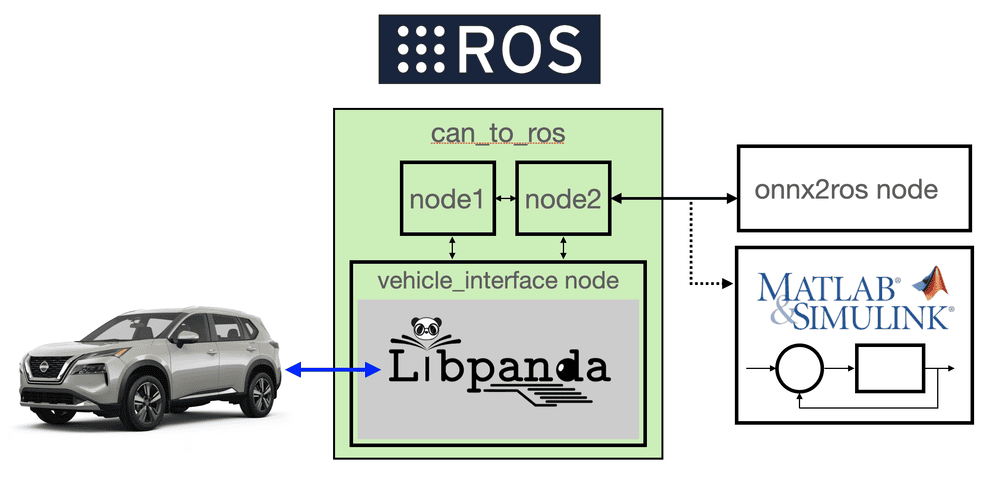}}{Diagram of vehicle control with ROS, showing the layout of can\_to\_ros, libpanda, and ONNX. \label{ros_sidebar_diagram}}

The Robot Operating System (ROS)~\cite{quigley2009ros} is an open-source framework that provides a suite of software libraries, tools, and capabilities for developing and controlling robotic systems. ROS is functionally a middleware, message passing framework where ROS \textit{nodes} act as individual programs, decoupling a system's components from a single executable into highly cohesive programs with dedicated tasks.  \textit{Message passing} operates as a publisher/subscriber framework, where messages are designated under certain topics.  Each ROS node send messages to other ROS nodes using message topics. We use this paradigm to provide a network where disparate software components can connect. OEM in-vehicle networks, RL neural networks, generated Simulink models, and other applications like live-tracking modules, all meet in the ROS network to exchange information.

ROS features utilities to record data in the \emph{bag} format, which are export as bagfiles.  These are timestamped data files of all messages in ROS topics.  This greatly aids developers by being able to replay prior tests into updated algorithms, giving the ability to validate message topics and verify system behavior.

Many open source drivers exist for a plethora of robotics-focused sensors, however for vehicle integration a custom solution was designed. To extend stock vehicles into experimental control vehicles, we build a custom software bridge from the in-vehicle network to the ROS network. The two primary components of this \textit{software bridge} are libpanda~\cite{bunting2021libpanda} and can\_to\_ros~\cite{elmadani2021can,nice2023middleware}. Libpanda is a broad-ranging software tool which facilitates the low-level vehicle integration of OEM network protocols, experimental control access, and reading/recording vehicle network data in real time. Can\_to\_ros is a model-based code generation framework that decodes and publishes critical vehicle sensor data in real-time, and features a vehicle\_interface node built on top of libpanda. This software bridge effectively abstracts the hardware system into the ROS network. By leveraging the open source ROS framework, the complex software system is untangled, and modularity increases. Figure~\ref{ros_sidebar_diagram} depicts how these components are connected. System modularity allows for complex system interactions such as interfacing with vehicle control to be simplified into just a few ROS topics.

MATLAB's Simulink \cite{MATLAB} has the ability to generate ROS C++ code from models.  This lets control designers focus on controller design and simulation, leveraging the tools available in MATLAB. This correct-by-contstruction code generation was used to test control algorithms meeting in ROS, and effecting experimental vehicle actuation. Other controllers can also be integrated in a language of choice.  For example, a package named onnx2ros~\cite{bhadani2023approaches} was also built to easily port models in ONNX to interface with ROS. This opened the pathway for the controller in this work to meet the rest of the software components in the ROS network.

By establishing ROS node interfaces upfront, dedicated teams can focus on different project aspects.  In this project, a hardware team focused on the development of vehicle integration while the controller team focused on RL.  The usage of ROS let expert teams focus on their expert development for project deliverables.

\end{sidebar}



\begin{figure}
    \centering
    \includegraphics[width=0.5\linewidth]{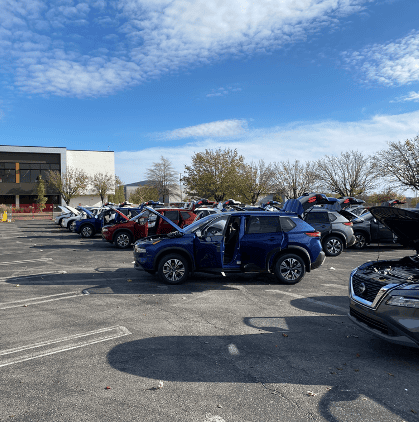}
    \caption{The Nissan Rogue vehicles with their hoods popped open to access custom equipment necessary to enable RL control.}
    \label{fig:car_hoods_up}
\end{figure}

\begin{figure}
    \centering
    \includegraphics[width=0.8\linewidth]{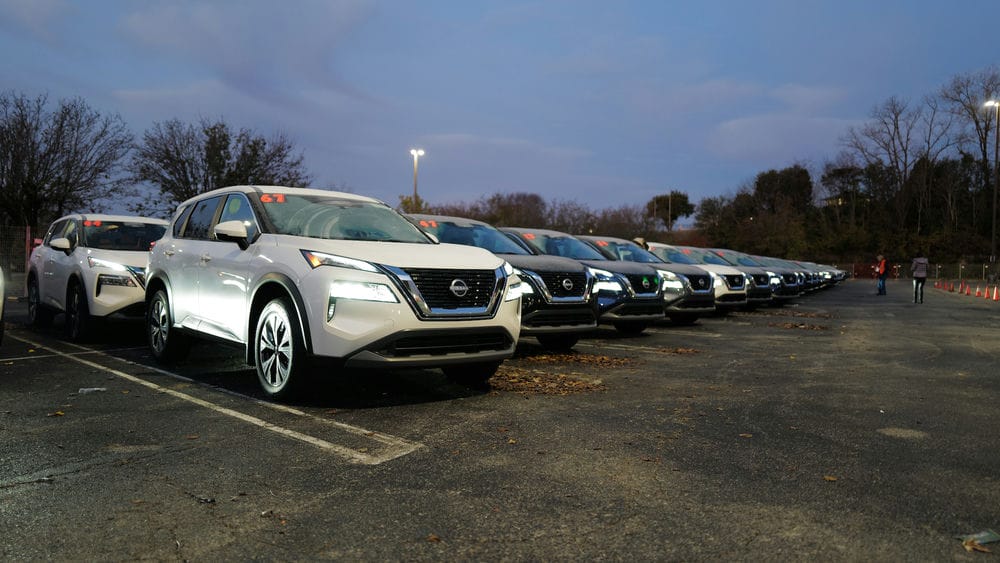}
    \caption{A shot of two rows of Nissan Rogue vehicles, which the RL controllers were run on. The vehicles were stored in a parking lot during the duration of the MVT.}
    \label{fig:rogue_rows}
\end{figure}


\subsection{Validation Work}
\label{sec:validation}

A chief safety issue concerned the migration of the controllers from software to hardware. While safety and feasibility of a controller can be evaluated in a simulation via unit tests and performance analysis, the measured efficacy of the controller may not be maintained when crossing into the real world. This phenomenon is commonly referred to in RL as the ``sim-to-real gap'' or ``reality gap,'' describing how the discrepancies between simulation and the real world can cause a policy to behave in unexpected ways \cite{jang2019simulation, pinto2017robust}. This reality gap challenge is increased by the nuanced complexity of underlying software and hardware platforms which enable the RL controller to be executed in a real vehicle. 

For more information on the hardware migration, including details on ROS, please refer to ''\nameref{sbar-ROS}'' or to "Vehicle Interfacing" in the MegaController paper 
\cite{lee2024traffic}.


Once the ONNX model is loaded into the vehicle software stack, a \textit{software bridge} via can\_to\_ros \cite{nice2023middleware} and libpanda \cite{bunting2021libpanda} decodes the Controller Area Network (CAN) bus sensor information for state data inputs, and interprets the ONNX outputs to make speed changes to the car through a \textit{hardware bridge}. The hardware bridge, consisting of an embedded computer (Figure~\ref{fig:rasp_pi}), a custom printed circuit board (PCB), and a custom CAN cable, provided access to the in-vehicle network. The custom PCB, managed by the embedded software, transforms ROS-based commands into voltages to change ACC setting.

The path from simulation to hardware is described as follows: 
\begin{enumerate}
    \item The form the controllers takes when trained in simulation is that of a neural net, wrapped as a PyTorch model. 
    \item The PyTorch model is then exported to ONNX format. ONNX (the Open Neural Network Exchange) is a machine learning model standard that is compatible with a variety of platforms \cite{onnx}.
    \item The ONNX model communicates with the \acronym{Robot Operating System} (ROS) via a customized package in order to exectue the ONNX model at runtime, the development of which is outside the scope of this paper.
    \item Once the ONNX model is loaded into the vehicle software stack, a \textit{software bridge} via can\_to\_ros \cite{nice2023middleware} and libpanda \cite{bunting2021libpanda} decodes the Controller Area Network (CAN) bus sensor information for state data inputs, and interprets the ONNX outputs to make speed changes to the car through a \textit{hardware bridge}.
    \item The hardware bridge, consisting of an embedded computer (Figure~\ref{fig:rasp_pi}), a custom printed circuit board (PCB), and a custom CAN cable, provided access to the in-vehicle network. The custom PCB, managed by the embedded software, transforms ROS-based commands into voltages to change ACC setting.
\end{enumerate}

\begin{figure}
    \centering
    \includegraphics[width=0.5\linewidth]{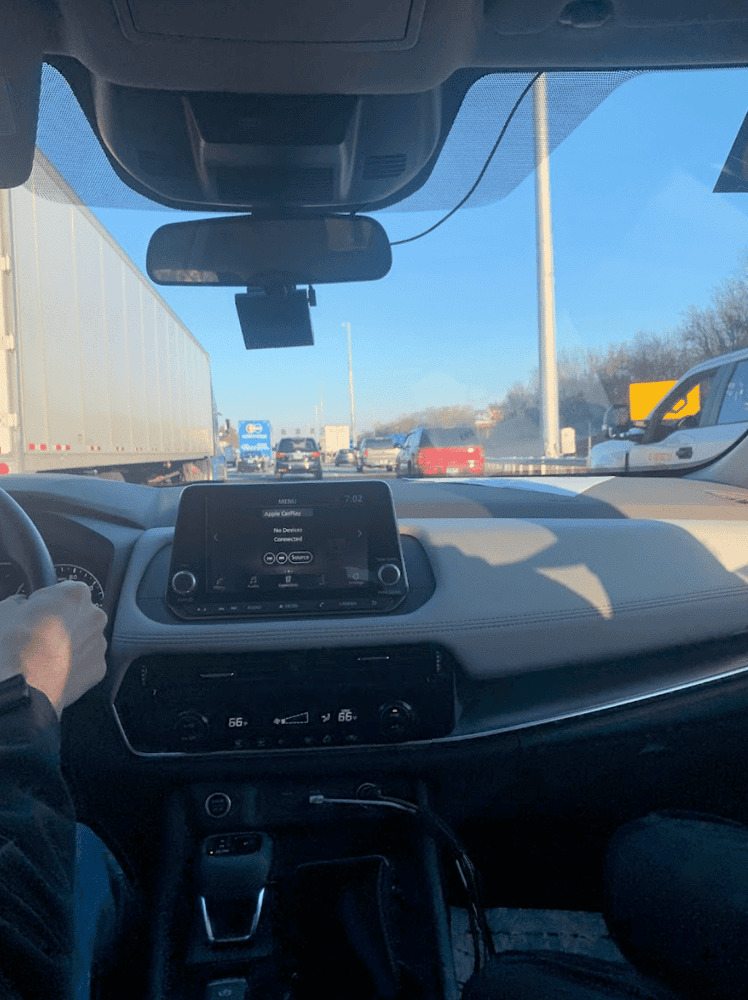}
    \caption{Performing a dry run in the Rogue on the I-24 with the RL-controller turned on, testing controller efficacy live with a connected laptop, before the MVT.}
    \label{fig:driving_pov}
\end{figure}


\begin{pullquote}
    Controllers can reduce stop-and-go waves even if onboard sensors do not provide anticipatory knowledge of these waves. 
\end{pullquote}

In the migration from software to hardware, an array of validation tools were used to guarantee and test for safety prior to deployment. These approaches are largely left to future papers, but include both extensive testing in simulation and test driving with live debugging.

After using these validation tools, our teams could 
compare the outputs from the ROS-based wrapper of the ONNX models 
to that of the neural net 
with the same inputs,
to ensure no loss of information between this migration.

\begin{figure*}[]
\centering
\begin{multicols}{2}
    \includegraphics[width=0.9\linewidth]{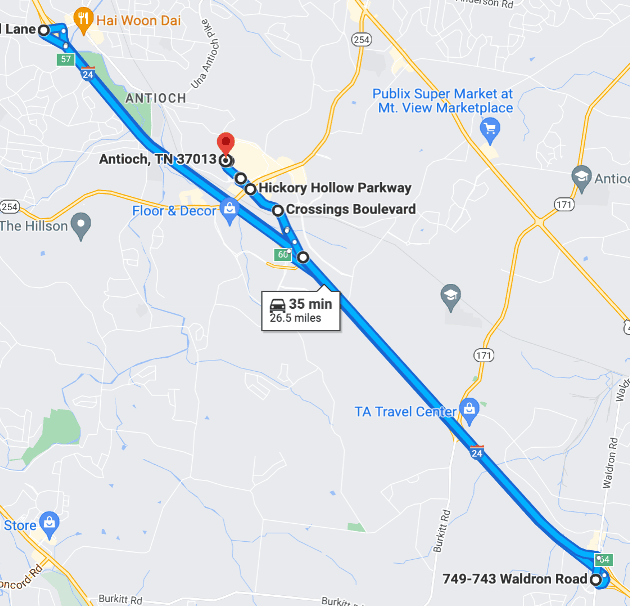}\par 
    \includegraphics[width=0.9\linewidth]{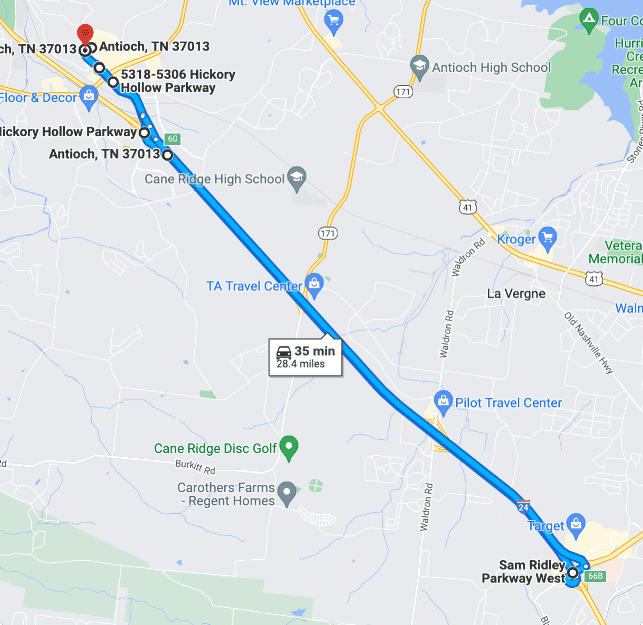}\par 
\end{multicols}
\caption{Depicted here are the two routes test drivers used during the MVT. Left: The orange route, roughly 7.2 miles one way. Right: The yellow route, roughly 6.6 miles one way, is further southeast on the I-24 than the orange route. For further details on routes and logistics, refer to 
\cite{ameli2024designing}.}
\label{fig:routes}
\end{figure*}

Following this, the next step of validation was a series of rigorous test drives, one of which is shown in Figure~\ref{fig:driving_pov}. Test drives occurred on any subset of either of the yellow or orange routes depicted in Figure~\ref{fig:routes}. Through test driving, we discovered and patched numerous bugs in our control scheme that led to some of the following features. 

\subsubsection{Activation when Westbound}
During both the actual MVT and the test drives, the controller is only activated on the westbound side of the highway. This is due to the fact that morning traffic, the focus of the MVT, only occurs for cars that are on the westbound side of the I-24 in the test corridor. 
A positional latitude-based westbound flag was initially used. This boolean function, returning true or false, is given by:
\begin{equation}
    w = x_{i+1} \geq x_i,
\end{equation}
where $i$ is time step, and $x$ is the latitude portion of the GPS coordinate. The initial rendition of the westbound flag was a position-based function, checking if the latitude was increasing from the previous timestep and returning True if so. Due to the frequent standstill traffic where vehicles would come to a full stop, as well as slight noise in the GPS reading, this flag was prone to false-negatives while at a stop during westbound standstill traffic, with the latitude occasionally reading incorrectly as decreasing. 

The final estimator fielded at the close of the MVT was a spline-based method. A spline of the entire westbound portion of I-24 
was extracted from OpenStreetMap (OSM). Each vehicle's ($X,Y$) GPS coordinates were compared with the spline mapping, with the most recent 10 comparisons ``voting'' to determine travel direction.

\subsubsection{Clipping Mechanism}

Due to driver feedback, a clipping mechanism was applied to the speed setting portion of the action tuple to ensure smooth, comfortable control. The lower and upper clip bounds are based on the average speed of the vehicle during the last one second (10 time steps):
\begin{align}
l &= \frac{1}{10}\sum_{i=1}^{10} v_i - 15 \text{mph}\;,\\
u &= \frac{1}{10}\sum_{i=1}^{10} v_i + 5 \text{mph}\;,    
\end{align}
where $l$ and $u$ are the lower and upper bounds of the clip, respectively. The final clamp is executed as:
\begin{equation}
    s = \min(\max(\min(\max(a, l), u), 20 \text{mph}), 73 \text{mph})
\end{equation}
where $s$ is the final speed setting command, and $a$ is the un-normalized output, or action, from the neural network. That is, the acceleration is capped between $u$ and $l$, the two bounds, and then capped between 20 mph and 73 mph. The values 20 and 73 mph were chosen to keep controls in a safe region.

\subsubsection{Acceleration Estimation}
One notable quantity that was not directly read from the CAN bus was acceleration. Since accelerations are an input to our controller, we used the speed history to estimate the accelerations via a sliding window approach. First, noisy accelerations are calculated based on the current velocity and the previous velocity from the last time step (0.1 seconds). Then, the final acceleration $a_t^\text{fixed}$ is estimated as as average of the last four noisy acceleration estimates.

\begin{align}
a_t^\text{noisy} &= \frac{v[t+1] - v[t]}{0.1}\;,\\
a_t^\text{fixed} &= \frac{1}{4}\sum_{i=0}^3 a_{t-i}^\text{noisy}\;,
\end{align}
where $t$ indicates time step. 

\subsection{Experimental Design}

The RL controller discussed in this article was deployed on 100 Nissan Rogue cars, which can be viewed in Figures~\ref{fig:car_hoods_up} and \ref{fig:rogue_rows}. 
This section will describe the design decisions of test week, including information on routes, vehicle fleet, and day by day controller choices. For a high-level overview of the schedule of test week, please refer to Table~\ref{table:schedule}. This will discuss exactly which version of the RL controller is being run each day (''Production''), which versions are being ran to test efficacy for the following day (''Experimental''), as well as changes and issues encountered during each day with either the Production or Experimental controllers.

\begin{table*}[]
\centering
\begin{tabular}{|m{2.3cm}|m{4cm}|m{4cm}|m{4cm}|}
\hline
\multicolumn{4}{|c|}{MVT Schedule} \\ \hline

\textbf{Date} & \textbf{Description of Events} & 
\textbf{Changes from Previous Day} & \textbf{Issues Encountered} \\ \hline

Monday 11/14 
& \textbf{Production:} $~80+$ Nissans running stock ACC. \break \textbf{Experimental:} 5 Nissans running RL controller with speed planner, westbound flag hardcoded to True. 
& n/a 
& (1) Latency in ACC Actuation. (2) Speed planner I/O, getting default values only. (3) Speed planner not publishing. (4) Westbound flag not robust. (5) Speed setting defaulting to 59 mph at full stop. \\ \hline

Tuesday 11/15 
& \textbf{Production:} None, due to inclement weather. \break \textbf{Experimental:} 8 Nissans running RL controller with speed planner, data fusion turned off. 
& Updated westbound flag calculation; dynamic lane assignment; new clamping logic; minor bugs. 
& n/a \\ \hline

Wednesday 11/16 
& \textbf{Production:} $~80+$ Nissans deployed with RL controller on westbound, stock ACC elsewhere. \break \textbf{Experimental:} 10 Nissans running RL controller on westbound and eastbound, stock ACC on side streets 
& Controls are allowed if server agrees and if not on a side street. 
& (1) Some locations identified I-24 East incorrectly as westbound. (2) Server for speed planner was offline $\approx 10$ minutes.\\ \hline

Thursday 11/17 
& \textbf{Production:} $~80+$ Nissans deployed with RL controller with server error (detailed right). \break \textbf{Experimental:} 10 Nissans running RL controller with fix to same bug.
& Fixed westbound bug from Wednesday. 
& (1) Server for speed planner responded slowly, or not at all, due to exponential growth in computationally-demanding speed planner query. \\ \hline

Friday 11/18 
& \textbf{Production:} 100 Nissans deployed with RL controller. \break \textbf{Experimental:} None. 
& Speed planner query periodically executed, and served from cache. This speeds up queries and cached results by over 100x. 
& n/a \\ \hline

\end{tabular}
\vspace{2mm}
\caption{A day-by-day description of the events of the MVT week, including which algorithms were run and changes from previous days. All the dates are in 2022. 
}
\label{table:schedule}
\end{table*}

\subsubsection{Route Plans}
\label{sec:route_plans}
There were two selected routes that drivers operated on during the MVT. Both of these routes share an origin. They are:
\begin{itemize}
    \item Orange Route: Starts at HQ, heads northwest, loops back southeast at Haywood Lane, heads southeast, loops back northwest at Waldron Road. Roughly 7.2 miles one way. Assigned to lanes 2 or 3.
    \item Yellow Route: Starts at HQ, heads southeast, and loops back northwest at Sam Ridley Parkway West. Roughly 6.6 miles one way. Assigned to lane 4.
\end{itemize}

Drivers are assigned with a 67 Orange--33 Yellow split. They are asked to drive the route and loop around as many times as they feel comfortable, or until they are asked to return to HQ when traffic had dissipated. Generally, driving would take place between 6:30 am and 9:30 am.

\subsubsection{Driver Information}
The CIRCLES Consortium
recruited 170 drivers to drive for the MVT that week. 702 drivers responded to our announcement, 308 were cleared, and 170 attended our in-person training session, leading to 100 on the road at a given time.
For further details on how logistics for the MVT operated, from driver recruitment to fleet management to route assignment, we direct you to the corresponding standalone article \cite{ameli2024designing}.  


\begin{figure}
    \centering
    \includegraphics[width=0.6\linewidth]{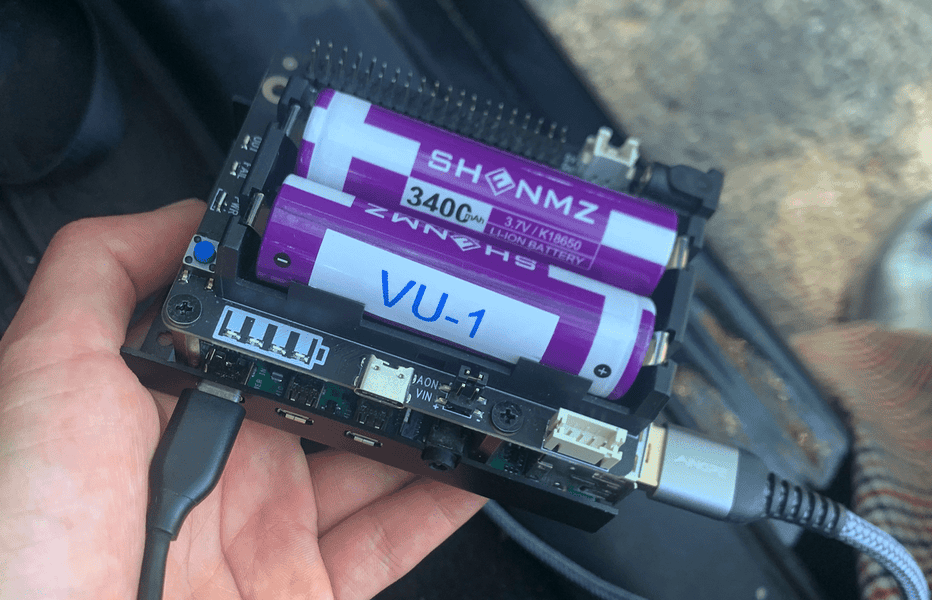}
    \caption{One of the many Raspberry Pis that contain the code for the RL controller. These are connected to the vehicle via a series of cables, which can be partially seen in Figure~\ref{fig:driving_pov}.}
    \label{fig:rasp_pi}
\end{figure}

\subsection{Result Analysis}



In this section, we present findings from our analysis of the MVT data~\cite{lee2024traffic}
collected from I-24. Data was captured using the I-24 MOTION testbed~\cite{gloudemans202324} where the 100 RL-controlled vehicles were deployed. The testbed uses a computer vision pipeline~\cite{gloudemans2021vehicle,gloudemans2023so,gloudemans2023interstate}, a trajectory post-processing pipeline~\cite{wang2022automatic,wang2023onlinemcf}, and a visualizaton package~\cite{10.1145/3576914.3587710} to produce  the trajectories of all vehicles on the designated highway portion during the experiment period. One challenge we face is the sheer volume of the data, which makes comprehensive analysis complex. In the subsequent discussion, we demonstrate our preliminary insights, suggesting that our AVs  effectively mitigate traffic waves. However, it is important to emphasize that given the vastness of the dataset, further analysis may reveal additional perspectives or correlations and potentially lead to a reevaluation of our initial conclusions. Nonetheless, the trends that we observe appear consistent across the several metrics that we measure.

\subsubsection{Data collected during the experiment}

Data is collected using the I-24 MOTION testbed~\cite{gloudemans202324,gloudemans2020interstate}, which consists of 276 high-resolution cameras mounted on poles along a 4.2 miles stretch of the I-24. Individual vehicle trajectories are extracted from the camera videos using computer vision algorithms (see ~\cite{gloudemans202324} for a detailed overview of the system, algorithms, and datasets). This is a challenging problem due to the raw volume of data and difficulty of stitching images from different cameras together as well as the presence of bridges, resulting in several segments of trajectories for a same vehicle. More information about how the data is collected, processed, formatted and accessible online can be found in \cite{gloudemans202324,gloudemans2023so} and 
\cite{lee2024traffic}.

The data collected during the experiment spans from 11/14/2022 to 11/18/2022, from 6am to 10am each day. In total, the resulting dataset is over 200GB, with about 600,000 vehicle trajectories each morning, or 3 million over the whole week, where each trajectory consists of 440 data points on average (with a large variance). Trajectories are collected on all lanes of the highway (ranging from 4 to 5 lanes) in both eastbound and westbound directions. The whole database contains over 1.3 billion rows, about 1\% of which (so about 13 million points) correspond to trajectories of our AVs. The average duration of a trajectory is 18.2s (with a standard deviation of 16.3s), and the average distance traveled per trajectory is 311.4m (with a standard deviation of 348.7m).

The vast volume of the data presents analytical challenges. Therefore, for many of our assessments, we choose a representative random sample to compute aggregate metrics. This sample is sized to ensure consistency across different subsets, while also being computationally manageable. Besides, as evidenced by the time-space diagrams in 
\cite{lee2024traffic},
our AVs experienced peak engagement in the period between 7am and 9am, registering the highest penetration rate of engaged AVs. Consequently, we focus our analysis during this time frame.

Finally, the data is pre-processed to include energy information: for each data point, and depending on the type of vehicle (extracted from the video data among one of 6 classes~\cite{gloudemans202324}), a corresponding calibrated energy model is leveraged to estimate the instantaneous fuel consumption of the vehicle. This allows us to compute energy metrics. See Sidebar~``\nameref{sbar-energy}'' and the corresponding standalone article 
\cite{khoudari2023reducing}
for more information. 

\subsubsection{Dampening of traffic waves}

It is a complex task to measure the energy impact of our AVs due to the absence of a counterfactual. Indeed, we can measure the energy consumption of vehicles on the highway during the experiment, but cannot know what would have happened in the absence of our controlled vehicles; comparing between different days does not appear to be a viable option, due to significant differences in day-to-day traffic patterns and dynamics. Instead, we consider a proxy by computing metrics as a function of the distance to the closest engaged AV in the traffic ahead (downstream). The idea is to show that vehicles that are close behind an AV experience smaller amplitude waves or less fuel consumption on average than vehicles far away behind any AV, which would suggest that our AVs have a smoothing effect.

\begin{figure}[t]
    \centering
    \includegraphics[width=0.7\linewidth]{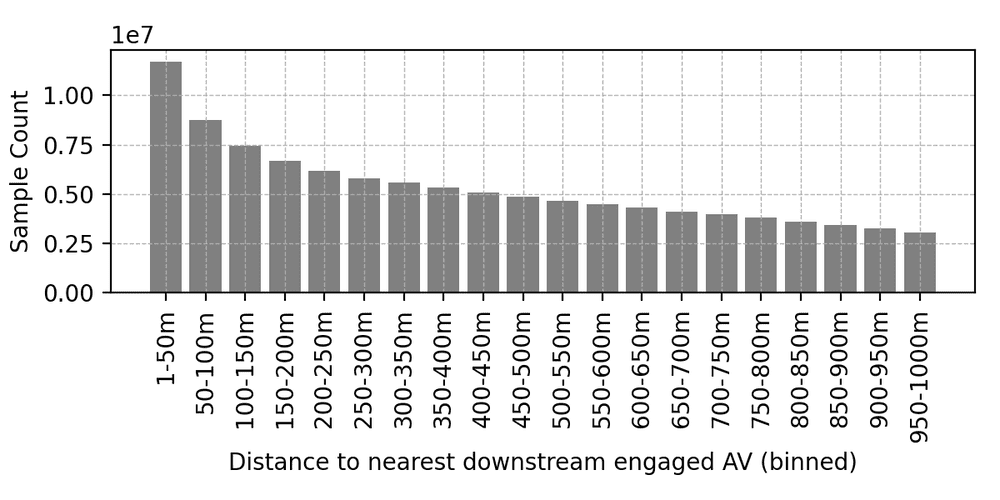}
    \caption{Number of data points as a function of distance to the nearest engaged AV in the downstream traffic, for all of the data collected between 7am and 9am on Wednesday 16, Thursday 17 and Friday 18, days on which our RL-controlled AVs were deployed.}
    \label{fig:count_by_dist_to_av}
\end{figure}

We consider the data collected from Wednesday 11/16/2022 to Friday 11/18/2022, between 7am and 9am each day. The amount of data this corresponds to is shown in Figure~\ref{fig:count_by_dist_to_av}. In this analysis, we bin the distances to the nearest engaged downstream AV by intervals of 50m, which allows us to get a smoother measurement and also account for errors of up to $\pm 10$m in the computation of these distances.

\begin{figure}[h!]
    \centering
    \includegraphics[width=0.5\linewidth]{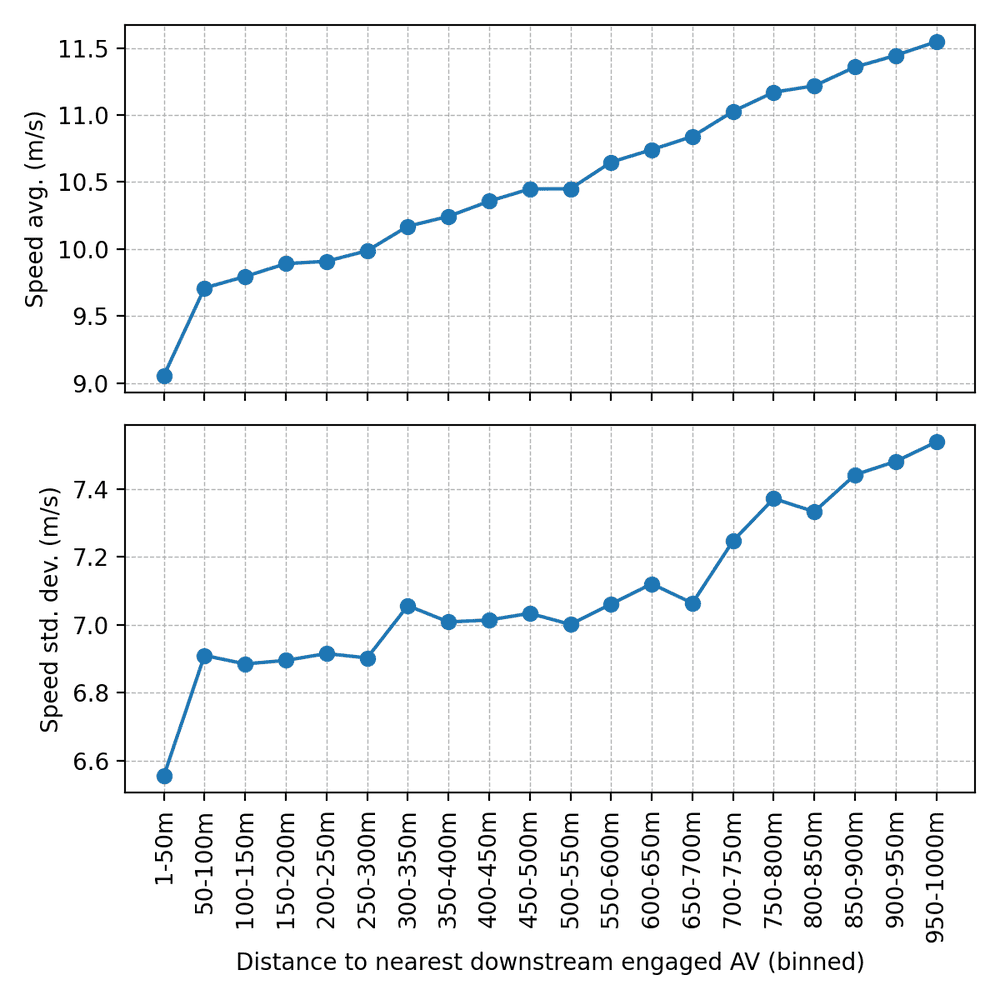}
    \caption{Speed average and standard deviation as a function of distance to the nearest engaged AV in the downstream traffic. The averages and standard deviations are computed over all of the data collected between 7am and 9am on Wednesday 16, Thursday 17 and Friday 18, days on which our RL-controlled AVs were deployed. This corresponds to over 2.5 million data points for each distance bin.}
    \label{fig:speed_by_dist_to_av}
\end{figure}

We observe in Figure~\ref{fig:speed_by_dist_to_av} that the average and standard deviation of the vehicles' speeds steadily increase as a function of distance to AV. The increase in speed is understandable: in order to smooth traffic, the AVs aim to open larger gaps used to absorb waves (as illustrated in Section ``\nameref{sec:accel_sim_res}''), which they do by driving at lower speeds. More importantly, we observe a smaller speed variance behind AVs than far away from them, which could suggest that the AVs are reducing the amplitude of the waves. 

\begin{figure}[h!]
    \centering
    \includegraphics[width=0.5\linewidth]{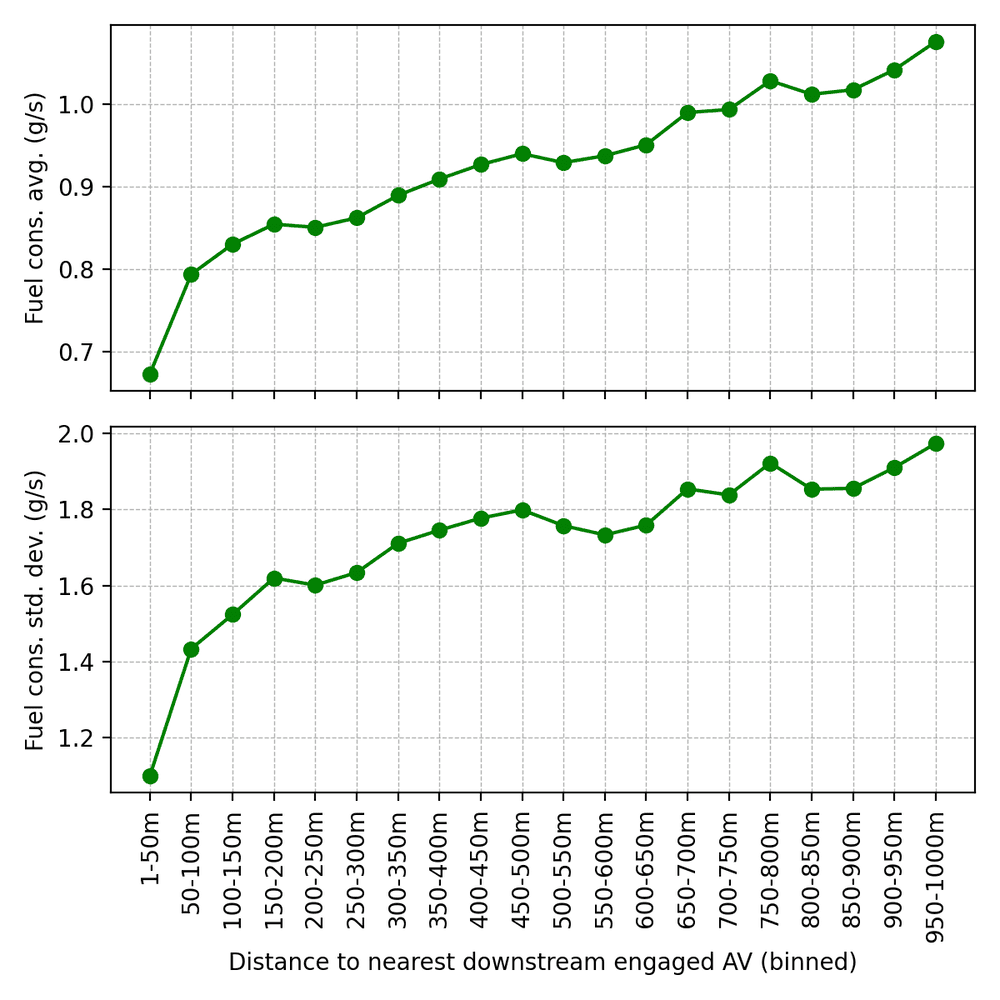}
    \caption{Instantaneous fuel consumption average and standard deviation as a function of distance to the nearest engaged AV in the downstream traffic. Instantaneous fuel consumption is calculated as a function of vehicle type, speed, instantaneous acceleration and road grade, using calibrated energy models available in  \cite{circles_energy_models}. The averages and standard deviations are computed over all of the data collected between 7am and 9am on Wednesday 16, Thursday 17 and Friday 18, days on which our RL-controlled AVs were deployed. This corresponds to over 2.5 million data points for each distance bin.}
    \label{fig:fuel_by_dist_to_av}
\end{figure}

Another metric pointing towards the AVs having a dampening effect is the fuel consumption analysis presented in Figure~\ref{fig:fuel_by_dist_to_av}. Similarly to the speed analysis, we observe that the instantaneous fuel consumption increases both in mean and variance as a function of distance to AV, indicating than vehicles closer behind AVs consume less fuel on average.

\subsubsection{Data in $v-a$ space}


\begin{figure}[h!]
    \centering
    \includegraphics[width=0.49\linewidth]{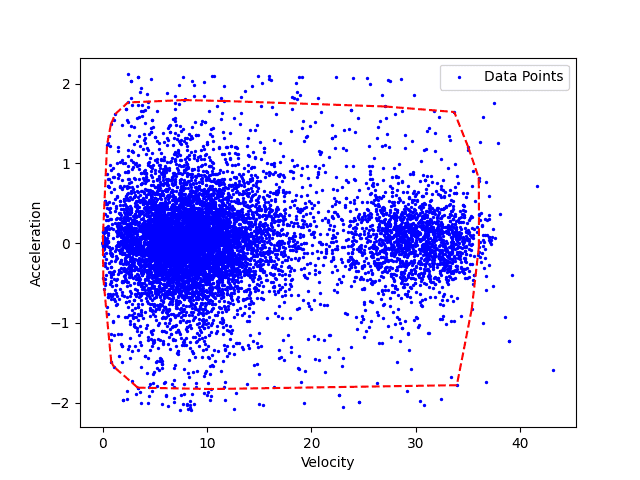}
    \includegraphics[width=0.49\linewidth]{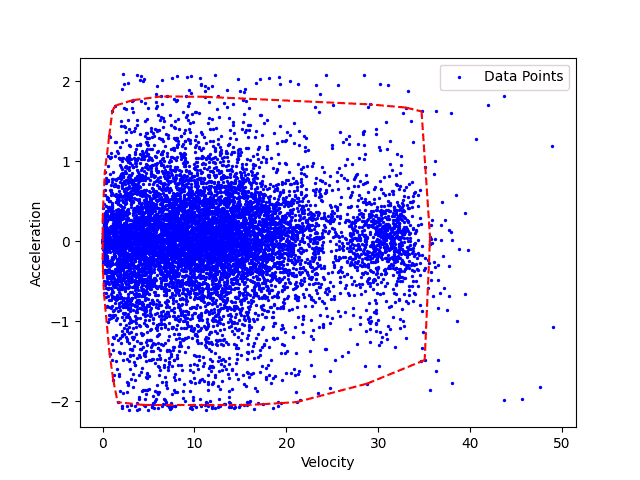}
    \caption{A scatterplot of velocity-acceleration pairs. Left: Data on Tuesday, November 15th, 2022. Right: Data on Friday, November 18th, 2022. Data from both days are collected from 7:00 am to 9:00 am CST, which we sample at a 1:1000 rate. For Tuesday, no AVs are deployed; we note that the convex hull bounding $99\%$ of the data points is larger in the absence of AVs, and that the data appear in two distinct clusters. For Friday, when AVs are deployed, the convex hull bounding $99\%$ of the
 data points is smaller with AVs, and that the data appears to be more spread out, indicating smoothing.}
    \label{fig:convex2}
\end{figure}

Figure~\ref{fig:convex2} represents the data taken on two separate days, Tuesday and Friday, of the MVT. Tuesday is considered the baseline of the MVT, as no AVs were run on that day; however, it is important to note that inclement weather on Tuesday may have affected the data. Data from 7:00 am to 9:00 am CST are represented as a scatterplot in velocity-acceleration space. Two observations may be made. First, the two distinct clusters that are seen in the left plot of Figure~\ref{fig:convex2} represent the two bounds of the stop-and-go waves, with one at around $10 m/s$ and another at around $30 m/s$. In Figure~\ref{fig:convex2}, which represents RL-controller AVs, the clusters are far less distinct. This is consistent with the results seen in Figure~\ref{fig:tsd_acc_sim}, where stop-and-waves are replaced with more consistent velocities. Second, it can also be observed that convex hull (ignoring $1\%$ of outliers) of the RL-controlled day is considerably smaller in size.

\begin{figure}[h!]
    \centering
    \includegraphics[width=0.49\linewidth]{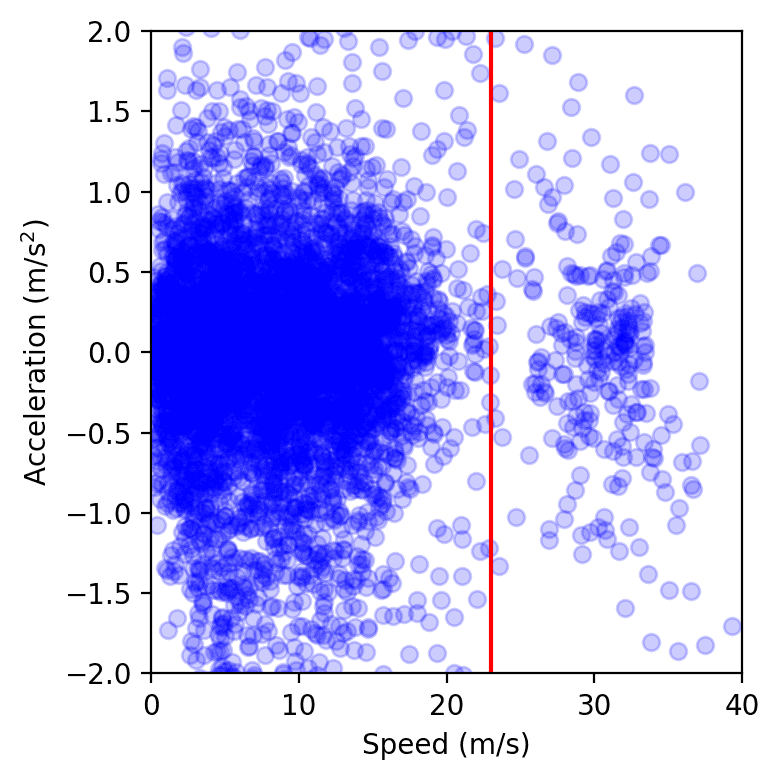}
    \includegraphics[width=0.49\linewidth]{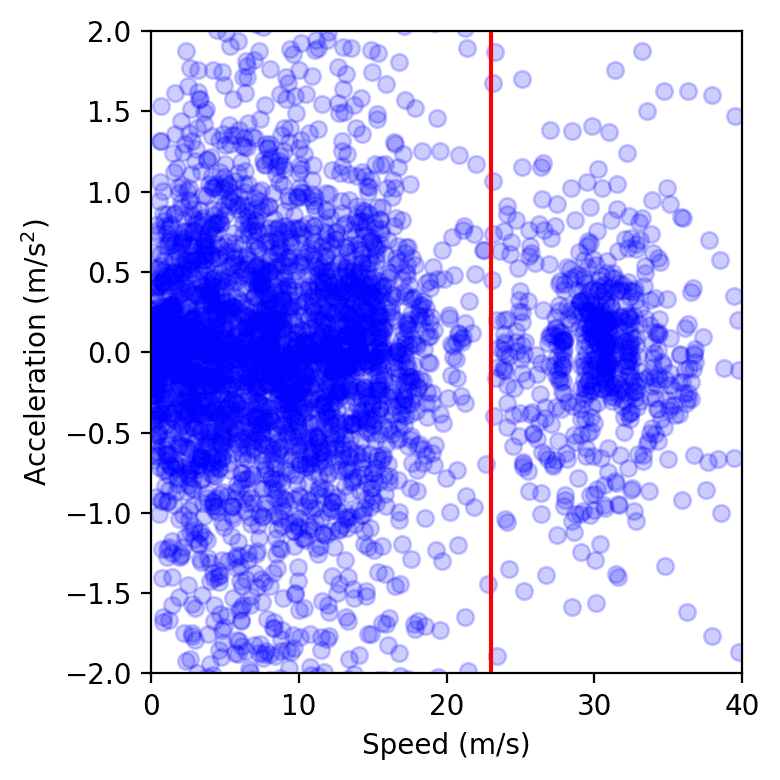}
    \caption{Data collected on Friday 18 between 7am and 8am, sampling 1 in 1000 data points, plotted in acceleration vs speed space. We plot points corresponding to vehicles that are up to 300m behind an AV (top), and from 300m to 600m behind an AV (bottom).}
    \label{fig:results_va_data}
\end{figure}

\begin{figure}[h!]
    \centering
    \includegraphics[width=0.49\linewidth]{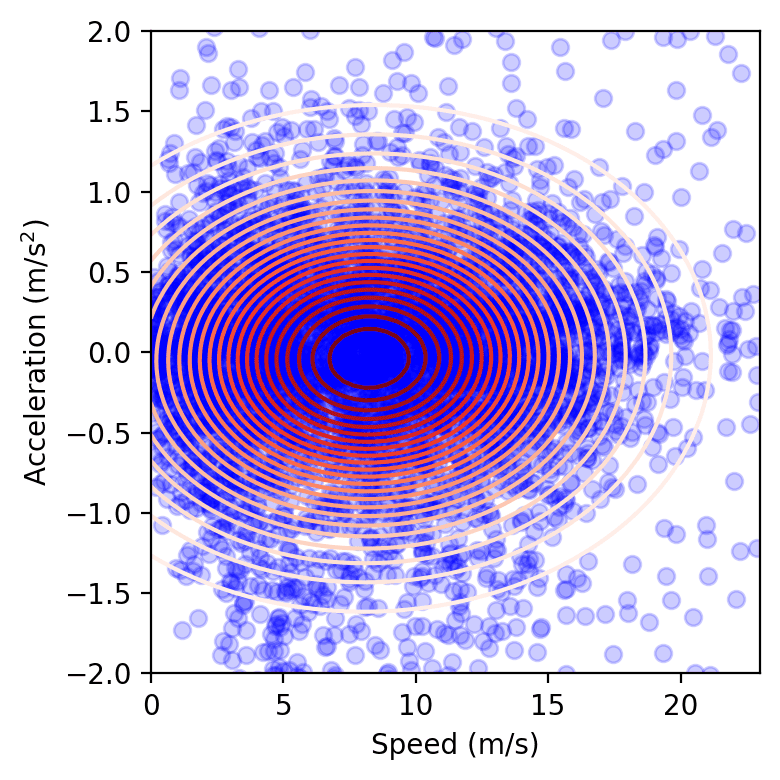}
    \includegraphics[width=0.49\linewidth]{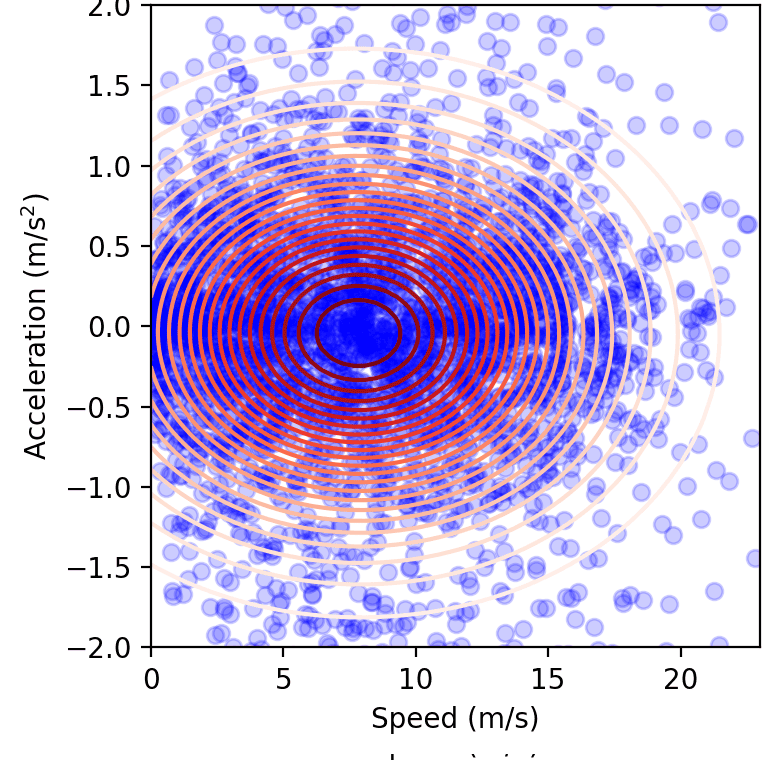}
    \caption{We would like to analyze how much "space" does our lower-speed cluster take. A naive approach would be to compute the area of the convex hull, but that is very sensitive to outliers. Instead, we fit the data with a Gaussian kernel, which gives us a variance matrix, and we compute the determinant of that matrix which is a way to measure the size of the Gaussian. Doing this procedure, we obtain a determinant of 10.23 for the data up to 300m behind an AV (top plot), and of 14.40 for the data from 300m to 600m behind an AV (bottom plot). This means that closer behind an AV, there is on average a smaller range in terms of speeds and/or accelerations, suggesting reduced waves compared to further away from the AV.}
    \label{fig:results_va_gaussian}
\end{figure}

Figure~\ref{fig:results_va_data} shows two samples of the data collected on Friday 11/18/2022, one for vehicles that are closer behind AVs, the other for vehicles further way. We can notice two clusters: a high-speed one likely corresponding to free flow, and a lower-speed one that should contain both congestion and waves. We are mostly interested in what happens in the lower speed cluster, which we split at a speed of 23$\frac{\text{m}}{\text{s}}$ corresponding to the vertical red line.
We would like to understand how much "space" does our lower-speed cluster take. Computing the area of the convex hull would be very sensitive to outliers; instead, we fit the data with a Gaussian kernel, which gives us a variance matrix, and we compute the determinant of that matrix which gives us an estimate of the size of the Gaussian. Doing this procedure, which is shown in Figure~\ref{fig:results_va_gaussian}, we obtain a determinant of 10.23 for the data up to 300m behind an AV (top plot), and of 14.40 for the data from 300m to 600m behind an AV (bottom plot). This means that closer behind an AV, there is on average a smaller range of speeds and accelerations, which is another indicator suggesting that the AVs have a smoothing effect.

\subsubsection{Histograms}

\begin{figure}[h!]
\centering
    \includegraphics[width=0.7\linewidth]{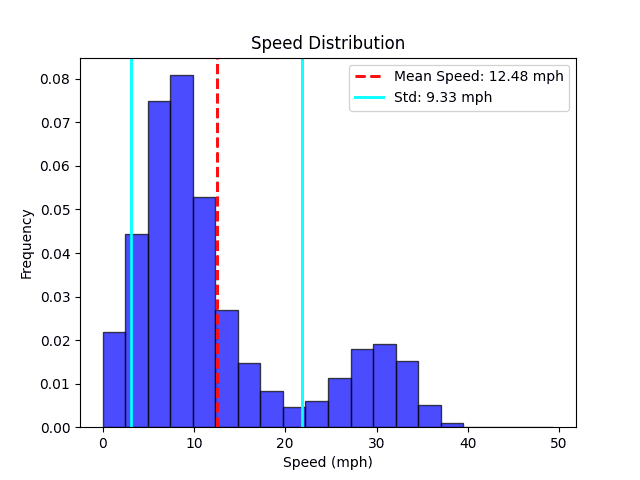}
\caption{The speed distribution of vehicles on day 2 of the MVT. We consider data on Monday, November 15th, 2022, from 7:00 am to 9:00 am CST, which we sample at a 1:1000 rate. Comparing with Figure~\ref{fig:histogram2}, we notice that the standard deviation tends to be higher and the speed bins are less spread out, with the highest frequency at $8\%$.}
\label{fig:histogram1}
\end{figure}

\begin{figure}[h!]
\centering
\begin{multicols}{2}
    \includegraphics[width=0.95\textwidth]{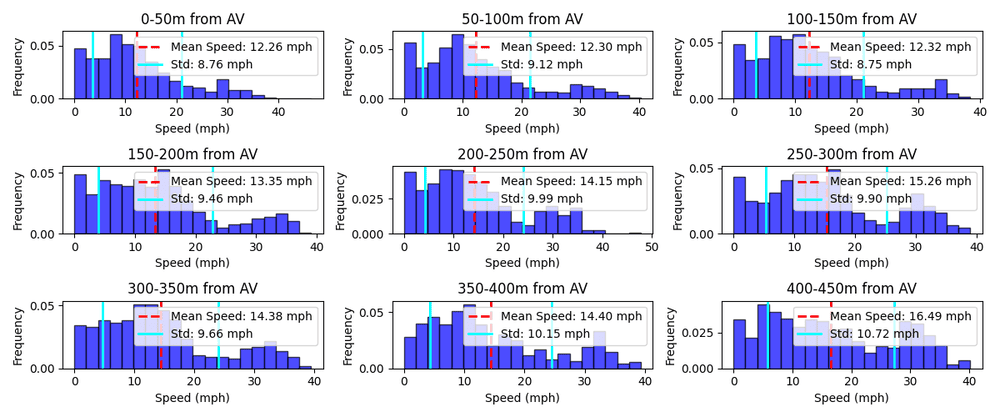}
\end{multicols}
\caption{The speed distribution of vehicles on day 1 of the MVT, when stock ACC was run on the AVs. We consider data on Monday, November 14th, 2022, from 7:00 am to 9:00 am CST, which we sample at a 1:1000 rate. Comparing with Figure~\ref{fig:histogram2}, we notice that the standard deviation tends to be higher and the speed bins are less spread out.}
\label{fig:histogram2}
\end{figure}

\begin{figure}[h!]
\centering
\begin{multicols}{2}
    \includegraphics[width=0.95\textwidth]{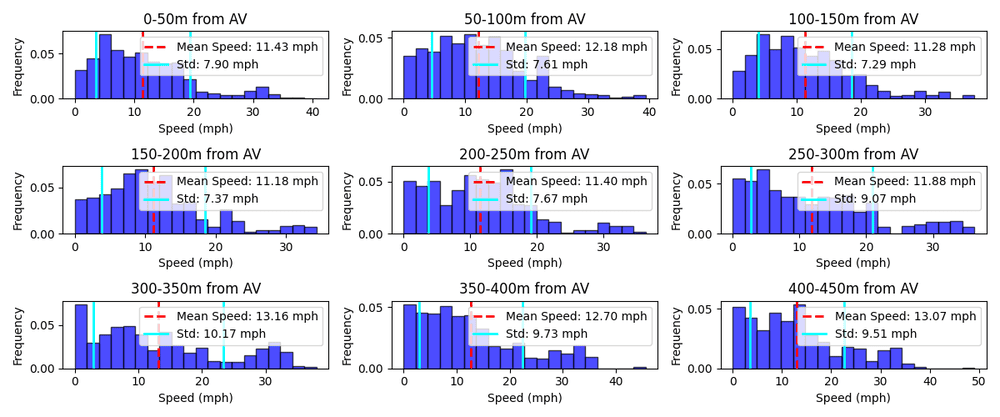}
\end{multicols}
\caption{The speed distribution of vehicles on day 5 of the MVT. We consider data on Friday, November 18th, 2022, from 7:00 am to 9:00 am CST, which we sample at a 1:1000 rate. Comparing with Figure~\ref{fig:histogram1}, we notice that the standard deviation tends to be higher and the speed bins are more spread out, indicating that the AVs are smoothing the clusters that form at low and high speeds.}
\label{fig:histogram3}
\end{figure}

The data are also represented in histograms with different speed bins. We compare three days: (a) Monday (Stock ACC running on AVs), shown in Figure~\ref{fig:histogram2} (b) Tuesday (No AVs), shown in Figure~\ref{fig:histogram1}, and (c) Friday (RL-controlled AVs)~\ref{fig:histogram3}. In general, we notice that the variance of the RL-controlled day is smaller than that of Monday or Tuesday, which indicates that traffic flow smoothing is being done. The difference is especially apparent between Tuesday and Friday, with frequencies of speed bins as high as $8\%$, indicating high magnitudes of stop-and-go waves.

\section{Conclusion}

This article has provided a comprehensive discussion of the technical intricacies surrounding the design and deployment of the RL controller at the MVT, the largest deployment of AVs to date designed to smooth traffic flow. By detailing the challenges and development of the controller design and training process, we have shed light on how RL may be utilized to influence autonomous systems in the future.

Across both the acceleration-based and ACC-based RL controllers that were presented in this article, we observe that the RL controller has a positive impact on flow-smoothing in a congested scenario. While the acceleration-based controller offers more flexibility in actuation, the ACC-based controller provides a smoother driving experience. We were able to produce $9\%$ energy gains in simulation via the ACC-based controller. Through a variety of methods, we are able to share how the RL controller stabilizes speeds between stop-and-go waves by yielding smaller deviations in accelerations, which is an essential component of achieving energy-efficient solutions. 

For a more detailed look at the other work that went into making the MVT happen, please look at 
\cite{lee2024traffic}
for a detailed look at how all the various components of the MVT worked together to produce this result; \cite{wang2024hierarchical} for an explanation of the speed planner that influenced the RL controller; \cite{ameli2024designing} for a deep-dive into another controller that was used for a different platoon of vehicles (Toyotas) during the MVT; and 
\cite{khoudari2023reducing}
for a look at the energy modeling that informed the RL algorithm.

\section{ACKNOWLEDGMENT}
This work was supported in part by the C3.ai Digital Transformation Institute under Grant Number 053483. This material is based upon work supported by the National Science Foundation under Grants CNS-1837244 (K.~Jang), CNS-1837244/1837481/1837652 (A.~Bayen/B.~Piccoli/D.~Work), CNS-2135579 (D. Work, A.~Bayen, J.~Lee, J.~Sprinkle).  Nathan Lichtlé is supported in part by the International Emerging Actions project SHYSTRA (CNRS). This material is based upon work supported by the U.S.\ Department of Energy’s Office of Energy Efficiency and Renewable Energy (EERE) under the Vehicle Technologies Office award number CID DE--EE0008872. The views expressed herein do not necessarily represent the views of the U.S.~Department of Energy or the United States Government.

\section{Author Information}

\begin{IEEEbiography}{{K}athy Jang}{\,}(kathyjang@berkeley.edu) is currently working on her Ph.D. in the Department of Electrical Engineering and Computer Sciences at U.C. Berkeley. Her research groups include Berkeley Artificial Intelligence Research (BAIR) and Berkeley Deep Drive (BDD).  Her focus is on robust reinforcement learning with applications to mixed-autonomy traffic.
\end{IEEEbiography}

\begin{IEEEbiography}{{N}athan Lichtlé}{\,}(nathan.lichtle@gmail.com) is currently working on his Ph.D. in the Department of Electrical Engineering and Computer Sciences at U.C. Berkeley, and from the CERMICS group at École des Ponts ParisTech. His research groups include Berkeley Artificial Intelligence Research (BAIR) and Berkeley Deep Drive (BDD). His focus is on reinforcement learning and mixed-autonomy traffic.
\end{IEEEbiography}

\begin{IEEEbiography}{Eugene Vinitsky}{\,}(vinitsky.eugene@gmail.com) is currently an assistant professor in Civil Engineering at New York University and is affiliated with the C2SMARTER center. He completed his Ph.D. in the Department of Mechanical Engineering at U.C. Berkeley. His focus is on multi-agent reinforcement learning and mixed-autonomy traffic.
\end{IEEEbiography}

\begin{IEEEbiography}{Adit Shah}{\,}(aditshah@berkeley.edu) is an undergraduate student in the Department of Electrical Engineering and Computer Sciences at U.C. Berkeley, affiliated with Berkeley Artificial Intelligence
Research (BAIR). He focuses on machine learning and its societally beneficial applications for climate and health.
\end{IEEEbiography}

\begin{IEEEbiography}{Matthew Bunting}{\,}(matthew.r.bunting@vanderbilt.edu)  received a Ph.D. in Electrical and Computer Engineering at the University of Arizona, and is a research scientist in the Institute for Software Integrated Systems, at Vanderbilt University. His focus is on hardware development of autonomous vehicles, democratizing vehicle data collection and control through open source projects.
\end{IEEEbiography}

\begin{IEEEbiography}{Matthew Nice}{\,}(matthew.nice@vanderbilt.edu)  received a M.Eng. in Cyber-Physical Systems, and is a doctoral candidate in the Department of Civil and Environmental Engineering, in the Institute for Software Integrated Systems, at Vanderbilt University. He has led experimental research on the impact of technology in connected and automated vehicles on transportation systems. His research has been supported by multiple Dwight D. Eisenhower Transportation Fellowship Awards from FHWA.
\end{IEEEbiography}

\begin{IEEEbiography}{Benedetto Piccoli}{\,}(piccoli@camden.rutgers.edu) is University Professor at Rutgers University-Camden. He also served as Vice Chancellor for Research. He received his Ph.D. degree in applied mathematics from the Scuola Internazionale Superiore di Studi Avanzati (SISSA), Trieste, Italy, in 1994. He was a Researcher with SISSA from 1994 to 1998, an Associate Professor with the University of Salerno from 1998 to 2001, and a Research Director with the Istituto per le Applicazioni del Calcolo “Mauro Picone” of the Italian Consiglio Nazionale delle Ricerche (IAC-CNR), Rome, Italy, from 2001 to 2009. Since 2009, he has been the Joseph and Loretta Lopez Chair Professor of Mathematics with the Department of Mathematical Sciences, Rutgers University–Camden, Camden, NJ, USA.
\end{IEEEbiography}
 
\begin{IEEEbiography}{Benjamin Seibold}{\,}(seibold@temple.edu) is a Professor of Mathematics and Physics, and the Director of the Center for Computational Mathematics and Modeling, at Temple University. His research areas, funded by NSF, DOE, DAC, USACE, USDA, and PDA, are computational mathematics (high-order methods for differential equations, CFD, molecular dynamics) and applied mathematics and modeling (traffic flow, invasive species, many-agent systems, radiative transfer).
\end{IEEEbiography}

\begin{IEEEbiography}{Dan Work}{\,}(dan.work@vanderbilt.edu) is a Professor of Civil and Environmental Engineering and the Institute for Software Integrated Systems, at Vanderbilt University. His interests are in transportation cyber-physical systems.
\end{IEEEbiography}

\begin{IEEEbiography}{Maria Laura Delle Monache}{\,}(mldellemonache@berkeley.edu) is an assistant professor in the Department of Civil and Environmental Engineering at the University of California, Berkeley. Dr.~Delle Monache’s research lies at the intersection of transportation engineering, mathematics and control and focuses on modeling and control of mixed autonomy large scale traffic systems.
\end{IEEEbiography}

\begin{IEEEbiography}{Jonathan Sprinkle}{\,}(jonathan.sprinkle@vanderbilt.edu)
is a Professor of Computer Science at Vanderbilt University since 2021. Prior to joining Vanderbilt he was the Litton Industries John M. Leonis Distinguished Associate Professor of Electrical and Computer Engineering at the University of Arizona, and the Interim Director of the Transportation Research Institute. From 2017-2019 he served as a Program Director in Cyber-Physical Systems and Smart \& Connected Communities at the National Science Foundation in the CISE Directorate. 
\end{IEEEbiography}

\begin{IEEEbiography}{Jonathan Lee}{\,} (jonny5@berkeley.edu)
received the B.S.~degree in engineering physics from the University of California, Berkeley and M.S.~and Ph.D.~degrees in mechanical engineering from Rice University. At Sandia National Laboratories (2011--13), he completed his postdoctoral appointment studying electrical and material properties via molecular dynamics simulations. He subsequently served as a senior data scientist and product manager on various teams at Uber Technologies, Inc.~(2014-2019). Since 2019, he has served as an engineering manager at the University of California, Berkeley and the program manager and Chief Engineer of CIRCLES.
\end{IEEEbiography}

\begin{IEEEbiography}{Alexandre Bayen}
    (bayen@berkeley.edu) is the Associate Provost for Moffett Field Program Development at UC Berkeley, and the Liao-Cho Professor of Engineering at UC Berkeley. He is a Professor of Electrical Engineering and Computer Science and of Civil and Environmental Engineering (courtesy). He is a visiting Professor at Google. He is also a Faculty Scientist in Mechanical Engineering, at the Lawrence Berkeley National Laboratory (LBNL). From 2014 - 2021, he served as the Director of the Institute of Transportation Studies at UC Berkeley (ITS). 
\end{IEEEbiography}

\bibliographystyle{IEEEtran}
\bibliography{references,CIRCLES-key-papers}


\endarticle

\end{document}